\documentclass[11pt,preprint]{aastex}

\newcommand{\ldl}{$\lambda/{\Delta}{\lambda}$}

\newcommand{\teff}{T$_{eff}$}
\newcommand{\lbol}{$\log{{\rm L}_{bol}/{\rm L}_{\sun}}$}

\newcommand{\meth}{CH$_4$}
\newcommand{\wat}{H$_2$O}

\newcommand{\kms}{km~s$^{-1}$}

\newcommand{\chisq}{$\chi^2$}
\newcommand{\vtan}{$V_{tan}$}

\slugcomment{Accepted for publiction to ApJ}

\shorttitle{Unresolved L/T Binaries I}
\shortauthors{Burgasser et al.}

\begin{document}

\title{SpeX Spectroscopy of Unresolved Very Low-Mass Binaries. I. Identification of Seventeen Candidate Binaries Straddling the L Dwarf/T Dwarf Transition}

\author{Adam J.\ Burgasser \altaffilmark{1,2,3},
Kelle L.\ Cruz \altaffilmark{3,4},
Michael Cushing\altaffilmark{5},
Christopher R.\ Gelino\altaffilmark{6},
Dagny L.\ Looper\altaffilmark{3,5},
Jacqueline K.\ Faherty\altaffilmark{7},
J.\ Davy Kirkpatrick\altaffilmark{3,6},
and
I.\ Neill Reid\altaffilmark{3,8}}

\altaffiltext{1}{Center for Astrophysics and Space Science, University of California San Diego, La Jolla, CA 92093, USA; aburgasser@ucsd.edu}
\altaffiltext{2}{Massachusetts Institute of Technology, Kavli Institute for Astrophysics and Space Research,
Building 37, Room 664B, 77 Massachusetts Avenue, Cambridge, MA 02139, USA}
\altaffiltext{2}{Visiting Astronomer at the Infrared Telescope Facility, which is operated by the University of Hawaii under Cooperative Agreement no. NCC 5-538 with the National Aeronautics and Space Administration, Science Mission Directorate, Planetary Astronomy Program.}
\altaffiltext{4}{Department of Physics and Astronomy, Hunter College, City University of New York, 695 Park Avenue, New York, NY 10021, USA}
\altaffiltext{5}{Institute for Astronomy, University of Hawaii, 2680 Woodlawn Drive, Honolulu, HI 96822, USA}
\altaffiltext{6}{Infrared Processing and Analysis Center, California Institute of Technology, Pasadena, CA 91125, USA}
\altaffiltext{7}{Department of Astrophysics, American Museum of Natural History, Central Park West at 79th Street, New York, NY 10034, USA ; Department of Physics and Astronomy, Stony Brook University, Stony Brook, NY 11794-3800, USA}
\altaffiltext{8}{Space Telescope Science Institute, 3700 San Marin Drive, Baltimore, MD 21218, USA}

\begin{abstract}
We report the identification of 17 candidate brown dwarf
binaries whose components straddle the L dwarf/T dwarf transition.  
These sources were culled from a large near-infrared spectral sample of
L and T dwarfs observed with the Infrared Telescope Facility SpeX spectrograph.
Candidates were selected on the basis of spectral ratios which segregate
known (resolved) L dwarf/T dwarf pairs from presumably single sources.  
Composite templates, constructed by combining 13581 pairs of 
absolute flux-calibrated spectra, are shown to 
provide statistically superior fits to the 
spectra of our seventeen candidates as compared to single templates.  Ten
of these candidates appear to have secondary components that are significantly brighter
than their primaries over the 1.0--1.3~$\micron$ band, indicative of rapid
condensate depletion at the L dwarf/T dwarf transition.  
Our results support prior indications of enhanced multiplicity amongst early-type
T dwarfs; 53$\pm$7\% of the T0--T4 dwarfs in our spectral sample are found to be either resolved or unresolved (candidate) pairs, although
this is consistent with an intrinsic (volume complete) brown dwarf binary fraction of only 15\%.
If verified, this sample of spectral binaries more than doubles the number of known L dwarf/T dwarf transition
pairs, enabling a broader exploration of this poorly-understood phase of brown dwarf atmospheric evolution.
\end{abstract}


\section{Introduction}

Among the outstanding problems in brown dwarf astrophysics today
is the depletion of photospheric condensate clouds at the transition between the 
two lowest-luminosity spectral classes of low-mass stars and
brown dwarfs, the L dwarfs and T dwarfs (\citealt{2005ARA&A..43..195K} and references therein).
The L dwarfs, with effective temperatures of 1300 $\lesssim$ {\teff} $\lesssim$ 2200~K \citep{2004AJ....127.3516G}, host a variety of mineral condensate species in their photospheres \citep{2002ApJ...577..974L}, as 
evident by their very red near-infrared colors, muted {\wat} absorption bands, weak
TiO and VO bands (gaseous precursors to condensates)
and presence of silicate grain features (e.g., \citealt{1996A&A...305L...1T, 1999ApJ...519..802K, 2004AJ....127.3553K,2006ApJ...648..614C}).
In contrast, the cooler T dwarfs (600 $\lesssim$ {\teff} $\lesssim$ 1300~K), which have 
blue near-infrared colors and strong molecular gas absorption, appear to have relatively
condensate-free atmospheres \citep{1996Sci...272.1919M, 1999ApJ...520L.119T}. 

One-dimensional, static cloud models have been able to explain these trends
through the confinement of condensates which are vertically bound by a balance of gravitational settling and vertical mixing
(e.g., \citealt{2001ApJ...556..872A,2002ApJ...575..264T,2006ApJ...640.1063B}).  In these models,
the cloud layer resides at the photosphere in the L dwarfs, but sinks below the photosphere and out of view in the T dwarfs.
While providing a robust, conceptual framework for understanding cloud evolution
at the L dwarf/T dwarf transition, these models have nevertheless failed to explain the surprisingly narrow
range of effective temperatures ($\Delta${\teff} $\approx$ 200--400~K) and luminosities ($\Delta${\lbol} $\approx$ 0.3~dex) that encompass this transition \citep{2000AJ....120..447K,2004AJ....127.3516G,2007ApJ...659..655B}.  
More remarkably, surface fluxes in the 1.0--1.3~$\micron$ region are observed to {\em increase} amongst early-type T dwarfs, the so-called ``$J$-band bump'' \citep{2002AJ....124.1170D,2003AJ....126..975T}.  
Careful tuning of cloud parameters (e.g, settling efficiency, cloud coverage or cloud top temperature)
can reproduce the characteristics of the L dwarf/T dwarf transition, but in an arguably ad-hoc manner (e.g., \citealt{2002ApJ...571L.151B, 2005ApJ...621.1033T,2008ApJ...678.1372C}).  

Binaries with L dwarf and T dwarf components have helped to clarify this
transition by providing coeval, co-spatial
laboratories for detailed atmospheric investigations.
Resolved photometry for a handful of these systems 
\citep{2006ApJS..166..585B,2006ApJ...647.1393L,2008ApJ...686..528L}
have confirmed the $J$-band bump to be an intrinsic aspect of brown dwarf evolution, rather than variations in surface gravity, metallicity
or unresolved multiplicity in heterogeneous samples \citep{2003ApJ...585L.151T,2006ApJ...640.1063B}.  
L/T binaries have also enabled better characterization of absolute magnitude/spectral type 
relations across the L dwarf/T dwarf transition, confirming the narrow luminosity range
over which this transition occurs.  Interpreted as a short timescale for 
the removal of condensates from brown dwarf photospheres
($\sim$500~Myr; \citealt{2007ApJ...659..655B}), this result also explains 
the relative rarity of early-type
T dwarfs \citep{2008ApJ...676.1281M} and
their high rate of multiplicity (nearly twice that of other brown dwarfs in magnitude-limited
samples; \citealt{2006ApJS..166..585B, 2006ApJ...647.1393L}).

Two factors conspire to hamper resolved imaging studies of L dwarf/T dwarf pairs:
their low space density and corresponding greater distances from the Sun,
and the inherently small physical separations of brown dwarf multiples in general 
($\sim$98\% have $a \lesssim 20$~AU; e.g., \citealt{2007ApJ...668..492A}).
Fortunately, the distinct and highly-structured near-infrared spectral energy distributions of L and T dwarfs makes it possible to both identify and characterize {\em unresolved} pairs from combined-light
near-infrared spectroscopy, so-called spectral binaries.
Indeed, one of the first L/T binaries found, 
2MASS~J0518-2828\footnote{Names of sources listed in the text are shortened to 
[PREFIX]~Jhhmm$\pm$ddmm, where [PREFIX] is the original discovery catalog:  2MASS (Two Micron All Sky Survey: \citealt{2006AJ....131.1163S}), SDSS (Sloan Digital Sky Survey: \citealt{2000AJ....120.1579Y}), DENIS (Deep Near-Infrared Survey of the Southern Sky: \citealt{1997Msngr..87...27E})
or IPMS (Infrared Proper Motion Survey; \citealt{2006ApJ...651L..57A}), and hhmm and $\pm$ddmm are the sexigesimal Right Ascension (hours and minutes) and declination (degrees and arcminutes) at equinox J2000.  Full source names and coordinates are given in Tables~\ref{tab_templates},~\ref{tab_known} and~\ref{tab_candidates}; additional source data can be found at the DwarfArchives site, \url{http://dwarfarchives.org}.}, was originally identified by its peculiar near-infrared spectrum \citep{2004ApJ...604L..61C}, and subsequently confirmed via high angular resolution imaging with the {\em Hubble Space Telescope} ({\em HST}; \citealt{2006ApJS..166..585B}).  
A more recent spectral binary discovery, 2MASS~J0320-0446 \citep{2008ApJ...681..579B}, was independently identified as an
M8.5 + T5 spectroscopic binary with an orbital period of 0.67~yr  \citep{2008ApJ...678L.125B}.
This case illustrates the increased potential of finding very closely-separated systems among unresolved binaries, systems that are more amenable to astrometric and/or spectroscopic orbital mass measurements that provide critical tests of brown dwarf evolutionary theories
(e.g., \citealt{2004ApJ...615..958Z,2008ApJ...689..436L,2009ApJ...692..729D}).  Tight binaries also have a higher probability of eclipsing (e.g., \citealt{2006Natur.440..311S}) and hence the potential to yield radius measurements to constrain as-yet untested structural models of evolved brown dwarfs (i.e., ages $\gtrsim$ 1~Gyr; \citealt{2009AIPC.1094..102C}).   Unresolved pairs can be identified at greater distances, enabling the construction of larger and more statistically robust samples; and provide a means of measuring the true binary fraction of brown dwarfs currently uncertain by a factor of 2-3 due to separation selection biases (e.g., \citealt{2007MNRAS.380..712L, 2008A&A...492..545J}).

In this article, we report the identification of seventeen unresolved, L dwarf/T dwarf binary candidates selected from a large sample of low-resolution near-infrared spectra of late-type dwarfs.  In Section~2 we
summarize the construction and basic characteristics of the sample, including new observations
of L and T dwarfs.
In Section~3 we describe our method of selecting L dwarf/T dwarf binary candidates through the use of spectral indices.
In Section~4 we detail our spectral template analysis technique, define metrics to assess the probability
of unresolved multiplicity, and determine component parameters.  
In Section~5 we present the results of our spectral analysis, examining each candidate in detail.  
In Section~6 we discuss our results in the context of the absolute magnitude relations for L and T dwarfs,
the frequency of multiples across the L dwarf/T dwarf transition, and 
the process of cloud depletion at this transition.  
The major conclusions of this study are summarized in Section~7.
As Sections 2 through 4 are devoted to our selection and analysis techniques, readers
interested primarily in the binary candidates can skip to Section~5 without loss of 
context.

\section{SpeX Spectral Sample}

Our analysis is based on a large sample of low-resolution, near-infrared spectra of L and T dwarfs 
obtained with SpeX \citep{2003PASP..115..362R},
a low-resolution, 0.8--5.4~$\micron$ spectrograph
mounted on the 3m NASA Infrared Telescope Facility.
The spectra were obtained using the prism-dispersed
mode of SpeX, which provides 0.75--2.5~$\micron$ continuous wavelength coverage
in a single order, with dispersion of 20--30~{\AA}~pixel$^{-1}$ and an average resolution 
{\ldl} $\approx$ 120 for the 0$\farcs$5 slit (some sources were also observed with the 0$\farcs$8 slit).   
The sample is neither complete nor volume-limited; 
it constitutes a representative subset of ultracool dwarfs in the 
local disk population, based on ongoing
SpeX spectroscopic follow-up by ourselves and other researchers
of all known L and T dwarfs down to an effective magnitude limit of $K$ $\approx$ 16,
including nearly all known L and T dwarfs within 20 parsecs of the Sun.  
Most of these sources observed are 
compiled on the Dwarf Archives site\footnote{\url{http://dwarfarchives.org}.}; the SpeX spectra
examined here are available at the SpeX Prism Spectral Libraries\footnote{\url{http://www.browndwarfs.org/spexprism}.}.

\subsection{New Observations}

Ninety-nine L and T dwarfs were observed by the authors over several runs spanning 2004 September through 2008 September, as summarized
in Table~\ref{tab_new}.  These data were acquired in a variety
of weather (clear to partly cloudy) and seeing conditions (0$\farcs$3--1$\farcs$2), but were
uniformly observed using the 0$\farcs$5 or 0$\farcs$8 slit aligned at parallactic angle and generally
at low airmass (sec z $\lesssim$ 2).  Each source was observed in multiple exposures, dithering 
in an ABBA pattern along the slit, with total integration times of 240--1200~s depending on source brightness and weather conditions.
Nearby A0~V stars were observed for flux calibration and telluric absorption correction,
and internal flat field and argon arc lamps were observed with each flux standard
for pixel response and wavelength calibration.  All data were reduced 
using the SpeXtool reduction package
\citep{2003PASP..115..389V,2004PASP..116..362C} using standard settings, as described in detail in \citet{2007AJ....134.1330B}.

\subsection{Data from the Literature}

In addition to our new observations, we compiled from the literature SpeX observations 
of 134 L and T dwarfs with published optical and/or near-infrared spectral types spanning
L3 to T8 and with data having median signal-to-noise ratios 
of 20 or greater over the 0.9--2.4~$\micron$ window. Those data\footnote{Data were drawn from \citet{2004AJ....127.2856B,2006ApJ...639.1095B, 2006ApJ...637.1067B,2007ApJ...658..557B,2008ApJ...681..579B,2008ApJ...674..451B,2004ApJ...604L..61C,2006AJ....131.1007B,2006AJ....131.2722C,2006AJ....132.2074M,2006ApJ...639.1114R,2007ApJ...659..655B,2007AJ....134.1330B,2007ApJ...658..617B,2007ApJ...655..522L,2007AJ....134.1162L,2008ApJ...686..528L,2007ApJ...654..570L,2007AJ....133.2320S} and \citet{2009AJ....137..304S}.}  were
acquired and reduced following similar procedures as described above.
Combining the new and previous observations results in a sample of 253 spectra for 233 L3--T8 dwarfs. 

\subsection{Initial Template Sample}

In order to identify and characterize unresolved binaries, we required a ``clean'' sample of 
spectral templates.  An initial sample was constructed by excluding known binaries
(\citealt{2007prpl.conf..427B} and references therein)
young cluster objects (e.g., \citealt{2007AJ....134..411M}), and sources specifically
noted to have peculiar spectra associated with low surface gravities, subsolar metallicites or unusual cloud properties (e.g., \citealt{2000MNRAS.314..858L, 2003ApJ...592.1186B,2004AJ....127.3553K, 2006AJ....131.2722C,2007AJ....133..439C,2009AJ....137.3345C,2008ApJ...686..528L}).
This left us with a sample of 189 spectra of 178 sources, whose properties are
summarized in Table~\ref{tab_templates}. 

Figure~\ref{fig_sptdist} shows the distribution of published spectral types for these spectra, based on optical data for L3--L8 dwarfs or near-infrared data for L9--T8 dwarfs and those L dwarfs without optical classifications.  The distribution is relatively flat, albeit with fewer T0--T3 dwarfs and T8 dwarfs reflecting the rarity of these sources in current search samples.
The literature classifications are based primarily on the \citet{1999ApJ...519..802K}, \citet{2002ApJ...564..466G} or \citet{2006ApJ...637.1067B} schemes.  However, many L dwarfs have never been typed in the near-infrared, and those with near-infrared types have been classified following different schemes (e.g., \citealt{2001AJ....121.1710R,2001ApJ...552L.147T,2003ApJ...596..561M,2004ApJ...607..499N,2007ApJ...659..655B}).  
To place the sample on a self-consistent, homogeneous basis, we calculated near-infrared spectral types for each source directly from the SpeX data, using the spectral indices
{\wat}-J, {\meth}-J, {\wat}-H, {\meth}-H, {\wat}-K and {\meth}-K defined in 
\citet{2006ApJ...637.1067B} and index/spectral types relations given in \citet{2007ApJ...659..655B}. As in the latter study, the set of indices used to classify each source is selected based on an initial (in this case, published) spectral type estimate. The {\wat}-J, {\wat}-H and {\wat}-K indices are combined for types L8 and earlier and the {\wat}-J, {\meth}-J, {\wat}-H, {\meth}-H and {\meth}-K indices are used for later types.   As there can be significant differences between published and near-infrared spectral types, particularly for the L dwarfs (e.g., \citealt{2002ApJ...564..466G}), we repeated the classification process three times for each source, using each iteration's classification to determine the indices used for the next iteration.  The spectral type adopted for each source is the mean of its individual index types.

Table~\ref{tab_templates} lists the index-based types, with uncertainties (standard deviation of index types) given for those sources with considerable scatter (default classification uncertainty is $\pm$0.5 subtypes). Figure~\ref{fig_sptdist} shows the distribution of these classifications, which is not significantly different from the literature distribution.  Figure~\ref{fig_compspt} compares the SpeX and literature classifications against each other.  Over the entire template sample, the standard deviation between classifications is 0.9 subtypes, although average deviations are generally larger among the L dwarfs
($\sigma$ = 1.1 subtypes) than the T dwarfs ($\sigma$ = 0.5 subtypes).  In most cases, the discrepancies are between optical and near-infrared types for L dwarfs. Only eleven sources had SpeX classifications that differed by more than two subtypes from the published classifications,
one of which is a candidate binary (2MASS~J1711+2232; see Section~3.2).   For completeness, we used both published and index-based spectral types in our analysis.
 
\section{Identification of Binary Candidates}

\subsection{Spectral Peculiarities in Known Binary Systems}

Prior identification of unresolved L dwarf/T dwarf binaries from SpeX data 
has been based largely 
on the presence of peculiar spectral features, such as the 
sharp dip at 1.6~$\micron$ observed in the spectra of 2MASS~J0320-0446, SDSS~J0805+4812 
\citep{2007AJ....134.1330B} and Kelu-1A \citep{2008arXiv0811.0556S} 
attributed to overlapping FeH 
and {\meth} absorption features from unresolved L dwarf and T dwarf components.
However, such specific features need not arise in all L/T binary combinations;
see for example Figure~4 in \citet{2007ApJ...659..655B}.  

In an effort to make the selection process more robust, we examined the near-infrared spectra of six resolved L/T  binaries (Table~\ref{tab_known}): 
2MASS~J0518-2828,
SDSS~J0423-0414 \citep{2005ApJ...634L.177B}, 
SDSS~J1021-0304 \citep{2006ApJS..166..585B}, 
2MASS~J1404+3159 \citep{2008ApJ...685.1183L},
SDSS~J1534+1615 \citep{2006ApJ...647.1393L},
and 2MASS~J2252-1730 \citep{2006ApJ...639.1114R}.
Figure~\ref{fig_known} compares the 
combined light spectrum of each of these systems
to its closest match in our template library (see Section~4.2). 
Spectral deviations between the binaries and templates range from pronounced (in the case of 2MASS~J0518-2828) to subtle (in the case of SDSS~J1021-0304), but several common trends stand out:
\begin{enumerate}
\item The 1.6~$\micron$ {\meth} absorption band is typically stronger relative to 2.2~$\micron$ band
in the binaries, the former occasionally appearing in the absence of the latter (e.g., 2MASS~J0518-2828);
\item The 1.27~$\micron$ flux peak is more pronounced and {\wat} and {\meth} absorption deeper at 1.1~$\micron$ in the binaries (e.g., SDSS~J1534+1615);
\item The 2.1~$\micron$ flux peak is slightly shifted toward the blue in the binaries (e.g., SDSS~J1021-0304); and
\item The overall near-infrared spectral energy distribution is generally bluer in the binaries (e.g., 2MASS~J2252-1730).
\end{enumerate}

These traits suggest qualitative means of differentiating unresolved binaries from single sources, and systems such as 2MASS~J0518-2828 readily stand out in direct spectral comparisons.  However, comparative methods implicitly assume that the templates themselves are ``normal'' single spectra, an assumption we cannot make given that we are searching for unresolved binaries within the same sample. To overcome this ambiguity, we used a suite of spectral indices to identify sources whose overall spectral properties are similar to known binaries and therefore indicative of  unresolved multiplicity.

\subsection{Spectral Index Selection}

Eight spectral indices, defined in Table~\ref{tab_indices}, were measured for each spectrum in our full spectral sample (templates and known binaries/peculiar sources). These include the six
classification indices used in Section~2.2, the $K/J$ flux peak ratio
defined in \citet{2006ApJ...637.1067B}, and an additional
index sampling the 1.6~$\micron$ feature noted above.  We compared all eight indices against each other for every source in our sample, and examined trends in individual indices and index ratios as a function of spectral type using both literature and index-based classifications.

From these combinations, we identified
six pairings that best segregated known L dwarf/T dwarf binaries from the bulk of the spectral sample,
as shown in Figure~\ref{fig_indices}.  These pairings reflect the peculiar traits outlined above, in particular 
the strength of the $H$-band {\meth} feature ({\meth}-K versus {\meth}-H and H-dip versus {\wat}-H),
the unusual brightness of the $J$-band peak ({\wat}-K versus {\wat}-J, {\wat}-J/{\wat}-H and {\wat}-J/{\meth}-K versus spectral type), 
and blue near-infrared colors (K/J versus {\meth}-H).  In these comparison spaces, we identified regions (Table~\ref{tab_criteria}) which clearly segregated known binaries, but were conservative enough so as not to over-select unusual spectra that presumably arise from non-multiplicity effects (e.g., variations in surface gravity, metallicity or cloud properties).  The criteria also
allowed us to retain enough 
non-binary-like late-L/early-T dwarf spectra for our spectral
template analysis (Section~4).  
These criteria clearly provide an incomplete selection of unresolved binaries, as they do
not select several known binary systems with either weak signatures of 
{\meth} absorption (e.g., 2MASS J0850+1057 and 2MASS J1728+3948; \citealt{2001AJ....121..489R, 2003AJ....125.3302G}) or systems with 
identical components (e.g., SDSS~J0926+584; \citealt{2006ApJS..166..585B}).
However, as a first attempt at identifying
unresolved pairs, we chose to maximize the reliability of our candidate selection over sample completeness.

In all, 46 sources from our full SpeX sample of 253 spectra satisfied at least one selection criterion.  These include nine known binaries and all six sources
listed in Table~\ref{tab_known}.
We refined our selection by requiring binary candidates to satisfy at least two criteria.  We also segregated ``strong'' candidates (satisfying at least three criteria) from ``weak'' candidates (satisfying only two criteria), a distinction intended to test the robustness of the spectral index criteria.  These 20 sources are listed in Table~\ref{tab_candidates}, and constitute our initial candidate pool.  All have near-infrared classifications from SpeX data spanning L8.5--T3; several have been previously noted in the literature as having peculiar or highly uncertain spectral types.
We discuss each candidate in detail in Section~5.

\section{Spectral Template Analysis}

\subsection{Flux Calibration}

The final test of the binary nature of our candidates and characterization of their components is a comparison of their spectra against synthetic composites generated from the template spectra. The templates, purged of binary candidates (reducing the sample to 170 spectra of 161 sources), were interpolated onto a common wavelength scale and set on an absolute flux scale (in $F_{\lambda}$ flux units) using the MKO\footnote{Mauna Kea Observatory filter system; see \citet{2002PASP..114..180T} and \citet{2002PASP..114..169S}.} $M_K$/spectral type relations of \citet{2006ApJ...647.1393L}.  We considered the two $M_K$ relations defined in that study, one constructed by rejecting known binaries (hereafter, the ``bright'' relation) and one
constructed from rejecting
known and candidate binaries (hereafter, the ``faint'' relation; see Figure~5 in \citealt{2006ApJ...647.1393L}).
These two relations encompass our current
best constraints on absolute magnitude trends across the L dwarf/T dwarf transition, despite diverging by nearly $\sim$1~mag
over spectral types L8--T5.  Absolute magnitudes were assigned according to either literature or index-based spectral types,
resulting in four independent flux scalings for the templates.   As our baseline calibration set, we adopted the faint calibration applied to literature classifications (see Section~6.1).

\subsection{Comparison to Single and Composites}

For each candidate, we first determined best matches to individual template spectra following a procedure similar to that outlined in \citet{2008ApJ...681..579B}.
All spectra were initially normalized to the 
maximum flux in the 1.2--1.3~$\micron$ region.  We then computed a weighted chi-square
statistic between each candidate ($C[\lambda]$) and 
template ($T[{\lambda}]$) spectrum:
\begin{equation}
\chi^2 \equiv \sum_{\{ \lambda\} }w[{\lambda}]\left[ \frac{C[{\lambda}]-{\alpha}T[{\lambda}]}{\sigma_c[{\lambda}]} \right]^2
\label{equ_chisq}
\end{equation}
(see \citealt{2008ApJ...678.1372C}).
Here, $w[{\lambda}]$ is a vector of weights satisfying $\sum_{\{ \lambda\} }{w[{\lambda}]} = 1$, 
$\alpha$ is a scaling factor that minimizes {\chisq} (see Equation~2 in \citealt{2008ApJ...678.1372C}),
$\sigma_{c}[{\lambda}]$ is the noise spectrum for the candidate,
and the sum is performed over the wavelength ranges $\{\lambda\}$ =
0.95--1.35~$\micron$, 1.45--1.8~$\micron$ and 2.0--2.35~$\micron$ in order
to avoid regions of strong telluric absorption.  
We adopted the same weighting scheme used in \citet{2008ApJ...678.1372C}, with each
pixel weighted by its spectral width (i.e., $w_i \propto \Delta\lambda_i$).

The candidate spectra were then compared to a larger set of synthetic composites, constructed by combining all possible pairs of flux-calibrated templates for which one source was of equal or later spectral type (using either literature or index-based classifications).  The resulting 13581 composites were also normalized and compared to the candidate spectra using the same {\chisq} statistic as Equation~\ref{equ_chisq}.  
We note that because we are comparing the candidate spectra against a finite sample of spectra of distinct sources, each of which are modulated by some degree of noise, our expectation is not to achieve {\chisq} $\approx$ 1 for our best-fit cases; indeed, this expectation is not even realized in spectral model fits (see Figure 4 in \citealt{2008ApJ...678.1372C}).  What matters is whether a binary template provides a significantly improved fit, as discussed below.

\subsection{Assessing Fit Quality and Parameter Estimation}

The number of composites used to fit the candidate spectra vastly outnumbers the number of single templates, so we are almost assured of finding a better fit (lower {\chisq}) with the former.  It is therefore necessary to assess the statistical significance of the fit improvement in order to rule out the null hypothesis; i.e., that the candidate is not a binary.  We used the one-sided F-test for this purpose, using as our distribution statistic the ratio 
\begin{equation}
\eta_{SB} \equiv \frac{{\rm min}(\{\chi^2_{single}\})/\nu_{single}}{{\rm min}(\{\chi^2_{composite}\})/\nu_{binary}} 
\label{equ_eta}
\end{equation}
where $\nu$ is the degrees of freedom in each fit.  For both single and composite fits, one might initially assume that $\nu_{single}$ = $\nu_{composite}$ = $\nu$ is the number of data points used in the fit ($N = 296$), minus one to account for the relative scaling between data and template spectra.  However, individual weighting of the spectral points implies that they do not contribute equally to the total {\chisq} value. We therefore define the effective number of data points,
\begin{equation}
N_{eff} \equiv \frac{1}{{\rm max}(\{w\})}\sum_{i=1}^Nw_i
\label{equ_nu}
\end{equation}
which reduces to $N$ for $w_i$ = constant.  
For our weighting scheme, $N_{eff}$ = 254 and hence $\nu$ = 253.  
To rule out our null hypothesis at the 99\% confidence level (CL), we required $\eta_{SB} > 1.34$ as our final selection criteria.  

Typically, multiple single or composite templates yield similar {\chisq} values for a given candidate.  We therefore computed mean values and uncertainties for the component parameters (i.e., spectral types and relative brightnesses) using a weighting scheme based on the F-distribution.   In effect, each fit's parameter set was weighted by the likelihood that that fit is equivalent to the best fit:
\begin{equation}
W_i \propto 1-F(\eta_{i0} \mid \nu,\nu).
\end{equation}
Here, $\eta_{i0}$ $\equiv$ {\chisq}$_i$/min([{\chisq]) is the ratio of chi-square residuals between the best-fit template and the $i^{th}$ template, and $F(\eta_{i0} \mid \nu,\nu)$ is the F-distribution probability distribution function.  Parameter means ($\bar{p}$) and uncertainties ($\sigma_p$) were then computed as
\begin{equation}
\bar{p} \equiv \frac{\sum_iW_ip_i}{\sum_iW_i}
\label{equ_meanp}
\end{equation}
and
\begin{equation}
\sigma_p^2 = \frac{\sum_iW_i(p_i-\bar{p})^2}{\sum_iW_i},
\label{equ_sigmap}
\end{equation}
where the sums are over all fits.

\subsection{Known L/T Transition Binaries}

Before proceeding to examine our binary candidates, we first assessed how well
our index selection and spectral fitting procedures identified and
reproduced the properties of the known binary systems listed in Table~\ref{tab_known}.  The best-fit templates for these sources are shown in Figure~\ref{fig_knownbfit}, while 
Table~\ref{fig_knownbfit} summarizes the component parameters based on the faint calibration scale.  For each system, best-fit composites provided a statistically significant better match (CL $>$ 99\%) to the combined light spectrum than the best-fit single template.  Visual inspection of these fits also indicates clear improvement in the cases of 2MASS~J2252-1730, SDSS~J0423-0414, 2MASS~J0518-2828 and 2MASS~J1404+3159.  The fits for SDSS~J1021-0304 and SDSS~J1534+1615 are more subtle improvements, but notably reproduce the blue-shifted peak at $K$-band in the spectrum of the former and the strong 1.1~$\micron$ {\wat}/{\meth} band in the spectrum of the latter.  

Resolved photometric measurements have been made for these systems, so we examined how well our technique reproduces reported relative magnitudes.  For the composites, synthetic MKO $JHK$ and {\em HST}/NICMOS $F110W$ and $F170M$ photometry were computed directly from the flux-calibrated SpeX data by integrating these and a Kurucz model spectrum of Vega with the respective filter profiles (see \citealt{2005ApJ...623.1115C}).
Mean relative magnitudes and their uncertainties were calculated as in Equations~\ref{equ_meanp} and~\ref{equ_sigmap}, and are also listed in Table~\ref{tab_knownbfit}.  In all six cases, synthetic photometry is consistent with measurements to within $2\sigma$, although uncertainties in the former are as high as 0.8~mag.  For the two sources with the most reliable photometry, SDSS~J0423-0414 and 2MASS~J1404+3159, the agreement is within 1.5$\sigma$. 

Finally, while none of the known binaries listed in Table~\ref{tab_known} have reported resolved spectroscopy, the inferred component types are generally consistent with 
estimated types from the literature ($\pm$1 subtype) with the exceptions of  2MASS~J0518-2828, for which we infer later-type components (L7.5+T5 versus L6:+T4:), and DENIS~J2252-1730, for which we infer an earlier primary and later secondary (L4.5+T4.5 versus L6+T2).  However, reported types for the former are highly uncertain \citep{2004ApJ...604L..61C,2006ApJS..166..585B} while our spectral fit to the latter is superior to that presented in \citet{2006ApJ...639.1114R}. 

In summary, we find that our method is capable of robustly identifying and characterizing known L dwarf/T dwarf transition pairs, providing a measure of confidence in our candidate binary selection and component characterization.

\section{Results}

Single and composite template fits for each of the candidate binaries and for all four component flux calibrations are summarized in Tables~\ref{tab_fitsstrong} and~\ref{tab_fitsweak}.  These include best-fit and mean component types and relative MKO $JHK$ and {\em HST}/NICMOS $F110W$ and $F170M$ magnitudes.  The best fits based on the faint flux calibration are shown in Figures~\ref{fig_fitsstrong} and~\ref{fig_fitsweak}.  In the following sections we discuss each candidate in detail.

\subsection{Strong Candidates}

\subsubsection{SDSS~J024749.90-163112.6} 

Originally identified in the SDSS by \citet{2006AJ....131.2722C}, SDSS~J0247-1631 is classified T2$\pm$1.5 in the near-infrared, consistent with our index-based classification of T2.5$\pm$1.  The uncertainty in both reflects large scatter in spectral indices, with stronger absorption in the 1.1~$\micron$ {\wat}/{\meth} band relative to 1.6 and 2.2~$\micron$ {\meth} bands.  SDSS~J0247-1631 satisfied all six spectral index selection criteria, and the best-fit composite is a clear improvement over the single template fit and statistically significant (CL $>$99\%). Average primary and secondary spectral types are inferred to be T0$\pm$0.2 and T7$\pm$0.3, the latter being the latest-typed secondary in our sample. 
The composition of this system indicates a secondary that is fainter than the primary across the near-infrared band ($\Delta{J}$ = 0.68$\pm$0.10).  The primary component of the best-fit composite, SDSS~J1520+3546, has discrepant literature (T0$\pm$1) and index-based (L7.5)
near-infrared classifications, due in part to weak {\wat} absorption 
at 1.1~$\micron$ and 1.3~$\micron$; and its near-infrared color (2MASS $J-K_s$ = 1.54$\pm$0.08) is somewhat blue for either type.  \citet{2009AJ....137....1F} have found that unusually blue L dwarfs
have a large tangential velocity ({\vtan}) distribution, indicative of older ages, higher surface gravities and possibly subsolar metallicities.  SDSS~J0247-1631 itself has a relatively large {\vtan} = 46$\pm$7~{\kms}, compared to a mean of 26$\pm$19~{\kms} for L0--L9 dwarfs \citep{2009AJ....137....1F}.  These characteristics suggest that SDSS~J0247-1631 may be an older and/or slightly metal-poor pair of brown dwarfs.
No high angular resolution imaging observations of this source have been reported to date.  

\subsubsection{SDSS~J035104.37+481046.8}

\citet{2006AJ....131.2722C} classified SDSS~J0351+4810 a T1$\pm$1 in the near-infrared, similar to our SpeX classification (T1.5$\pm$1).  Again, 
the uncertainties arise from spectral peculiarities, in this case unusually weak {\meth} absorption at 2.2~$\micron$.
SDSS~J0351+4810 satisfied all six selection criteria, and the best-fit composite is a clear improvement over the best-fit single template and statistically significant.  Average primary and secondary spectral types are L6.5$\pm$0.7 and T5$\pm$0.7, with a secondary that may be slightly brighter than the primary at the 1.05~$\micron$ and 1.27~$\micron$ spectral flux peaks ($\Delta{J}$ = 0.31$\pm$0.31).  The primary in the best-fit composite, 2MASS~J0103+1935 (L6, 2MASS $J-K_s$ = 2.14$\pm$0.10), has a fairly steep red spectral slope, indicative of either low surface gravity (i.e., youth) or unusually thick
condensate clouds (e.g., \citealt{2008ApJ...686..528L}).  On the other hand, the best-fit secondary 
has a normal color for its spectral type.  This combination suggests a system with an unusually cloudy primary and normal secondary.  No high angular resolution imaging observations of SDSS~J0351+4810 have been reported to date.

\subsubsection{SDSS~J103931.35+325625.5} 

\citet{2006AJ....131.2722C} classified SDSS~J1039+3256 a T1 in the near-infrared (our SpeX classification is T1.5), and its spectrum  satisfied all six selection criteria.  The best-fit composite is a clear improvement over the best-fit single template and statistically significant.
Average primary and secondary spectral types are L7$\pm$0.2 and T4$\pm$0.2,
and the primary appears to be brighter than the secondary at all wavelengths ($\Delta{J}$ = 0.26$\pm$0.09).
The components of the best-fit composite appear to be fairly normal
for their respective spectral types.
No high angular resolution imaging observations of SDSS~J1039+3256 have been reported to date.

\subsubsection{2MASS~J11061197+2754225}

\citet{2007AJ....134.1162L} classified 2MASS~J1106+2754 a T2.5 in the near-infrared; we classified the same SpeX spectrum T2 using spectral indices.
Satisfying three selection criteria, 2MASS~J1106+2754 is subtly but significantly better fit by composites with average component types of T0$\pm$0.2 and T4.5$\pm$0.2.
These types are similar to those inferred for 2MASS~J1404+3159, and predict a secondary that is substantially brighter at the 1.05~$\micron$ and 1.27~$\micron$ spectral flux peaks ($\Delta{J}$ = -0.37$\pm$0.06).
Again, the best-fit primary is SDSS~J1520+3546, a somewhat peculiar blue late-L/early-T dwarf.

2MASS~J1106+2754 is one of only two strong candidates imaged at high angular resolution, in this case by \citet{2008ApJ...685.1183L} using the Keck AO system.  $K_s$-band images show only a single point source with a point spread function full width at half maximum of 68~mas.  Based on the relative magnitudes inferred from this analysis ($\Delta{K}$ = 1.16$\pm$0.09 for the faint calibration), this observation rules out a resolved binary system with projected separation $\gtrsim$1.5~AU at the time of the observation, assuming a distance of 22~pc \citep{2007AJ....134.1162L}.  
The null imaging result indicates one of three possibilities: 2MASS~J1106+2754 may be a single source, a binary observed close to line-of-sight alignment (e.g., Kelu-1; \citealt{1999Sci...283.1718M,2005ApJ...634..616L,2006PASP..118..611G}), or a tightly-separated system.  The excellent fit of this source's spectrum to composites supports one of the two latter possibilities, and 1.5~AU is not a particularly stringent constraint on the separation given that several brown dwarf multiple systems are known to have even smaller separations (e.g., \citealt{1999AJ....118.2460B,2006Natur.440..311S, 2007ApJ...666L.113J}).  Nevertheless, second-epoch imaging and/or spectroscopic monitoring are required to verify the binary nature of this source.

\subsubsection{2MASS~J13243559+6358284}

Discovered independently by \citet{2007AJ....134.1162L} and \citet{2008ApJ...676.1281M}, 2MASS~J1324+6358 was classified as a peculiar T2 in the former study based on its unusually red near-infrared spectral energy distribution (our SpeX classification is T2$\pm$1).  2MASS~J1324+6358 is also a red outlier in optical, near-infrared 
and mid-infrared colors \citep{2007AJ....134.1162L,2008ApJ...676.1281M,2009AJ....137....1F}.
Looper et al. specifically examined the possibility that this source is an unresolved binary,
finding a good match to an L9 + T2 composite.  We also find this source to be a likely binary, with statistically
significant better fits to composites with mean component types of L8$\pm$0.2 and T3.5$\pm$0.2.
The secondary of this system appears to be
brighter than the primary at the 1.05~$\micron$ and 1.27~$\micron$ flux peaks ($\Delta{J} = -0.05{\pm}0.06$). 
The primary component of the best-fit composite, 2MASS~J1043+2225 (2MASS $J-K_s$ = 1.97$\pm$0.08), has been noted as one of the reddest optically-classified L8 dwarfs known \citep{2007AJ....133..439C}, while the secondary has normal colors for its spectral type.  Like SDSS~J0351+4810, this combination suggests a system with an unusually 
cloudy primary and cloud-free T dwarf secondary.
No high angular resolution imaging observations of 2MASS~J1324+6358 have been reported to date.

\subsubsection{SDSS~J141530.05+572428.7}

\citet{2006AJ....131.2722C} classified SDSS~J1415+5724 a T3$\pm$1 (same as our SpeX classification), with an uncertainty driven by weak 2.2~$\micron$ {\meth} absorption relative to the pronounced 1.6~$\micron$ band.  Both \citet{2006AJ....131.2722C} and \citet{2009AJ....137....1F} note this source as being unusually red for its spectral type, and the former attempted (unsuccessfully) to reproduce its spectrum as a binary. SDSS~J1415+5724 satisfied all six selection criteria, and composites (average types L8$\pm$0.5 and T5$\pm$0.3) provide visually obvious and statistically significant better fits to its spectrum. 
The secondary component appears to be brighter than the primary
at the 1.05~$\micron$ and 1.27~$\micron$ spectral peaks ($\Delta{J} = -0.13{\pm}0.20$), and
both components in the best-fit composite have normal colors and spectral energy distributions for their respective spectral types.
No high angular resolution imaging observations of SDSS~J1415+5724 have been reported to date.

\subsubsection{SDSS~J143553.25+112948.6}

\citet{2006AJ....131.2722C} classified SDSS~J1435+1129 a T2$\pm$1 in the near-infrared (consistent with our T2.5$\pm$1 SpeX classification), with the uncertainty driven by weak 2.2~$\micron$ {\meth} absorption relative to 1.1~$\micron$ and 1.6~$\micron$ bands.  Its spectrum satisfied all six selection criteria, and the best-fit composite is a visually obvious better match than the best-fit single template, reproducing in particular the broad $K$-band peak and relatively strong 1.6~$\micron$ {\meth} absorption.  
The average component types, L7.5$\pm$0.4 and T6$\pm$0.3, indicate a secondary that is slightly fainter than the primary throughout the near-infrared ($\Delta{J} = 0.41{\pm}0.12$).  
The components of the best-fitting composite appear to have normal
colors and spectral energy distributions for their respective spectral types.
No high angular resolution imaging observations of SDSS~J1435+1129 have been reported to date.

\subsubsection{SDSS~J143945.86+304220.6}

\citet{2006AJ....131.2722C} classified SDSS~J1439+3042 a T2.5, identical to our index-based classification.  Its spectrum satisfied only four selection criteria, but the best-fit composite is a statistically significantly (albeit subtle) better match than the best-fit single template,
providing a better fit to the strong 1.6~$\micron$ {\meth} band and
blue-shifted $K$-band peak.
Average component types of T1$\pm$0.2 and T5$\pm$0.6 indicate a secondary
component that may be slightly brighter than the primary at the 1.27~$\micron$ spectral flux peak ($\Delta{J}$ = 0.06$\pm$0.24).
The best-fit binary components have normal colors and spectral energy distributions for their respective types.
No high angular resolution imaging observations of SDSS~J1439+3042 have been reported to date.

\subsubsection{SDSS~J151114.66+060742.9}

\citet{2006AJ....131.2722C} classified SDSS~J1511+0607 a T0$\pm$2 in the near-infrared (our SpeX type is T0.5$\pm$2), with the large uncertainty arising from  substantial scatter in index subtypes (L7--T2) on the \citet{2002ApJ...564..466G} scheme.  Its spectrum is clearly peculiar, with strong {\meth} absorption at 1.1~$\micron$ and 1.6~$\micron$, but very weak absorption at 2.2~$\micron$; these features are similar in nature to the resolved binary 2MASS~J0518-2828.  The spectrum of SDSS~J1511+0607 satisfied all six selection criteria, and the best-fit composite is a visually obvious and statistically significant better match than the best-fit single template.  Average component types, L5.5$\pm$0.8 and T5$\pm$0.4, indicate a secondary
component that is fainter than the primary across the near-infrared band
($\Delta{J}$ = 0.54$\pm$0.32).  
The primary component of the best-fit composite, SDSS~J0624-4521 (L5, 2MASS $J-K_s$ = 1.88$\pm$0.04) is somewhat red for its spectral type, while the secondary
has normal colors.
Hence, this may be another system with an unusually cloudy primary but normal secondary.
No high angular resolution imaging observations of SDSS~J1511+0607 have been reported to date.

\subsubsection{SDSS~J151603.03+025928.9}

\citet{2004AJ....127.3553K} classified SDSS~J1516+0259 a T0$\pm$1.5 on the near-infrared scheme of \citet{2002ApJ...564..466G}; our SpeX classification is L9.5$\pm$1.  The uncertainties in both arise from weak {\meth} absorption at 2.2~$\micron$ compared to the 1.1~$\micron$ and 1.6~$\micron$ bands.  
The spectrum of this source satisfied four of the spectral index criteria, and the best-fit composite is a significantly better match than the best-fit single template, 2MASS~J0328+2302, which may itself be a unresolved binary (see Section~6.1).
The average component types inferred for SDSS~J1516+0259, L7.5$\pm$1.1 and T2.5$\pm$2.2, are rather poorly constrained, as are the relative brightnesses ($\Delta{J}$ = 0.30$\pm$0.65).
Notably, the best-fit composite combines a somewhat red L7 (2MASS~J0318-3421, 2MASS $J-K_s = 2.06{\pm}0.07$) and a fairly normal T2.5,
again suggesting a system with a cloudy primary but normal secondary.
No high angular resolution imaging observations of SDSS~J1516+0259 have been reported to date.

\subsubsection{2MASS~J1711457+223204}

2MASS~J1711+2232 has 
been studied extensively in the literature, originally identified in the 2MASS survey and classified L6.5
in the optical with no detectable lithium absorption or H$\alpha$ emission \citep{2000AJ....120..447K}.  We derive a later but poorly constrained near-infrared classification of L9$\pm$3 based on strong {\wat} absorption and an indication of weak {\meth} absorption at 1.6~$\micron$.  2MASS~J1711+2232 was also examined in the near-infrared by \citet{2004ApJ...607..499N}, who noted the presence of {\meth} at 2.2~$\micron$ but not at 1.6~$\micron$; our SpeX spectrum in fact shows the opposite, although the $K$-band data have somewhat lower signal-to-noise.  \citet{2004ApJ...607..511T} commented on the unusual 1.6~$\micron$ spectral morphology of this source, suggesting that it may be due to absorption from FeH and possibly some other unidentified species.  

That other species is likely to be {\meth} from an unresolved T dwarf companion, as the unusual feature at 1.6~$\micron$ seen in the SpeX spectrum of 2MASS~J1711+2232 (which satisfied all six selection criteria) is similar to those noted in the spectra of 2MASS~J0320-0446, 
SDSS~J0805+4812 and Kelu-1A.
The best-fit composite is a significantly better match than our best-fit single template, reproducing the  1.6~$\micron$ feature in detail.  Mean component types of  L5$\pm$0.4 and T5.5$\pm$1.2 are similar to those inferred for SDSS~J0805+4812 (L4.5$\pm$0.7 and T5$\pm$0.6; \citealt{2007AJ....134.1330B}).
The secondary is considerably fainter than the primary across the near-infrared ($\Delta{J}$ = 0.92$\pm$0.32, $\Delta{K}$ = 3.05$\pm$0.60).

2MASS~J1711+2232 has been imaged at high angular resolution with {\em HST}/WFPC2 \citep{2003AJ....125.3302G} and reported to be unresolved in both F814W ($\lambda_c = 0.79~\micron$) and F1042M ($\lambda_c = 1.02~\micron$) bands.  However, the sensitivity of these observations were insufficient to detect the T dwarf component; the primary/combined source was marginally detected at F1042M, where the much fainter secondary ($\Delta$F1042M $\approx$ 2--3) would have been brightest.  Hence, in the absence of sufficiently sensitive near-infrared {\em HST} or AO observations, we cannot determine conclusively whether this system is a tight (unresolveable) binary).  2MASS~J1711+2232 is the only binary candidate in our sample with a parallactic distance measurement ($d$ = 30$\pm$4~pc; \citealt{2004AJ....127.2948V}).  Its component absolute magnitudes are discussed further in Section~6.1.

\subsubsection{2MASS~J21392676+0220226}

2MASS~J2139+0220 was classified T0 by \citet{2008AJ....136.1290R} based on its red optical spectrum, and T1.5 in the near-infrared by \citet{2006ApJ...637.1067B} based on the SpeX spectrum examined here (indices yield a T2.5$\pm$1 spectral type).  Its near-infrared color is somewhat red for its spectral type: 2MASS $J-K_s$ = 1.68$\pm$0.07 as compared to $\langle{J-K_s}\rangle$ = 1.31 for T1 dwarfs and $\langle{J-K_s}\rangle$ = 1.02 for T2 dwarfs \citep{2009AJ....137....1F}. 
The spectrum of 2MASS~J2139+0220 satisfied only three selection criteria, but this source nevertheless appears to be a strong binary candidate as its best-fit composite is a significant 
improvement over the best-fit single template.
However, the fit is not perfect, failing to reproduce the weak 1.6~$\micron$ {\meth} band relative to strong 1.1~$\micron$ and 2.2~$\micron$
absorptions.  The inferred average component types, L8.5$\pm$0.7 and T3.5$\pm$1.0, are also somewhat poorly constrained.  
Given its unusual pattern of {\meth} features even for an unresolved binary, the components of 2MASS~J2139+0220 may themselves have unusual properties. Its best-fitting primary, 
the L9$\pm$1 SDSS~J0830+4828, has a large tangential velocity ({\vtan} = 79$\pm$4~{\kms}; \citealt{2009AJ....137....1F}) although its absolute brightness and infrared colors appear to be normal \citep{2002ApJ...564..466G,2004AJ....127.2948V,2007ApJ...655.1079L}.
The inferred composition of this system indicates a secondary that is substantially brighter than the primary at the 1.05~$\micron$ and 1.27~$\micron$ spectral peaks ($\Delta{J}$ = -0.14$\pm$0.21).
No high angular resolution imaging observations of 2MASS~J2139+0220 have been reported to date. 

\subsection{Weak Candidates}

\subsubsection{SDSS~J011912.22+240331.6}

\citet{2006AJ....131.2722C} classified SDSS~J0119+2403 a T2 in the near-infrared (similar to our index-based T2.5 classification), and its spectrum exhibits a slightly enhanced $J$-band peak and stronger 1.6~$\micron$ {\meth} absorption relative to the T2 spectral standard, SDSS~J1254-0122.   The best-fit composite is a subtle but significant improvement over the best-fit single template, with 
inferred component types of T0$\pm$0.7 and T4$\pm$0.4, similar to those of the resolved binary 2MASS~J1404-3159. Like that source, the secondary of SDSS~J0119+2403 appears to be substantially brighter than the primary at the 1.05~$\micron$, 1.27~$\micron$ and possibly 1.6~$\micron$ spectral peaks ($\Delta{J}$ = -0.42$\pm$0.19).
The best-fit primary, SDSS~J0858+3256 (T1, MKO $J-K$ = 1.61$\pm$0.04) is noted to be a red outlier for its spectral type with a large tangential velocity ({\vtan} = 66$\pm$3~{\kms}; \citealt{2009AJ....137....1F}), while the best-fit secondary is an unexceptional T4 dwarf.  Hence, this system may again be composed of an unusually cloudy primary and cloud-free secondary.
No high angular resolution imaging observations of SDSS~J0119+2403 have been reported to date.

\subsubsection{SDSS J075840.33+324723.4}

\citet{2004AJ....127.3553K} classified SDSS~J0758+3247 a T2$\pm$1 on the \citet{2002ApJ...564..466G} scheme, with an uncertainty driven
by strong {\wat}/{\meth} absorption at 1.1~$\micron$ (our SpeX classification is T2.5).
This source is in fact a near-clone to SDSS~J1254-0122, and our binary
templates actually provide significantly {\em worse} fits (higher {\chisq} values) than this best-fit single template.  
While SDSS~J1254-0122 also satisfied one of our binary selection criteria,
it is unresolved in {\em HST} imaging observations \citep{2006ApJS..166..585B} so we cannot exclude the possibility that both sources are single brown dwarfs.  
No high angular resolution imaging observations of  SDSS~J0758+3247 have been reported to date.

\subsubsection{SDSS J090900.73+652527.2}

\citet{2006AJ....131.2722C} classified SDSS~J0909+6525 a T1.5 in the near-infrared, identical to our SpeX classification, and its spectrum
is fairly similar to that of the bright T2.5 IPMS~J0136+0933.  The MKO $J-K = 0.62{\pm}0.04$ color of SDSS~J0909+6525 is somewhat blue for its spectral type, and \citet{2009AJ....137....1F} specifically note it as a blue outlier with a normal
tangential velocity ({\vtan} $\approx$ 28~{\kms}).
The best-fit composite is a subtle but significantly better
match to the spectrum of SDSS~J0909+6525, indicating average component types of T1.5$\pm$0.5 and T2.5$\pm$0.3
and a secondary that is somewhat brighter than the primary at the 1.05~$\micron$ and 1.27~$\micron$ spectral peaks ($\Delta{J}$ = -0.12$\pm$0.10).  The best-fitting composite components have normal colors and spectral energy distributions for their respective spectral types.
No high angular resolution imaging observations of SDSS~J0909+6525 have been reported to date.

\subsubsection{2MASS J09490860-1545485}

\citet{2005AJ....130.2326T} classified 2MASS~J0949-1545 a T1$\pm$1 in the near-infrared (our SpeX classification is T1.5), the uncertainty arising from strong 1.6~$\micron$ {\meth} absorption but weak 1.1~$\micron$ {\wat}/{\meth} absorption as compared to spectral standards.  
The spectrum of this source is also similar to that
of IPMS~J0136+0933, but the best-fit composite is a significantly better match 
with components similar to
SDSS~J0909+6525, T1$\pm$0.2 and T2$\pm$0.2.  Again, the secondary is inferred to be somewhat brighter than the primary at the 1.05~$\micron$ and 1.27~$\micron$ spectral peaks ($\Delta{J}$ = -0.07$\pm$0.05) despite having comparable spectral types.  The best-fit composite components are normal for their respective classifications.
No high angular resolution imaging observations of 2MASS~J0949-1545 have been reported to date.

\subsubsection{SDSS J120602.51+281328.7}

\citet{2006AJ....131.2722C} classified SDSS~J1206+2813 a T3 in the near-infrared (identical to our SpeX classification), and it exhibits somewhat blue near-infrared colors for its spectral type, MKO $J-K$ = 0.13$\pm$0.04 versus $\langle{J-K}\rangle$ = 0.56$\pm$0.22 for other T2.5--T3.5 dwarfs \citep{2006AJ....131.2722C}.  Its spectrum is nearly identical to that of the 
T3.5 SDSS~J1750+1759, and while our best-fit composite provides a slight reduction in 
{\chisq}, at 92\% confidence it falls just short of our significance threshold.   It is possible that SDSS~J1206+2813 fails the binary significance criteria because 
SDSS~J1750+1759 is itself an unresolved binary, as suggested by its apparent overluminosity compared to other early-type T dwarfs \citep{2004AJ....127.2948V,2006ApJ...647.1393L}.  However,  SDSS~J1750+1759 was unresolved in {\em HST} images \citep{2006ApJS..166..585B} and it did not satisfy any of our spectral index selection criteria. Hence, we cannot rule out that both sources are single brown dwarfs.  No high angular resolution images of SDSS~J1206+2813 have been reported to date.

\subsubsection{SDSS~J120747.17+024424.8}

\citet{2002AJ....123.3409H} classified SDSS~J1207+0244 an L8 based on red optical data, while \citet{2004AJ....127.3553K} classified it T0 in the near-infrared (identical to our SpeX classification).  The spectrum of SDSS~1207+0244 is very similar to that of the resolved binary SDSS~J0423-0414, and as a consequence it was chosen to replace the latter as the T0 spectral standard
in the classification scheme of \citet{2006ApJ...637.1067B}.  However, it appears that this source may also be a binary system, as the best-fit composite provides a significantly better match to its spectrum.  The mean component types, L6.5$\pm$0.7 and T2.5$\pm$0.5 are similar to those inferred for SDSS~J0423-0414 (Table~\ref{tab_knownbfit}),
the secondary being fainter than the primary throughout the near-infrared
($\Delta{J}$ = 0.48$\pm$0.28).
The components of the best-fitting composite are normal, although the T3
2MASS~J1209-1004 satisfied one of our selection criteria.  The possibility that SDSS~J1207+0244 is a binary system with very different component types is potentially problematic for the currently-defined T dwarf classification scheme. High angular resolution images of this source have yet to be reported.

\subsubsection{SDSS~J151643.01+305344.4}

\citet{2006AJ....131.2722C} classified SDSS~J1516+3053 a T0.5$\pm$1 while our SpeX classification is T1.5$\pm$2. The uncertainties in both arise from this source's unusually red spectral energy distribution and anomalously strong {\wat} absorption bands.  \citet{2007ApJ...655.1079L} also
note this source as being unusually red in mid-infrared colors, and comparison to model colors suggest the presence
of strong vertical mixing and thick condensate clouds in its atmosphere \citep{2001ApJ...556..872A,2008ApJ...689.1327S}.  \citet{2006AJ....131.2722C} attempted
to fit their spectral data to a binary model, but did not find an adequate match.  Our analysis concurs with that result, as the best-fit composite is only a marginal improvement  over the best-fit single template (87\% confidence level), and notably fails to reproduce the strong {\wat}/{\meth} absorption feature at 1.1~$\micron$.   We therefore cannot rule out the possibility that this source is a single brown dwarf with unusual atmospheric properties.
No high angular resolution imaging observations of SDSS~J1516+3053 have been reported to date.

\subsubsection{SDSS~J205235.31-160929.8}

\citet{2006AJ....131.2722C} classified SDSS~J2052-1609 a T1$\pm$1 (our SpeX classification is T0$\pm$1), the uncertainty arising in part from strong absorption at 1.1~$\micron$ as compared to weak {\wat} and {\meth} features at longer wavelengths.  Despite satisfying only two selection criteria, this source appears to be a strong binary candidate, as the best-fit composite is a subtle but significantly better match to the data than the best-fit single template.  Average component types of L7.5$\pm$0.6 and T2$\pm$0.2
are similar to those of SDSS~J0423-0414, although in this case the secondary is inferred to be slightly brighter than its primary at the 1.05~$\micron$ and 1.27~$\micron$ spectral peaks ($\Delta{J}$ = 0.04$\pm$0.18).  The components of the best-fit composite, SDSS~J1155-0559 (L7; 2MASS $J-K_s$ = 1.54$\pm$0.11) and 2MASS~J1122-3512 (T2; 2MASS $J-K_s$ = 0.64$\pm$0.08), both have unusually blue near-infrared colors for their respective spectral types
($\langle{J-K_s}\rangle$ = 1.81 and 1.02 for L7 and T2 dwarfs, respectively; 
\citealt{2009AJ....137....1F}), suggesting that SDSS~J2052-1609 may be an older or slightly metal-poor system.   Indeed, its large proper motion ($\mu$ = 0$\farcs$483$\pm$0$\farcs$022~yr$^{-1}$; J.\ Faherty, in preparation)
implies a fairly large tangential velocity 
({\vtan} = 48$\pm$8~{\kms}), even
without accounting for unresolved multiplicity in the photometric distance estimate (21$\pm$3~pc).
No high angular resolution imaging observations of this source have been reported to date.

\subsubsection{Summary}

Based on our analysis, we find that 17 of the 20 binary candidates identified
by our spectral index selection criteria have a high probability of being
(as yet) unresolved binaries, as indicated by the statistically significant better fits provided by composites.  
These include all of our strong candidates but only 63\% (5/8) of our weak candidates, suggesting that at least three index selection criteria are preferred to identify robust binary candidates.
The inferred properties of these systems 
are summarized in Table~\ref{tab_summary}.  The twelve systems with component types between L7 and T5 more than doubles the number of known binaries whose components
span the L dwarf/T dwarf transition,\footnote{Known or suspected L/T transition binaries include the six sources listed in Table~\ref{tab_known} plus 
2MASS J0850+1057 \citep{2001AJ....121..489R,2002AJ....124.1170D}, 
2MASS J0920+3517 \citep{2001AJ....121..489R,2006ApJ...637.1067B}, 
2MASS J1728+3948 \citep{2003AJ....125.3302G},
$\epsilon$ Indi BC \citep{2004A&A...413.1029M}, 
and Gliese 337CD \citep{2005AJ....129.2849B}.}
although we emphasize that these systems remain candidates pending verification via high angular resolution imaging and/or high resolution spectroscopic monitoring.

\section{Discussion}

\subsection{Absolute Magnitudes Across the L Dwarf/T Dwarf Transition}

As discussed in Section~1, there remains considerable uncertainty in absolute magnitude trends
across the L dwarf/T dwarf transition due to the presence of unresolved multiples, gravity and metallicity
effects, and ultimately the behavior of condensate clouds as they disperse out of the photosphere.  With a large number of candidate binaries in our sample whose components straddle this transition, we considered whether it was possible to use these systems to better constrain these trends.

We first considered the quality of fits provided by the composites for the two flux calibration
relations of \citet{2006ApJ...647.1393L}, which differ by nearly a full magnitude between types L7 and T5.  For the seventeen sources identified as promising binary candidates, we find that for 65\% (11/17) of these the bright relation provides a lower {\chisq} value for the best-fit composite (the same fraction is derived whether literature or index-based classifications are used).  However, for only one case---SDSS~J1207+0244---does the bright relation provide a statistically significant better fit; i.e., CL $>$99\% in the ratio of minimum {\chisq} values between the best-fit composites for the bright and faint relations.  On the other hand, for SDSS~J1039+3256 and 2MASS~J1324+6358 the bright relation provides a statistically significant {\em worse} fit (CL $<$1\%).  Figure~\ref{fig_fntvbrt} compares the best-fit composites based on both faint and bright flux calibrations for these three sources, and indeed the lower {\chisq}
fits are clear improvements.  All three of these systems also contain a T2--T4 secondary,
within the spectral type range where the faint and bright relations of \citet{2006ApJ...647.1393L} differ the most.
Given the inconsistency in which relation provides better fits for our candidate binaries, we cannot definitely select one over the other on this analysis alone.

A more direct assessment 
can be made by focusing on individual
brown dwarfs (singles and binary components) with parallax distance measurements.  
Figure~\ref{fig_absmag} displays absolute MKO $J$ magnitudes\footnote{Photometry shown in Figure~\ref{fig_absmag} is based either on direct measurements \citep{2002ApJ...564..466G,2002ApJ...564..452L,2004AJ....127.3553K,2006AJ....131.2722C} or synthetic filter corrections from 2MASS  photometry as computed directly from the SpeX data; see Section~4.4.} 
for 58 L and T dwarfs with measured parallaxes and absolute magnitude errors 
$\leq$0.3~mag. Eighteen of these sources are known binaries; we include component magnitudes of the only candidate
binary in our sample with a parallax measurement, 2MASS~J1711+2232, along with measured component magnitudes for SDSS~J0423-0414, SDSS~J1021-0304 and $\epsilon$ Indi BC \citep{2004A&A...413.1029M}. 
Nearly all of the unresolved sources and binary components follow the faint $M_J$ relation up to spectral type $\sim$T2; only
the L9.5 2MASS~J0328+2302 stands out as clearly overluminous in this range, with
$M_J$ magnitudes comparable to combined light photometry for Gliese~337CD.
Between types T2 and T5, there are only two unresolved sources with parallax measurements,
SDSS~J1750+1759 (T3.5) and 2MASS~J0559-1404 (T4.5), both of which lie well above the faint
relation; indeed, the latter lies $\sim$0.6~mag above the {\em bright} $M_J$ relation.  This could indicate a dramatic brightening over this spectral type range, as suggested by the component brightnesses of the resolved binary 2MASS~J1404+3159 \citep{2008ApJ...685.1183L} and our unresolved candidate SDSS~J0119+2403 (Section~5.2.1).  On the other hand, SDSS~J1750+1759 and 2MASS~J0559-1404 have both been suggested as unresolved multiples \citep{2004AJ....127.3516G,2006ApJ...647.1393L, 2009ApJ...702..154S}, although neither satisfy any of our selection criteria.  We therefore cannot rule out the possibility that they are young or otherwise peculiar sources.

The paucity of absolute magnitude measurements in the T2--T5 range 
makes it impossible to determine at this point
whether absolute $J$  magnitudes remain relatively flat across the entire L/T transition (a ``$J$-band plateau''; \citealt{2006ApJ...647.1393L}), favoring of modest  flux reversal and more gradual cloud depletion; or undergo a sharp increase over types T2--T5 (a ``$J$-band spike''; \citealt{2008ApJ...685.1183L}), requiring very rapid cloud depletion.  Indeed, neither of these scenarios may encompass the behavior of all brown dwarfs.   Increasing the number of T2--T5 dwarfs with parallax measurements, and obtaining resolved photometry of the candidates presented here, are both necessary if we are to properly constrain trends and distributions of absolute magnitudes across the L/T transition.

\subsection{The Frequency of Brown Dwarf Binaries}

Higher rates of multiplicity 
amongst early-type T dwarfs have been observed in resolved imaging studies
\citep{2006ApJS..166..585B}, although statistics in these samples remain poor due to small numbers \citep{2008A&A...490..763G}.
As our candidate sample includes sources that might be undetectable in high resolution imaging, it is relevant to assess how their inclusion in multiplicity statistics modifies brown dwarf
binary fraction trends across the L/T transition, and what constraint can be made 
on the intrinsic binary fraction.  

As discussed in Section~2, our spectral sample is neither complete nor volume-limited, and selection biases are uncertain given that sources were identified from different search programs using different color and magnitude selection criteria.  However, as a first approach we made the simplifying assumption that the sample is roughly magnitude-limited, 
and estimated the fraction of L8--T6 dwarfs with SpeX data that are either resolved or candidate binaries.  These numbers are summarized in Table~\ref{tab_bf}, and indicate the same high multiplicity rate amongst early-type T dwarfs as found in
resolved imaging studies.  For T0--T4 dwarfs, we deduce a binary fraction of 53$\pm$7\% (23/43 sources), peaking above 60\% for T0--T2 dwarfs alone.  Over half of the early-type T dwarfs in this spectral sample appear to be multiples, compared to $\sim$30\% in resolved imaging samples \citep{2007ApJ...659..655B}.
Moreover, the fraction inferred here is likely to be a lower limit, given that we are unable to identify unresolved binaries with identical components or combinations that fall outside our spectral index selection criteria (see Section~3.2). 

Such a high {\em apparent} rate
of multiplicity does not necessary translate into a high {\em intrinsic} (i.e., volume-limited) rate,
however. 
As discussed in \citet{2007ApJ...659..655B}, the enhanced binary fraction of early-type T dwarfs arises predominantly from the small change in luminosity across the L dwarf/T dwarf transition, which induces both a sharp minimum in the luminosity function of brown dwarfs at type T0 (roughly three times rarer in number than early- and mid-type L dwarfs) and makes late-type L and early-type T dwarfs comparably bright in the near-infrared.
Combined with the preferential selection of binaries in magnitude-limited
samples, these effects collude to 
amplify the apparent multiplicity rate of early-type T dwarfs.
To infer the intrinsic rate, we reproduced 
the Monte Carlo population simulations of \citet{2007ApJ...659..655B}, using the baseline parameters defined in that study and assuming a magnitude-limited sample.
We find that an intrinsic rate of only 15\% can produce an apparent binary fraction of 56\% for T0--T4 dwarfs in a magnitude-limited sample, comparable to the fraction estimated above.
This intrinsic rate is in fact on the low end of current estimates that attempt to correct for unresolved multiplicity (e.g., \citealt{2006AJ....132..663B, 2007MNRAS.380..712L, 2008A&A...492..545J}), 
and remains  consistent with a brown dwarf multiplicity fraction that is significantly lower than those of more massive stars (e.g., \citealt{1991A&A...248..485D, 1992ApJ...396..178F}).
We reiterate, however, that our result constitutes a lower limit, 
and a robust measure requires detailed modeling
of selection effects (e.g., component spectral type sensitivity range, 
mass ratio sensitivity as a function of age)
that are beyond the scope of this paper.

\subsection{Flux Reversals and Cloud Evolution}

The physical feature that ties both absolute magnitude trends and enhanced binary fractions across the L/T transition is the ``sudden''  depletion og condensate clouds from the photosphere.  This is particularly evident in the increased surface fluxes at the 1.05~$\micron$ and 1.27~$\micron$ flux peaks where cloud particles are a dominant source of opacity
\citep{2001ApJ...556..872A}.
Amongst the binary  candidates identified here, 59\% (10/17) of the entire sample and 82\% (9/11) of candidates with L7--T5 components show similar flux reversals in the $J$-band region.  While these reversals are dictated in part by our adopted flux calibration (which are, however, tied to the $K$-band flux; see Section~4.1), the quality of the fits coupled with observed absolute magnitude/spectral type trends (Section~6.1) suggest that these flux reversals are real.   

With a relatively large sample of possible L dwarf/T dwarf pairs, we can also assess how these flux reversals vary between systems with similar spectral compositions.
Consider the candidates SDSS~J0119+2403 and
SDSS~J1106+2754, which have component types of
T0+T4 and T0+T4.5 and relative brightnesses of  $\Delta{J}$ =
-0.42$\pm$0.19 and -0.37$\pm$0.06, respectively.  Examination of the best-fit composites for these sources (Figures~\ref{fig_fitsstrong} and~\ref{fig_fitsweak}) reveals that the former has a far more pronounced flux reversal
than the latter, with the secondary of SDSS~J0119+2403 being potentially brighter even at the 1.6~$\micron$ flux peak.  The best-fit primary of this system was inferred to be an unusually cloudy source, which translates in greater opacity in the $J$-band region.
In contrast, the primary of 2MASS~J1106+2754 was inferred to be slightly blue for its spectral type,
which could be attributed to thinner condensate clouds (e.g., \citealt{2004AJ....127.3553K,2008ApJ...674..451B}).  If clouds are the main driver for the unusual colors of the primary components, and their secondaries are roughly equivalent, the different flux reversals can be tied directly to variations in cloud properties.  Specifically, an initially cloudy brown dwarf should lose more opacity in the $J$-band region, and experience a greater increase in brightness, as it evolves across the L/T transition as compared to a brown dwarf that initially has thinner clouds.  This correlation between initial cloud content and degree of flux reversal suggests that absolute magnitude trends may have more spread across the L/T transition than can be discerned in the currently sparse dataset; spread has also been suggested to arise from age/surface gravity effects (e.g., \citealt{2006ApJ...651.1166M,2007ApJ...657.1064M}) and metallicity variations (e.g., \citealt{2006ApJ...640.1063B}).
Again, resolved studies of confirmed binaries in our sample would provide insight into how atmospheric  properties modulate the removal of condensate clouds.

\subsection{Limitations of this Study}

Despite the large number of potential L dwarf/T dwarf pairs uncovered in this investigation, there are clearly limitations as pertaining to the completeness of the sample and robustness of the component characterizations.  
We reiterate that the 17 candidates listed in Table~\ref{tab_summary} may not represent
the full complement of potential binaries in our spectral sample.
Our selection criteria were specifically chosen to be conservative so as to not eliminate
too many spectral templates, particularly over the L8--T5 spectral type range.
As such, we failed to select known binary systems with early-type T dwarf
classifications (Section~3.2) and a 
number of suspected binary pairs such as 2MASS~J0328+2302. 
We are also unable to select systems with similar spectral 
component types, and are likely biased against higher-order
multiples such as Kelu~1 \citep{2008arXiv0811.0556S}
and DENIS-P~J020529.0-115925 \citep{2005AJ....129..511B}.
The retention of unresolved binaries in our template sample may also skew the statistical
significance of our candidates.  For instance, our rejected candidate
SDSS~J1206+2813, whose spectrum is a near clone to that of SDSS~J1750+1759, may
prove to be a binary after all should the latter be determined as such.
On the other hand, incomplete sampling of single L dwarf/T dwarf transition objects 
in our spectral template library may cause us to overselect candidates, a possibility for the
poorly-fit candidate 2MASS~J2139+0220.
It is in this vein that we encourage follow-up high resolution imaging (or second epoch imaging in the cases of 2MASS~J1106+2754 and 2MASS~J1711+2232) and spectroscopy of the candidates listed in Table~\ref{tab_summary} to both verify their binary nature and better characterize their components.

\section{Summary}

We have identified 17 candidate brown dwarf
binaries whose components straddle the L dwarf/T dwarf transition.
Their unresolved multiplicity is inferred from similarities in near-infrared
spectral indices
to known binary systems, and statistically significant better fits to
near-infrared spectra by composite templates as compared to single templates.  
Ten of these systems appear to 
have secondaries that are brighter than their primaries over the 1--1.3~$\micron$ region,
by up to $\sim$0.4~mag, consistent with a nonequilibrium
depletion of condensate clouds across the L dwarf/T dwarf transition.
This is despite the fact that our analysis of these systems is based largely on the faint absolute magnitude relations of \citet{2006ApJ...647.1393L}.  We cannot rule out an even more pronounced $J$-band flux reversal for the sparsely sampled T3--T5 dwarfs.
We find that 53$\pm$7\% of the T0--T4 dwarfs in our SpeX spectral sample are resolved
or unresolved (candidate) binaries, a rate that is consistent with an intrinsic
brown dwarf binary fraction of at least 15\%, assuming the sample is magnitude-limited.
We find some evidence of flux reversal variations between similarly-classified pairs that may arise from cloud effects, although these must be verified through more accurate characterization of the components.
While the sample of candidate binaries presented here does require verification through follow-up
high resolution imaging and/or spectroscopic monitoring, it constitutes a promising collection of coeval laboratories for understanding the atmospheric processes that drive the still poorly-understood L dwarf/T dwarf transition.

\acknowledgments

The authors would like to thank telescope operators Bill Golisch, Dave Griep and Paul Sears, and instrument specialist John Rayner,
for their assistance during our many IRTF runs.
This publication makes use of data 
from the Two Micron All Sky Survey, which is a
joint project of the University of Massachusetts and the Infrared
Processing and Analysis Center, and funded by the National
Aeronautics and Space Administration and the National Science
Foundation.
2MASS data were obtained from the NASA/IPAC Infrared
Science Archive, which is operated by the Jet Propulsion
Laboratory, California Institute of Technology, under contract
with the National Aeronautics and Space Administration.
This research has also made use of the SIMBAD database,
operated at CDS, Strasbourg, France; the M, L, and T dwarf compendium housed at DwarfArchives.org and maintained by Chris Gelino, Davy Kirkpatrick, and Adam Burgasser; the SpeX Prism Spectral Libraries, maintained by Adam Burgasser at \url{http://www.browndwarfs.org/spexprism}; and the VLM Binaries Archive maintained by Nick Siegler at \url{http://www.vlmbinaries.org}. 
The authors wish to recognize and acknowledge the 
very significant cultural role and reverence that 
the summit of Mauna Kea has always had within the 
indigenous Hawaiian community.  We are most fortunate 
to have the opportunity to conduct observations from this mountain.

Facilities: \facility{IRTF (SpeX)}

\clearpage

\begin{figure}
\centering
\epsscale{1.1}
\plottwo{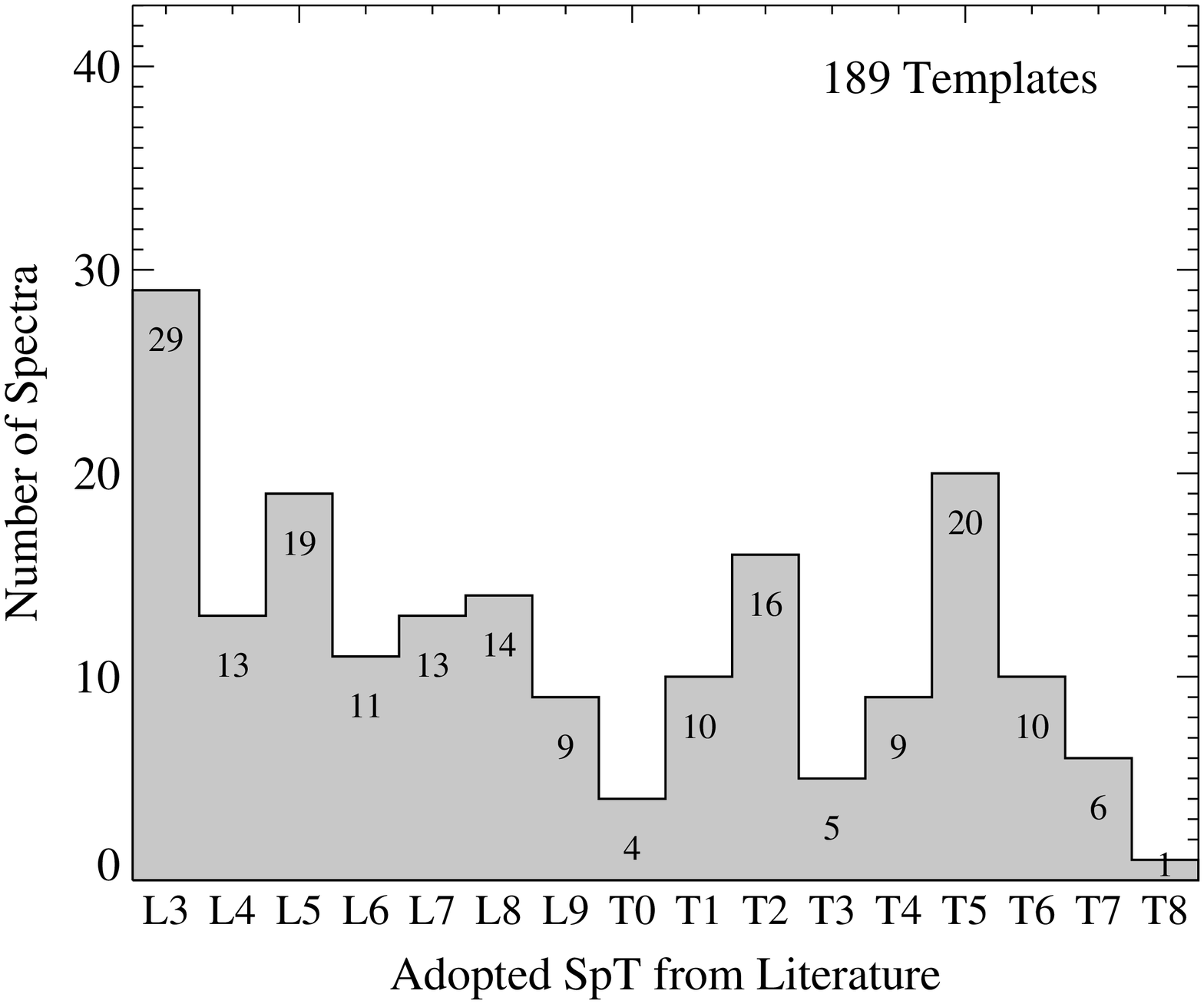}{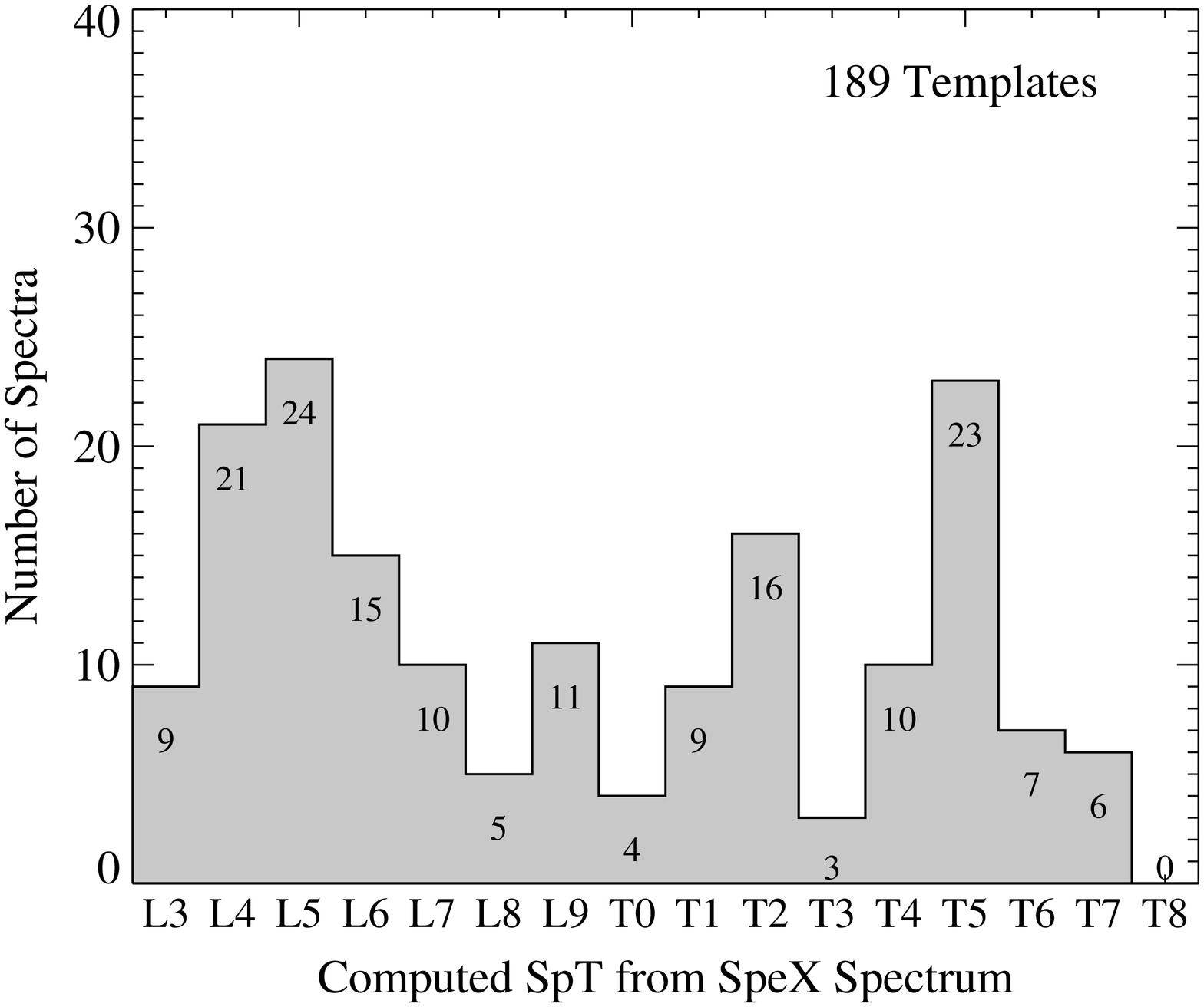}
\caption{Distribution of spectral types for the template sample based on classifications from the literature (left) and calculated from the SpeX data (right).  The former are based on optical types for L3--L8 dwarfs (where available) and near-infrared types for L9--T8 dwarfs.  The distributions are roughly consistent with each other and relatively flat for the small number of sources available per subtype.
\label{fig_sptdist}}
\end{figure}

\clearpage

\begin{figure}
\centering
\epsscale{0.7}
\plotone{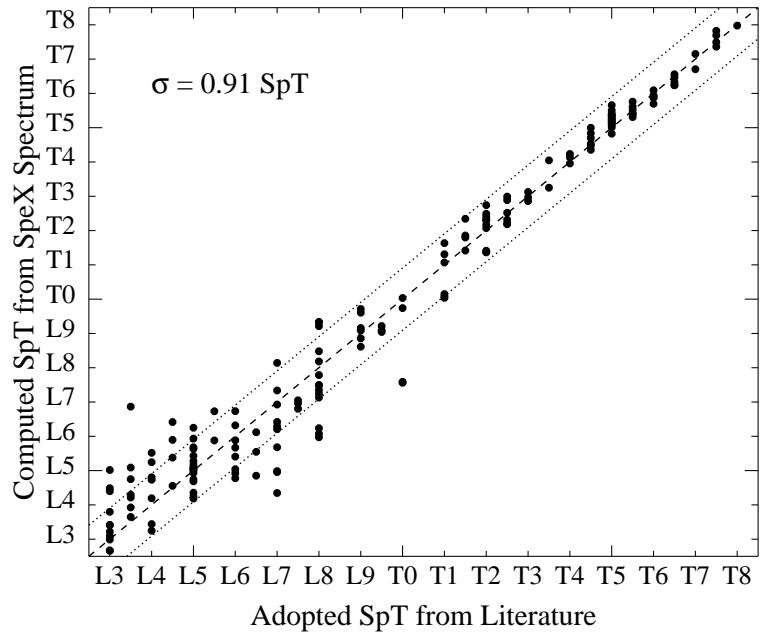}
\caption{Comparison of spectral types from the literature and those calculated directly from the SpeX data.  Overall, classifications are consistent to within 0.9 subtypes, although there is considerably more scatter among the L dwarfs
($\sigma$ = 1.1) than the T dwarfs ($\sigma$ = 0.5) 
\label{fig_compspt}}
\end{figure}

\clearpage

\begin{figure}
\centering
\epsscale{0.9}
\includegraphics[width=0.45\textwidth]{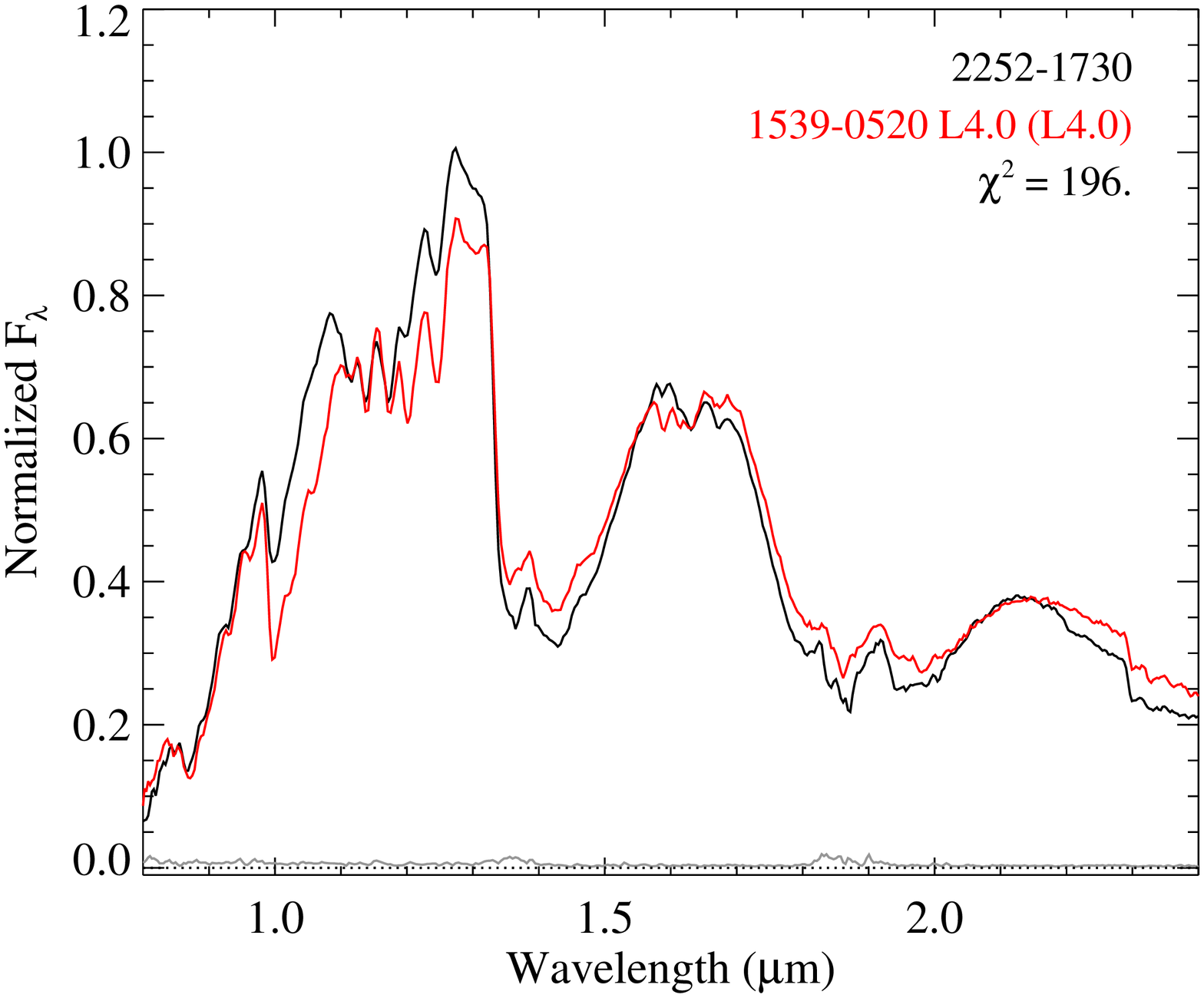}
\includegraphics[width=0.45\textwidth]{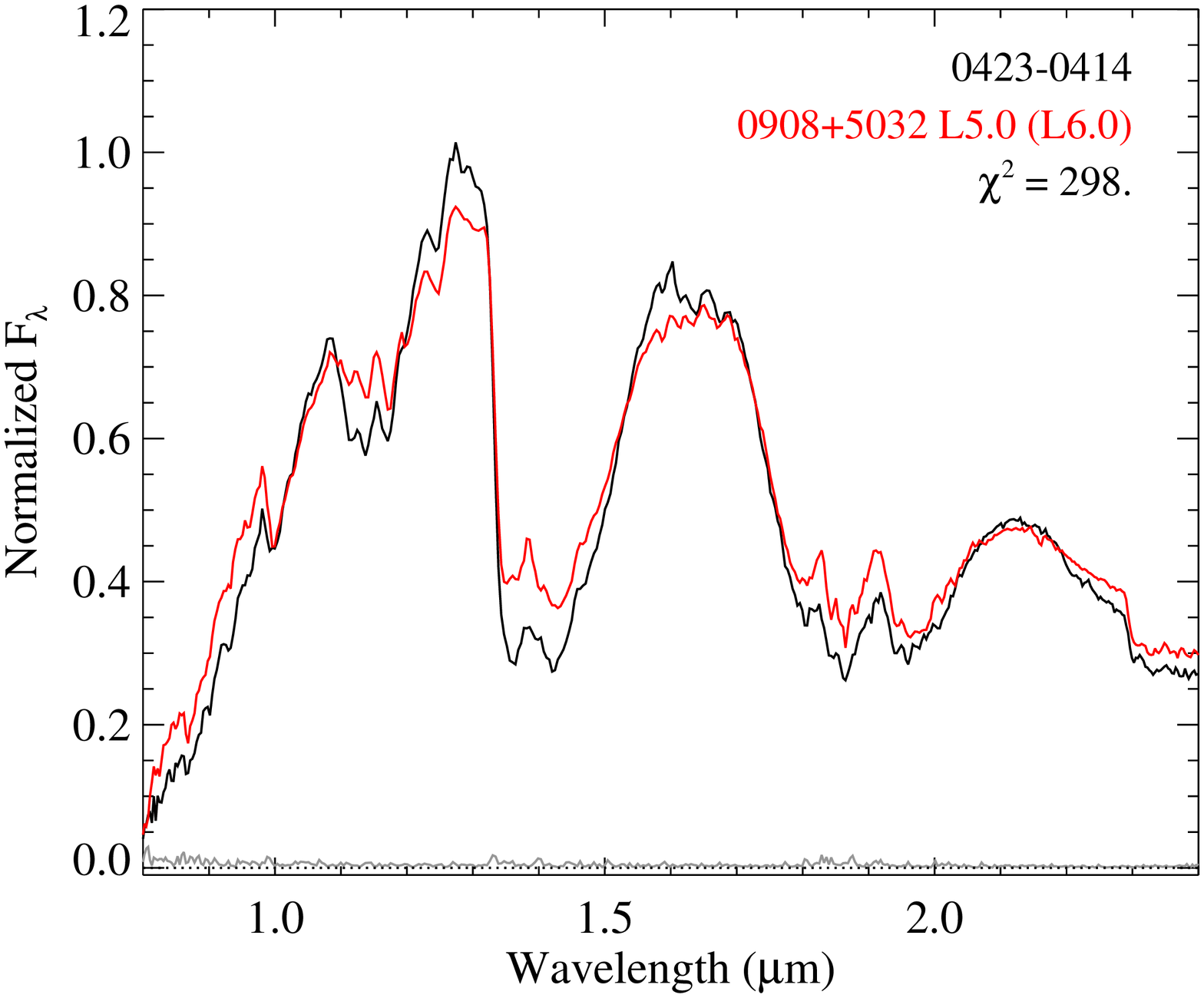}
\includegraphics[width=0.45\textwidth]{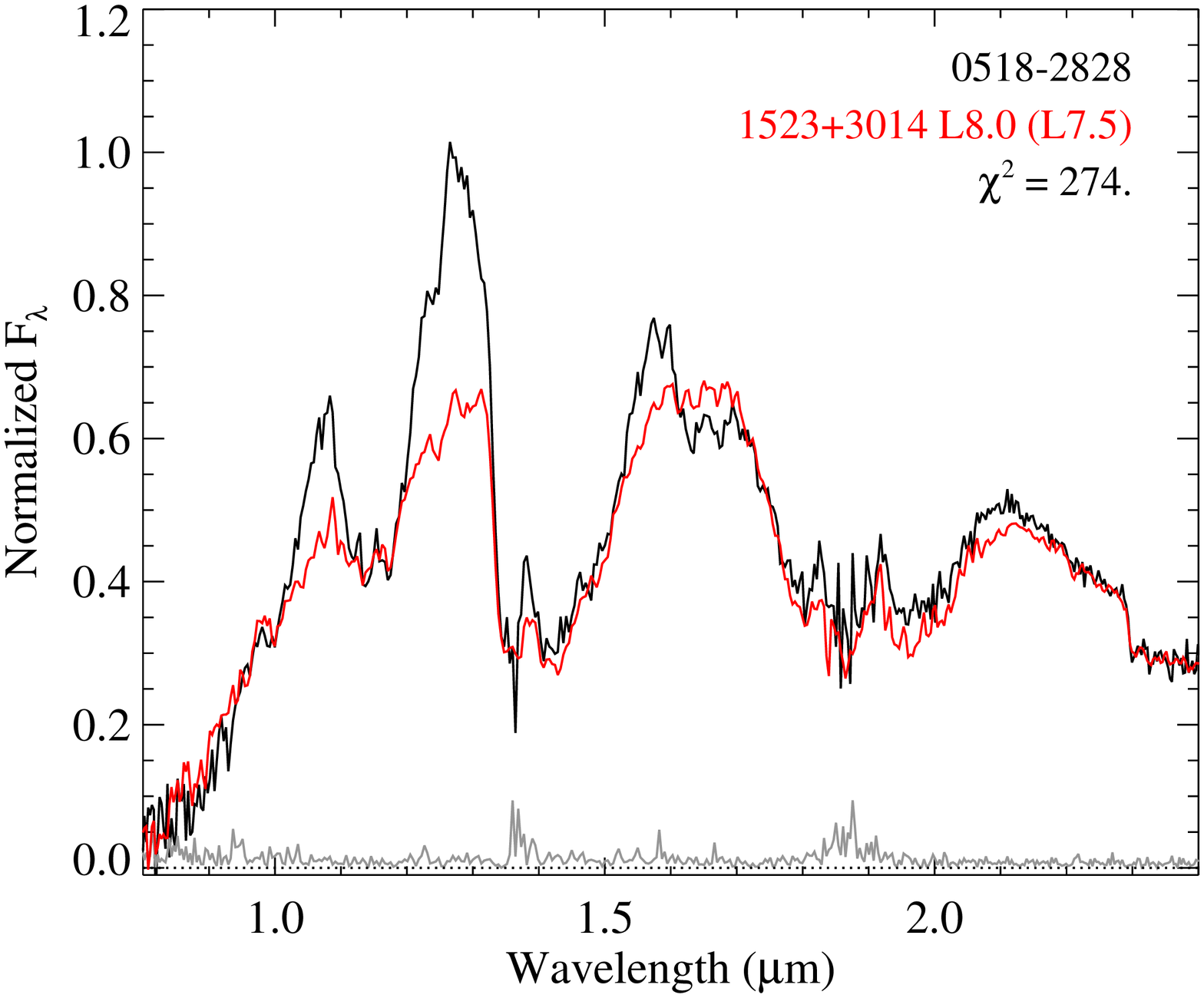}
\includegraphics[width=0.45\textwidth]{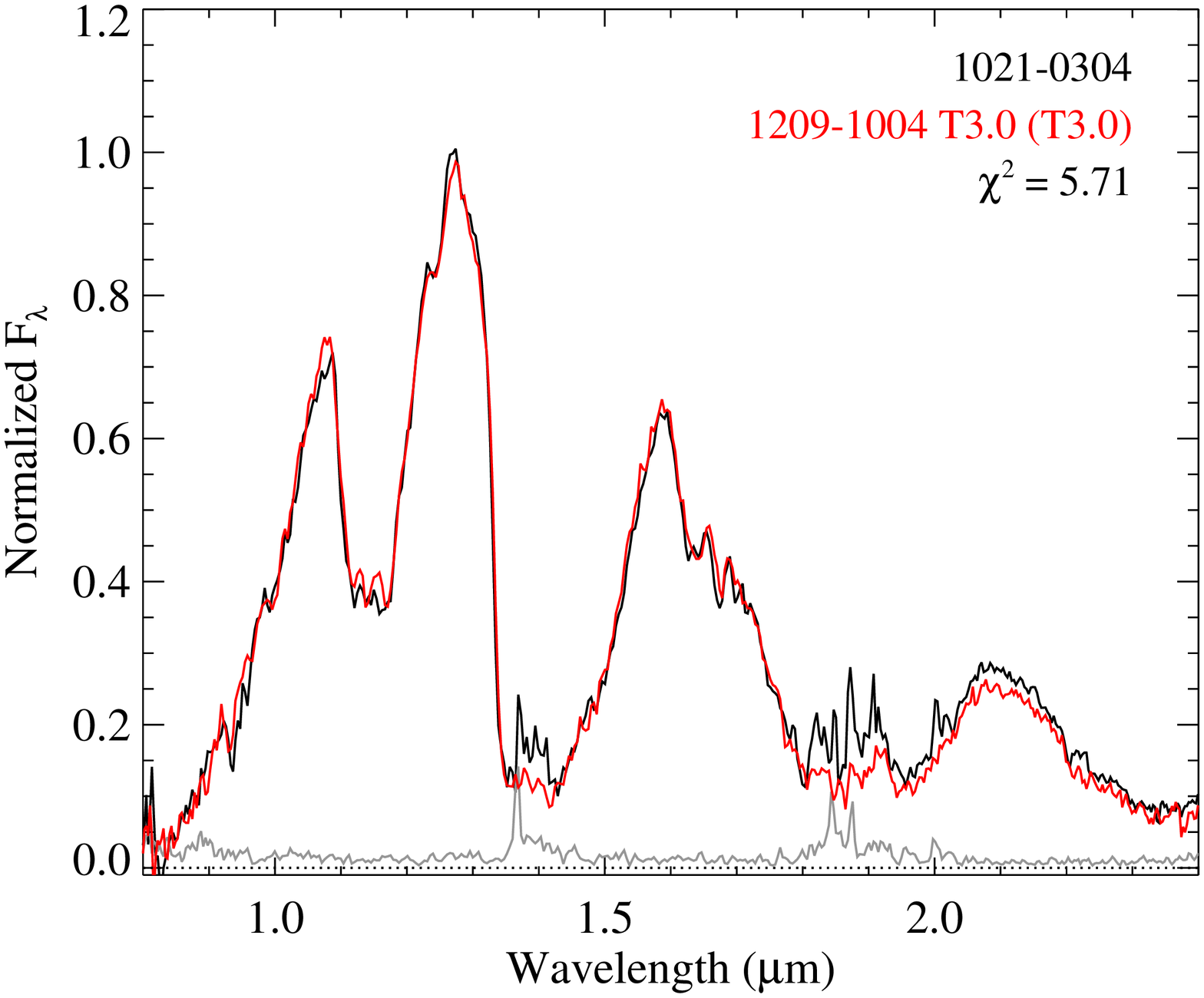}
\includegraphics[width=0.45\textwidth]{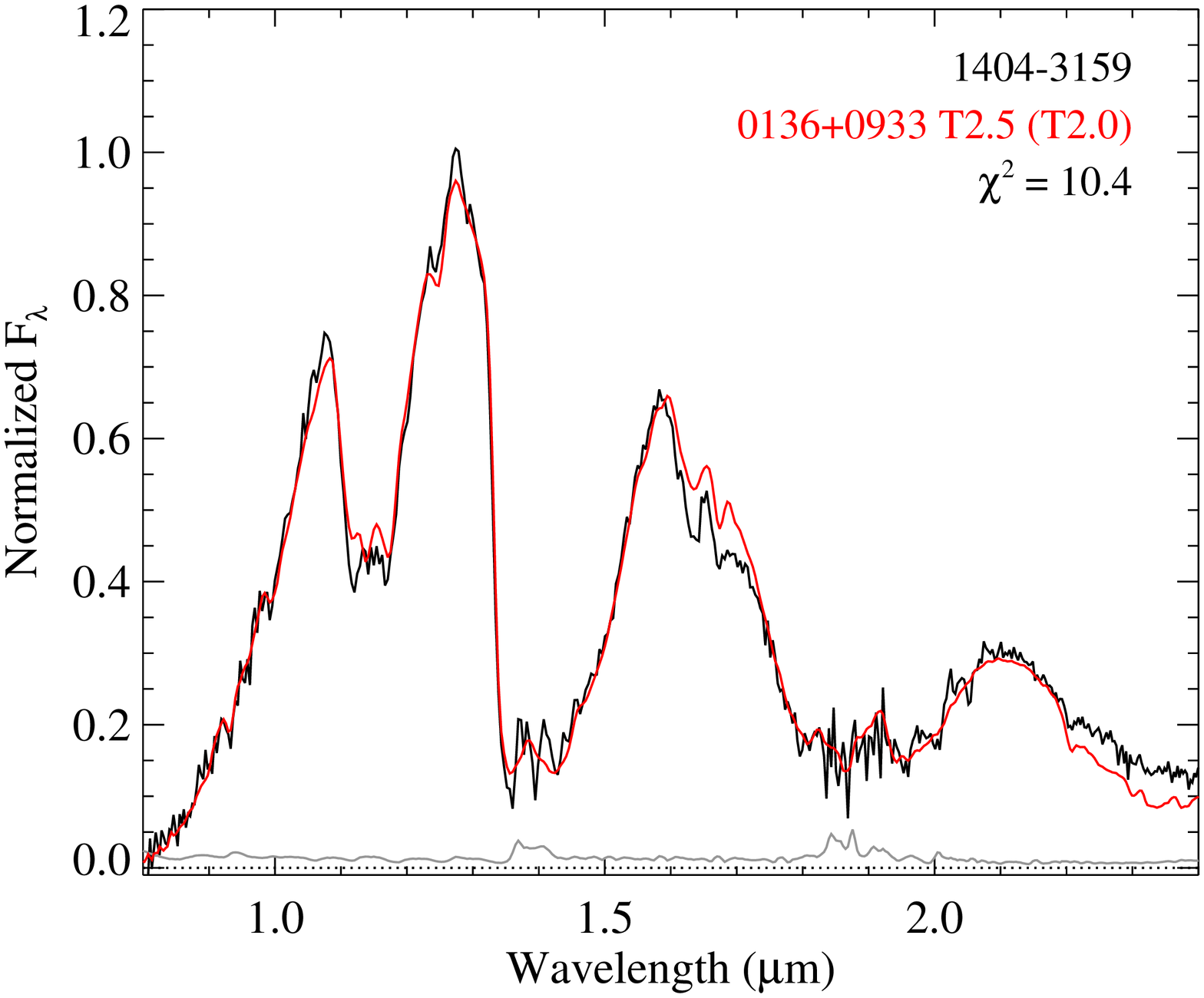}
\includegraphics[width=0.45\textwidth]{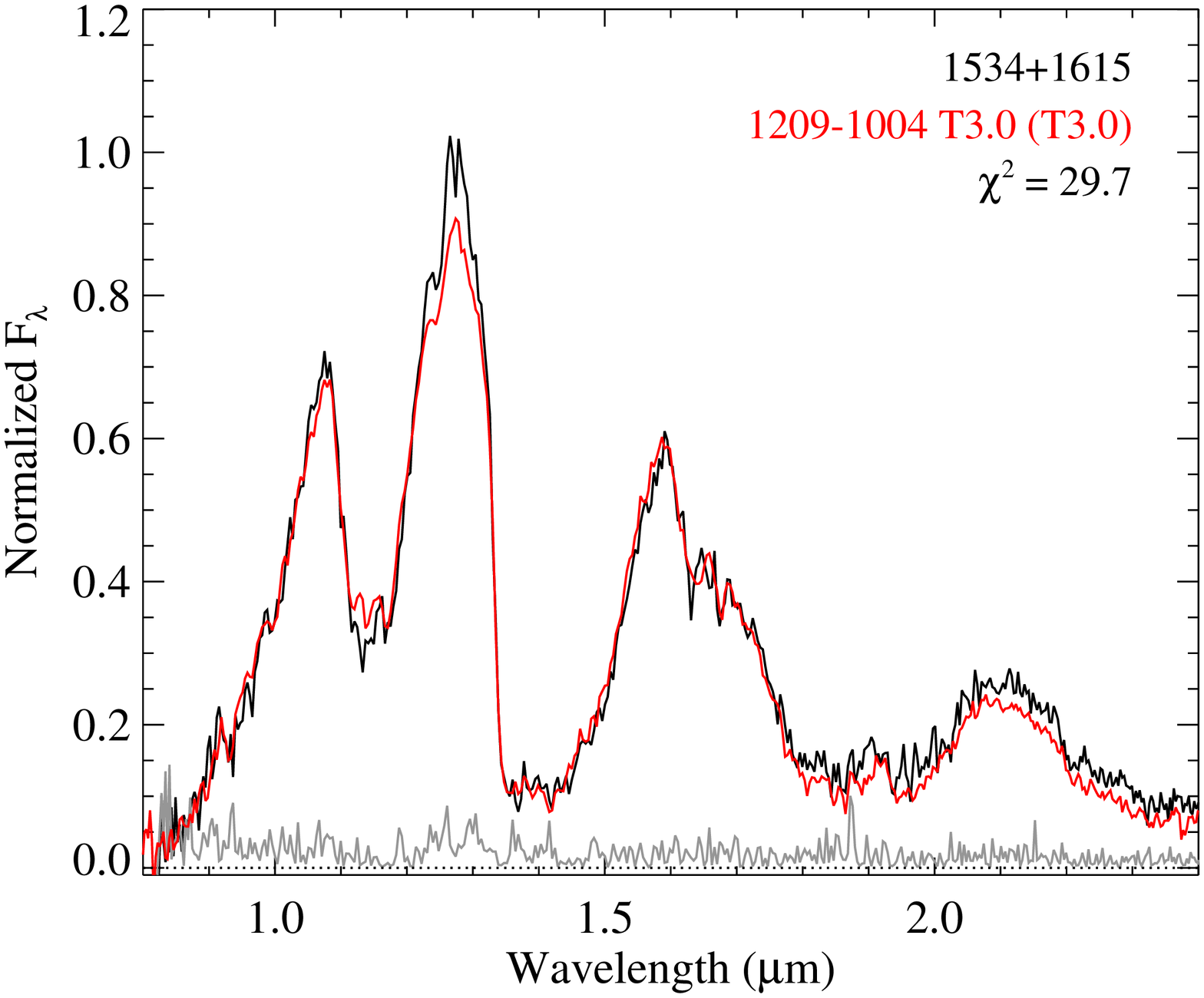}
\caption{\small Comparison of known (resolved) binary spectra (Table~\ref{tab_known}; black lines) to best-fit single templates (red lines).  Observed data (in $F_{\lambda}$ flux units) have been normalized to the peak flux in the 1.0--1.3~$\micron$ range; templates are relatively normalized to minimize {\chisq} deviations.  Source names for both the binary and best-fit template are listed in top right corner of each panel, with both published classifications and index-based classifications (in parentheses) indicated.  Also shown are the noise spectra (uncertainty in flux as a function of wavelength; grey lines) for the binaries.
\label{fig_known}}
\end{figure}

\clearpage

\begin{figure}
\centering
\epsscale{0.85}
\includegraphics[width=0.45\textwidth]{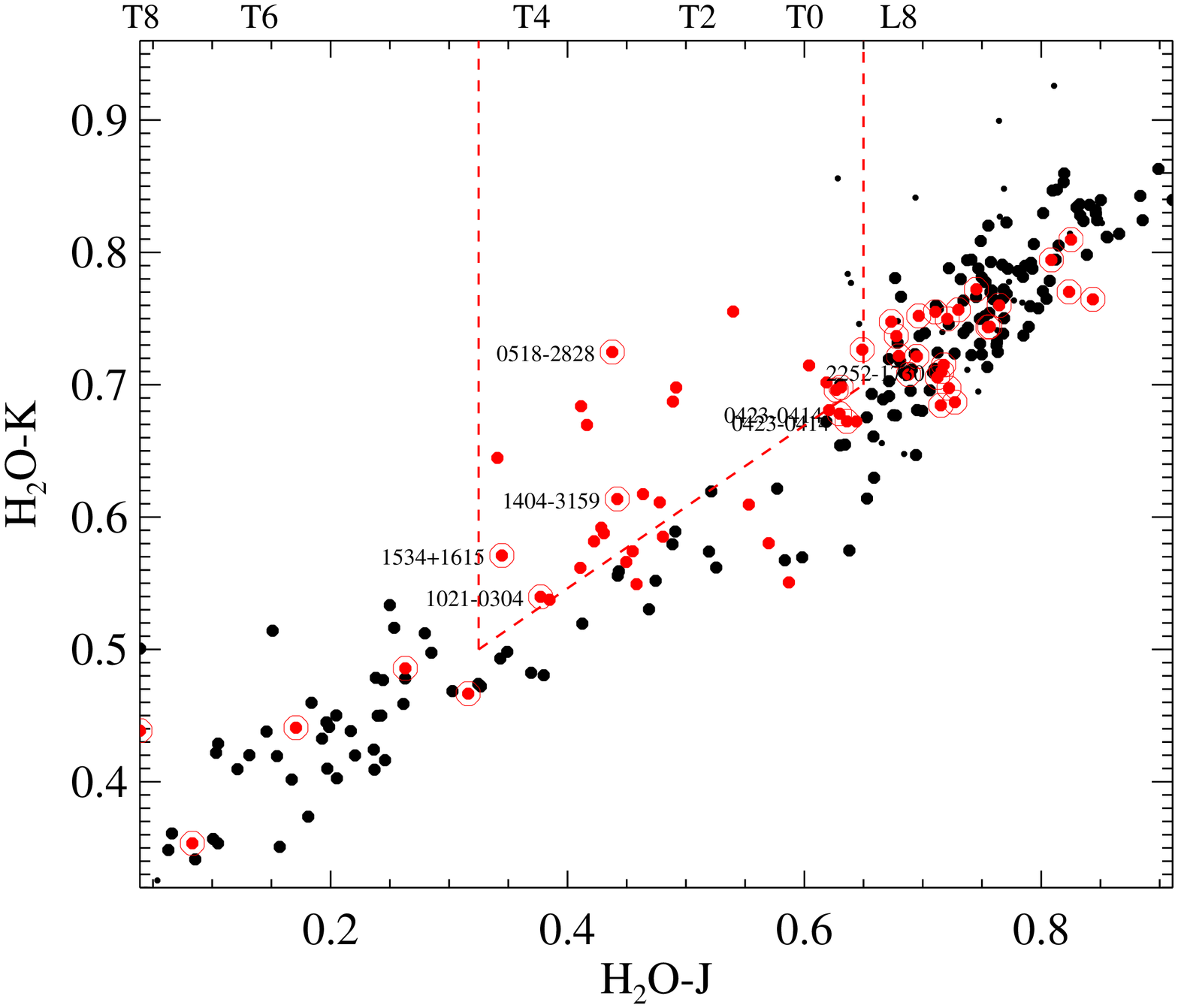}
\includegraphics[width=0.45\textwidth]{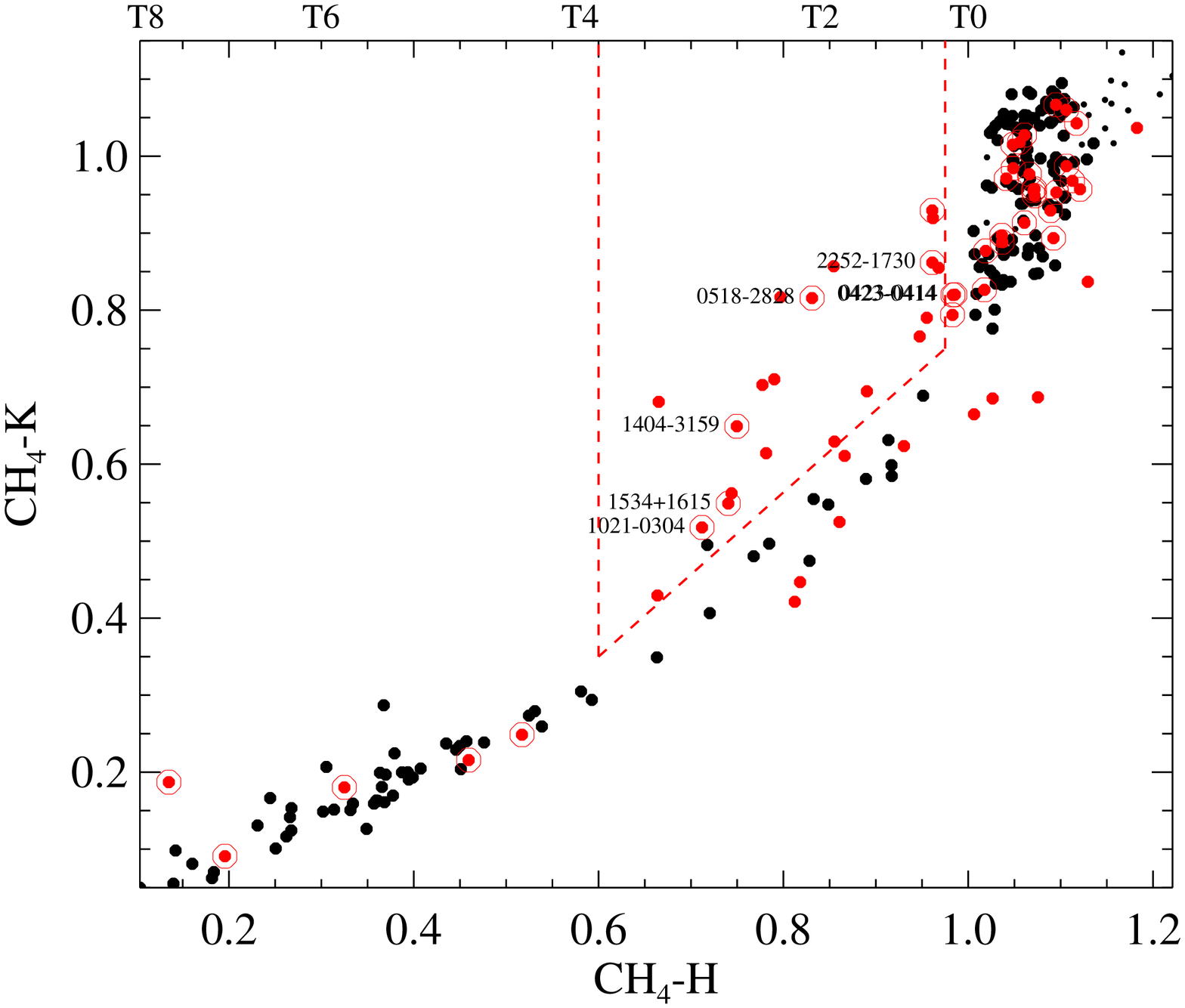}
\includegraphics[width=0.45\textwidth]{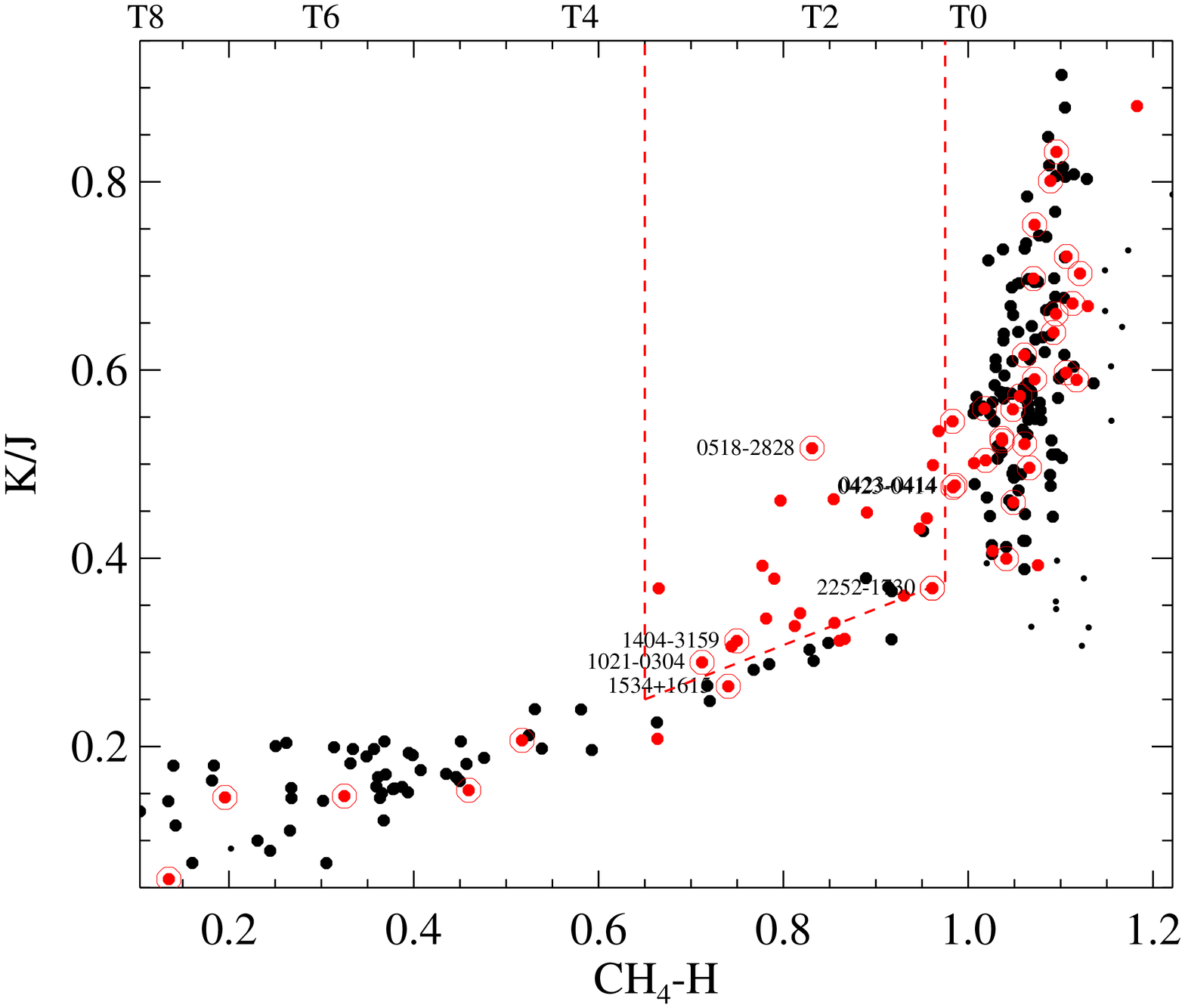}
\includegraphics[width=0.45\textwidth]{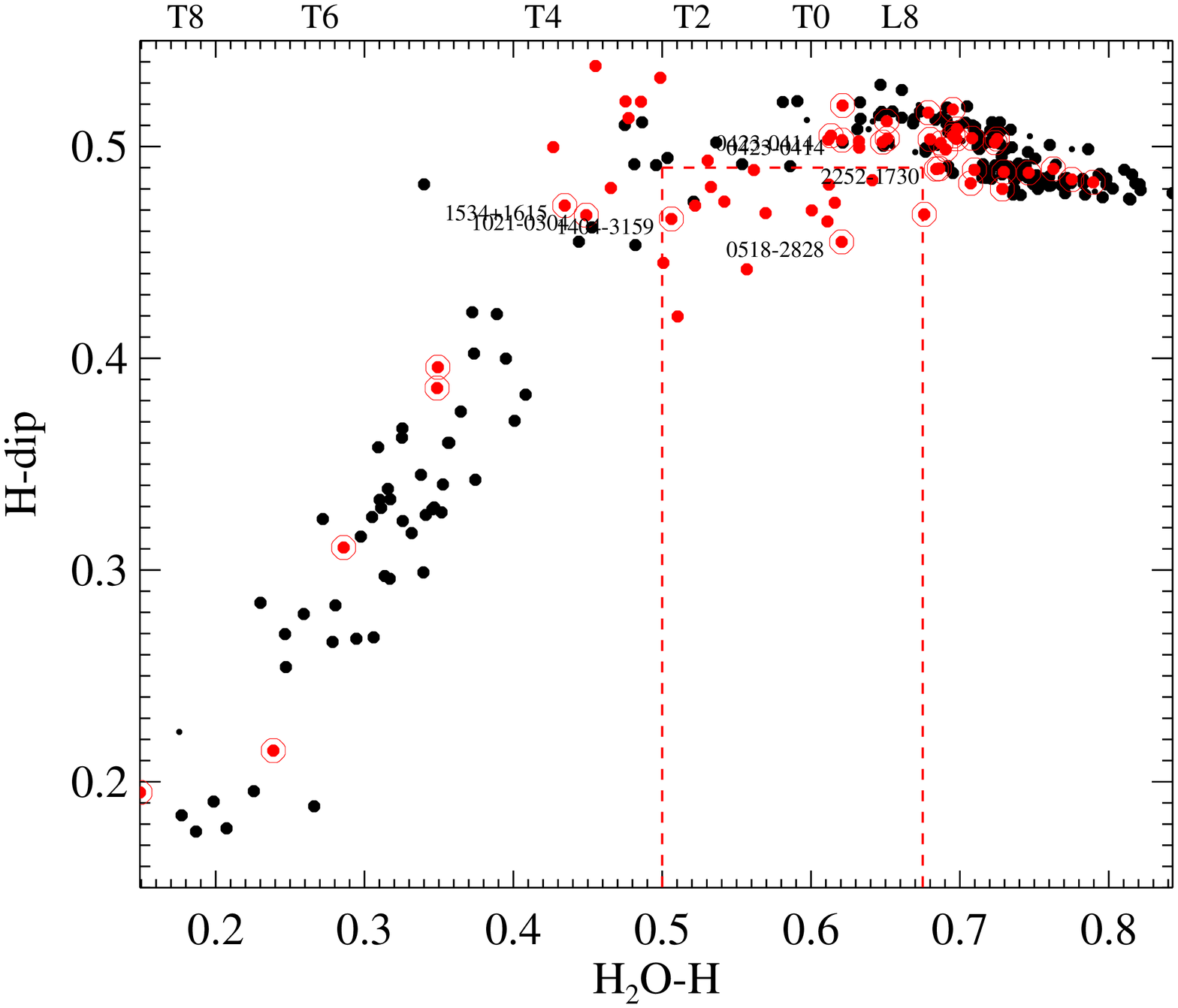}
\includegraphics[width=0.45\textwidth]{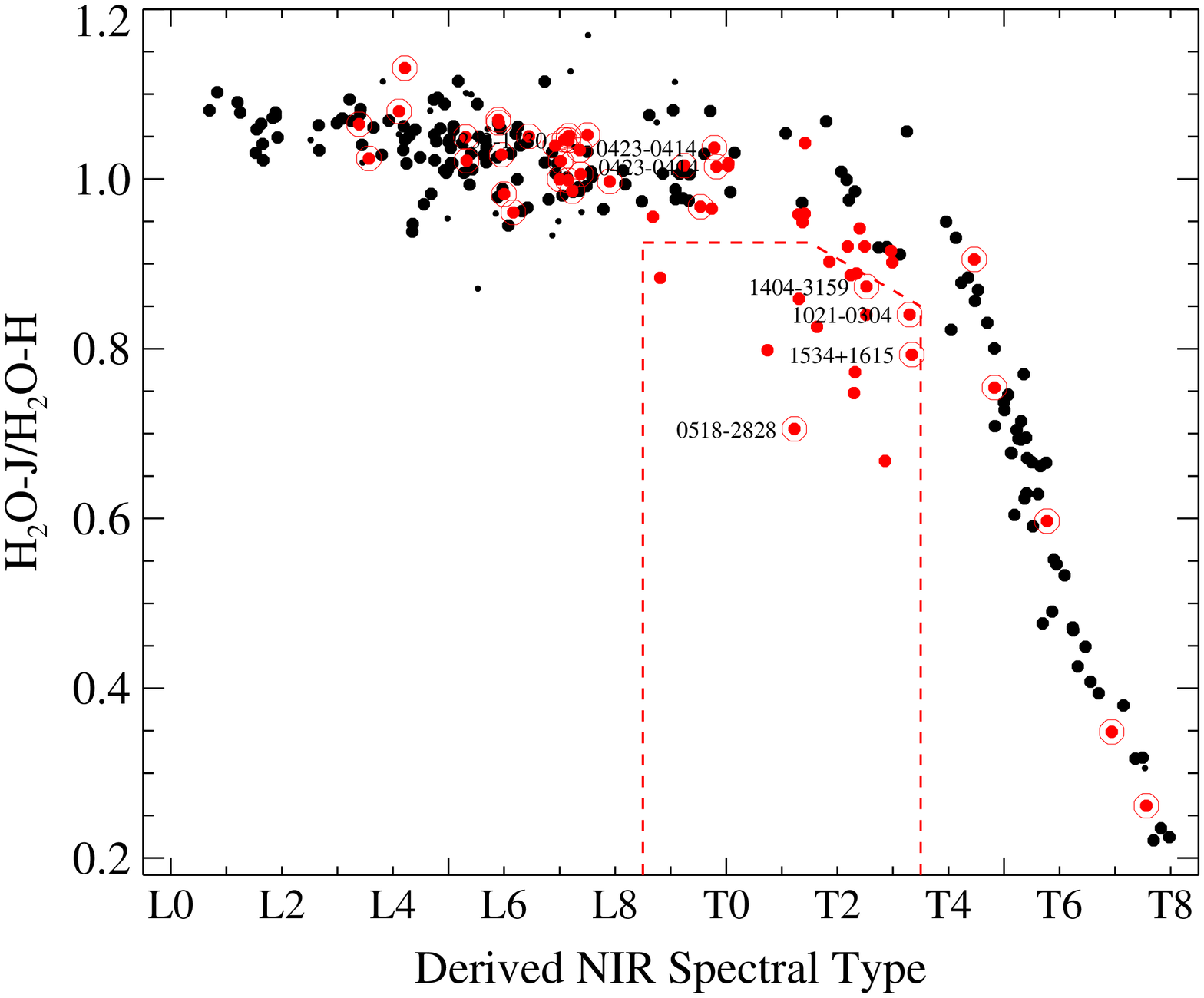}
\includegraphics[width=0.45\textwidth]{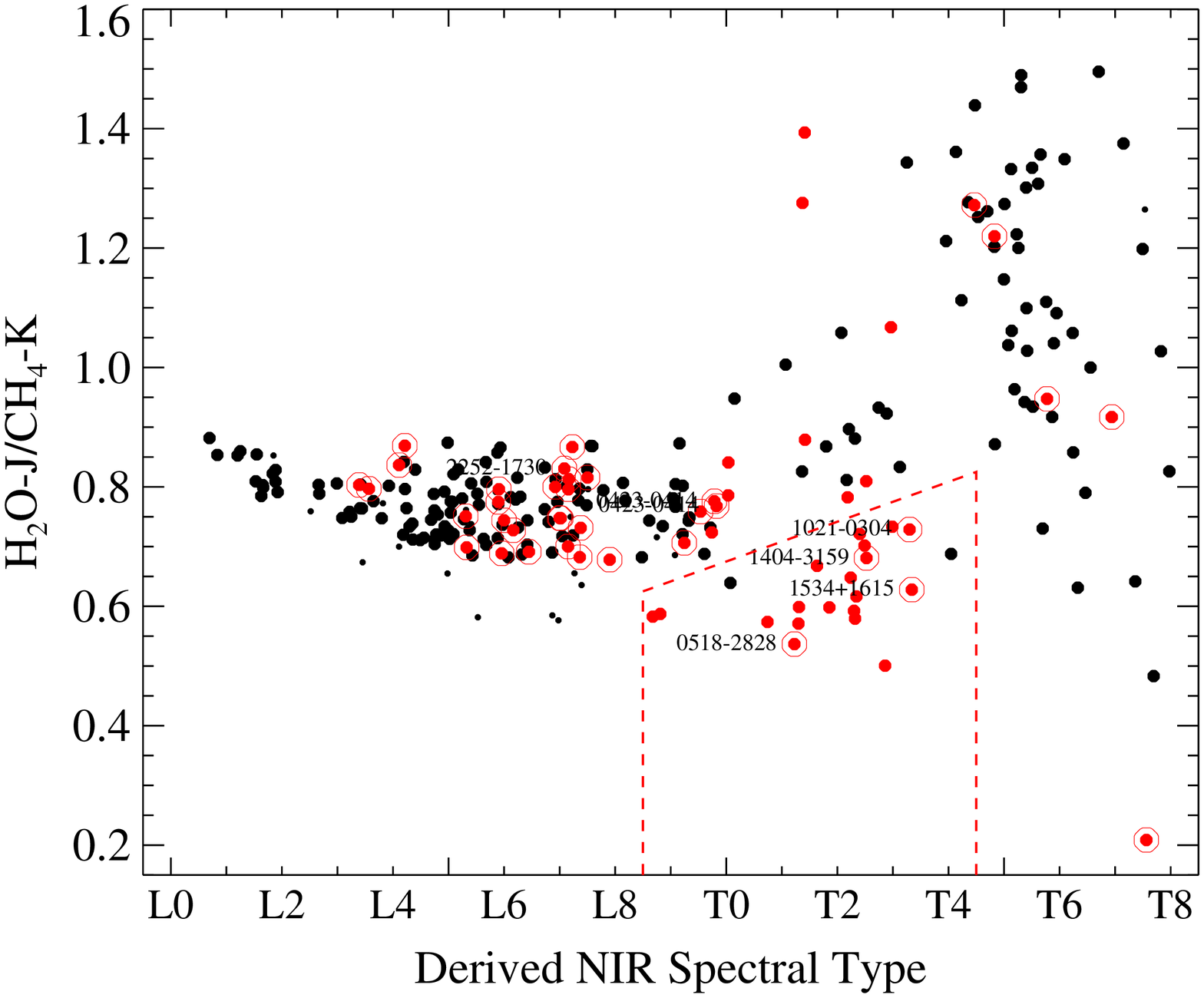}
\caption{\small Spectral index selection criteria for binary candidates. Top four panels compare pairs of index
values---{\wat}-K versus {\wat}-J, {\meth}-K versus {\meth}-H, K/J versus {\meth}-H and H-dip versus {\wat}-H---with
estimated spectral types for each abscissa indicated at top.
Bottom two panels compare index ratios {\wat}-J/{\wat}-H and {\wat}-J/{\meth}-K to near-infrared spectral
type as derived from the SpeX data.  In all panels, templates are indicated by large black dots, known 
peculiar sources by small black dots, candidate binaries by red dots and known (resolved) binaries by 
encircled red dots (those listed in Table~\ref{tab_known} are also labeled).  Selection spaces, with inflection points listed in Table~\ref{tab_criteria}, are indicated by dashed lines
\label{fig_indices}}
\end{figure}

\clearpage

\begin{figure}
\centering
\epsscale{0.85}
\includegraphics[width=0.45\textwidth]{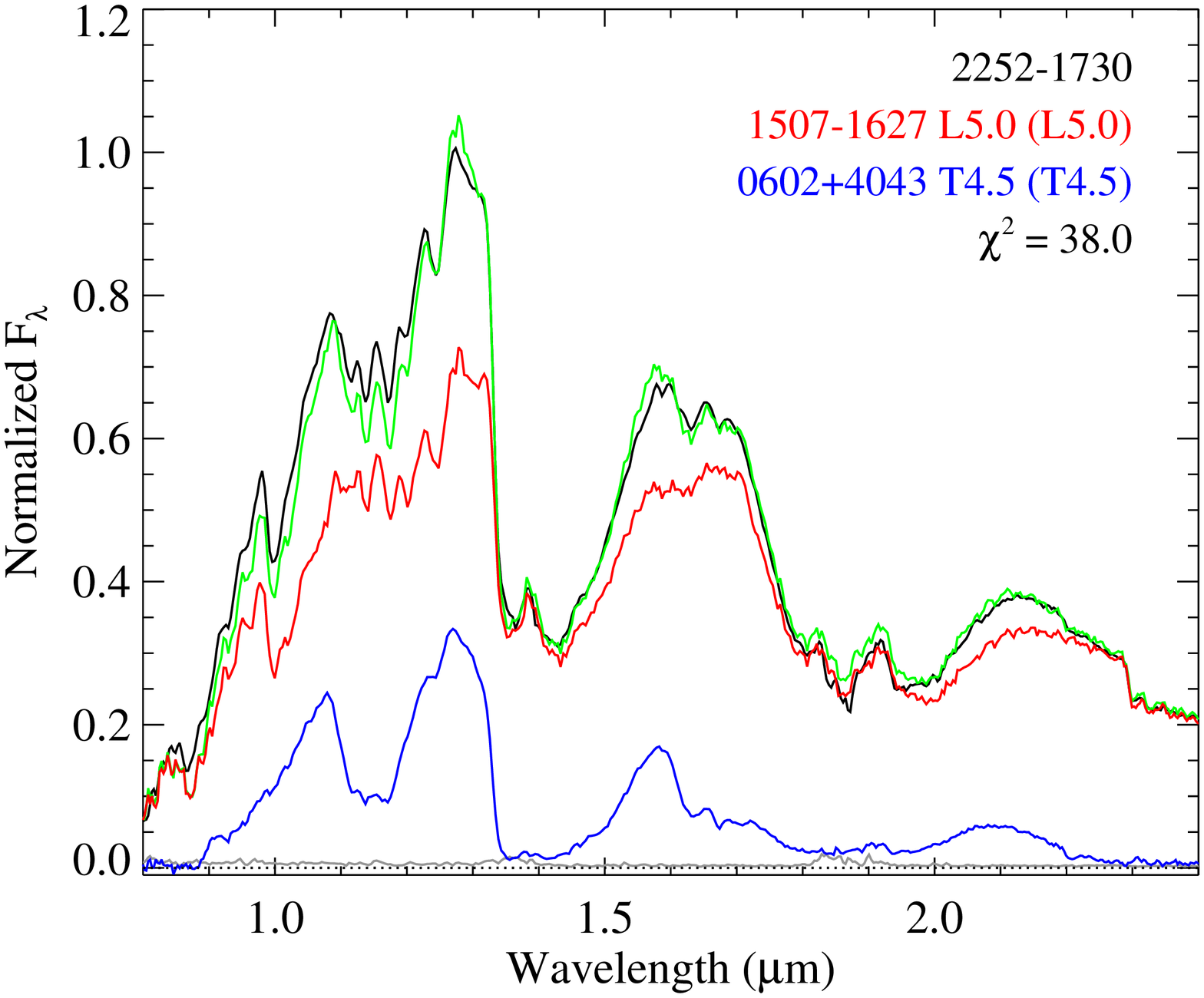}
\includegraphics[width=0.45\textwidth]{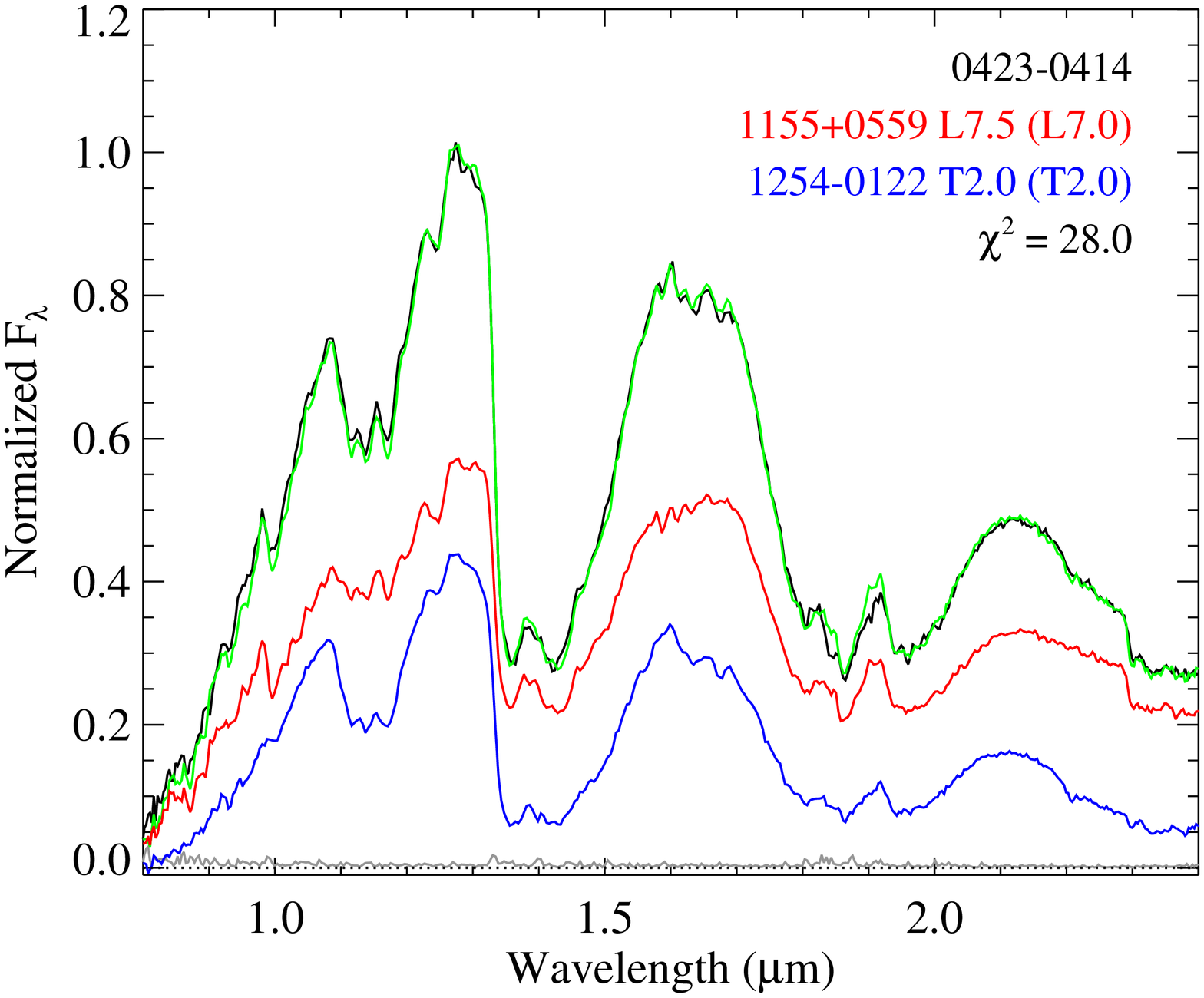}
\includegraphics[width=0.45\textwidth]{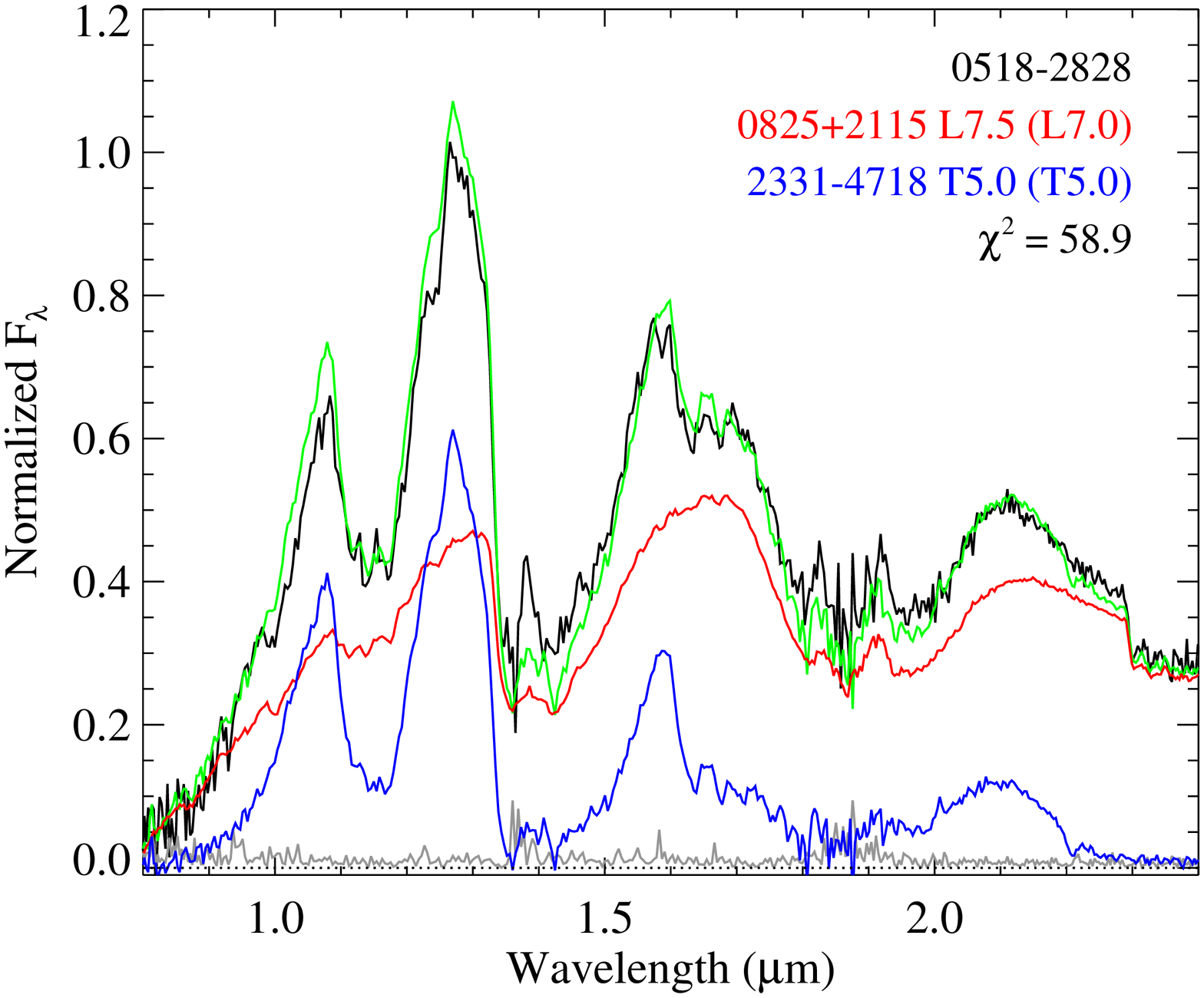}
\includegraphics[width=0.45\textwidth]{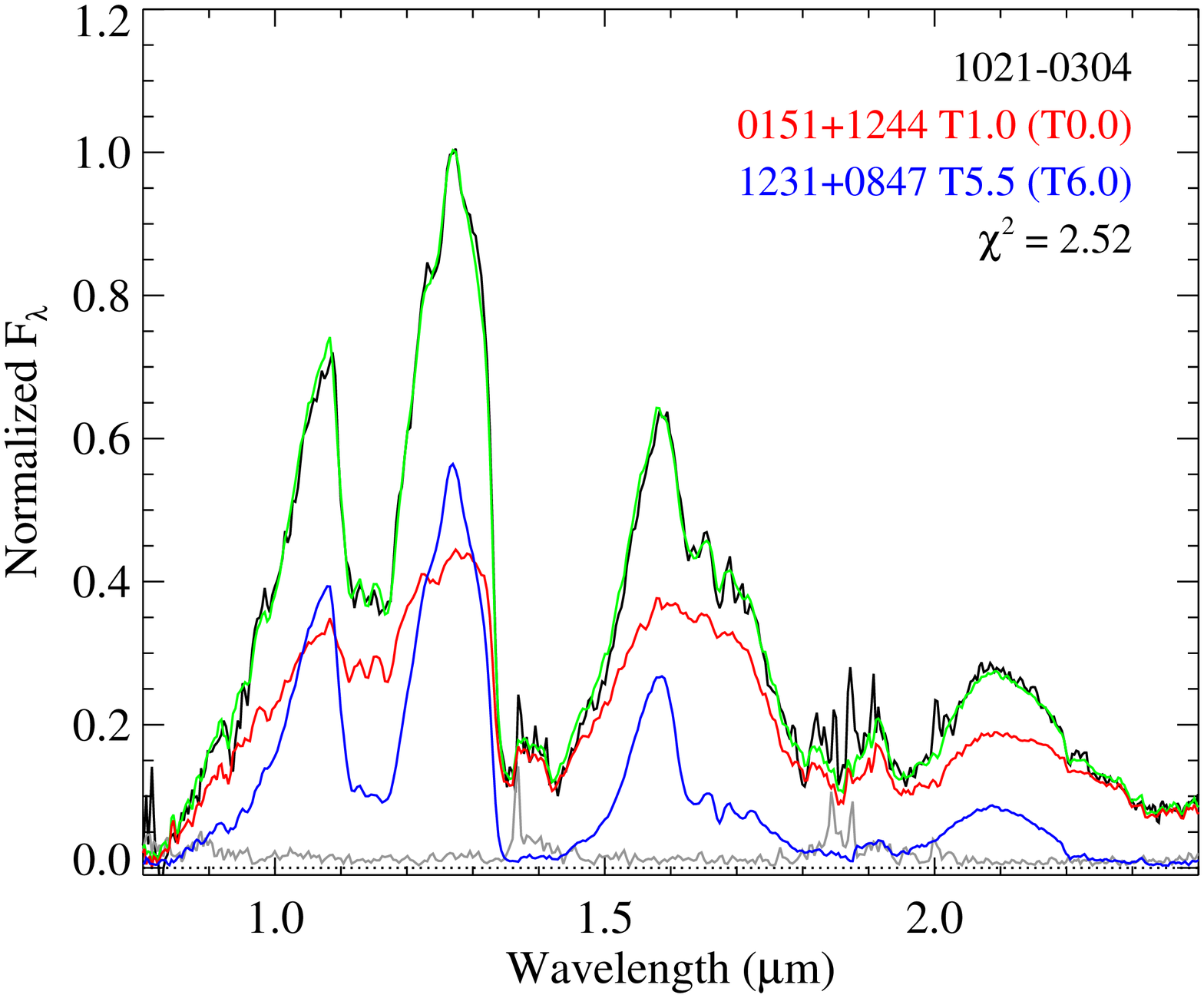}
\includegraphics[width=0.45\textwidth]{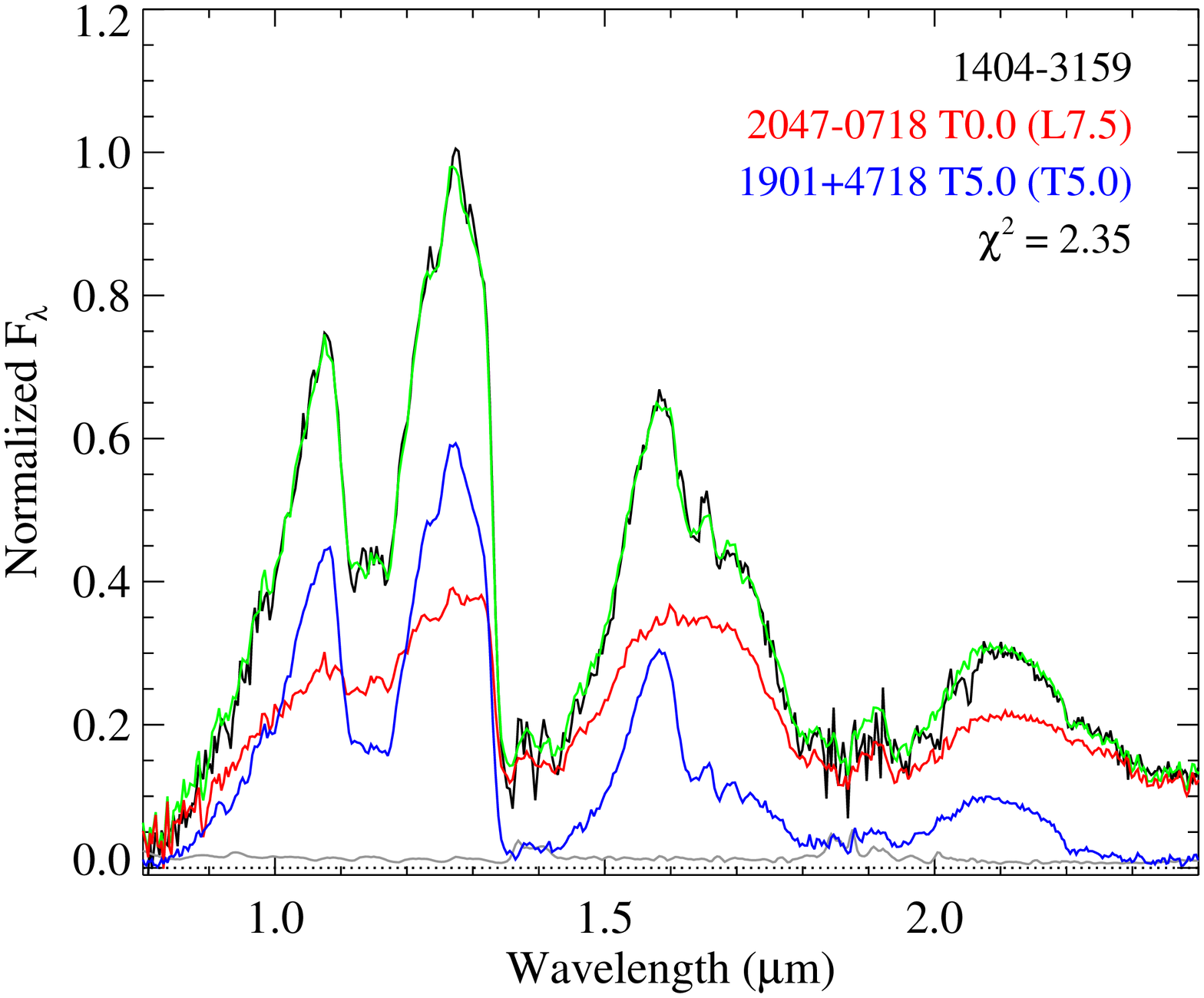}
\includegraphics[width=0.45\textwidth]{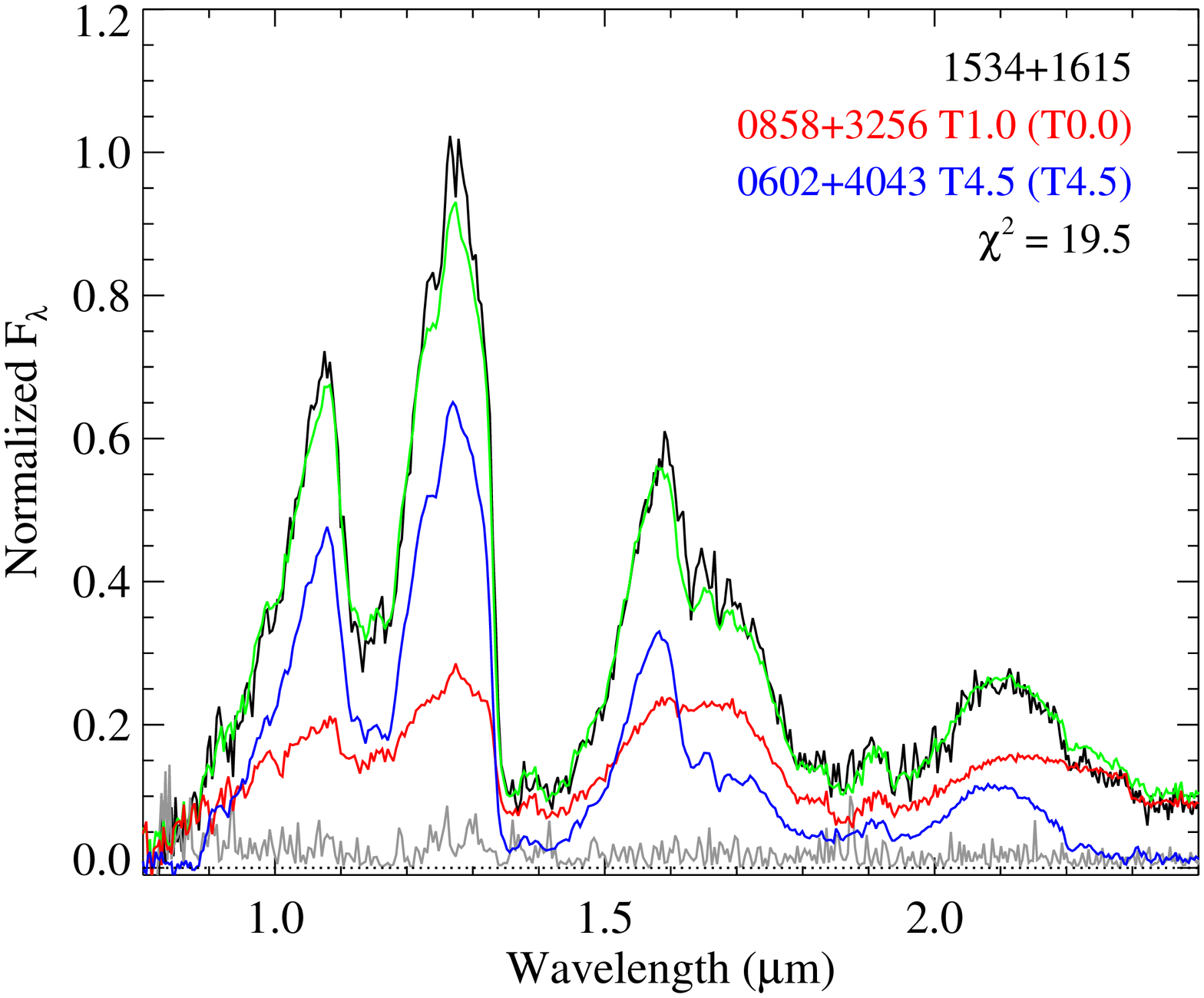}
\caption{\small Best-fit composites for the resolved L dwarf/T dwarf binaries  
shown in Figure~\ref{fig_known}.  In each panel, source spectra (black lines, in $F_{\lambda}$ flux units) are shown normalized to the peak flux in the 1.0--1.3~$\micron$ range; composites (green lines) are relatively normalized to minimize {\chisq} deviations; primary (red lines) and secondary (blue lines) component spectra are normalized to their respective contribution to the composite (based on the faint \citealt{2006ApJ...647.1393L}
relation; see Section~4.1).  
Noise spectra for candidate data are also indicated by grey lines. 
Component names, literature and index-based spectral types (the latter in parentheses) and {\chisq} deviations are listed in the top right corner of each panel.
\label{fig_knownbfit}}
\end{figure}

\clearpage

\begin{figure}
\centering
\epsscale{0.85}
\includegraphics[width=0.45\textwidth]{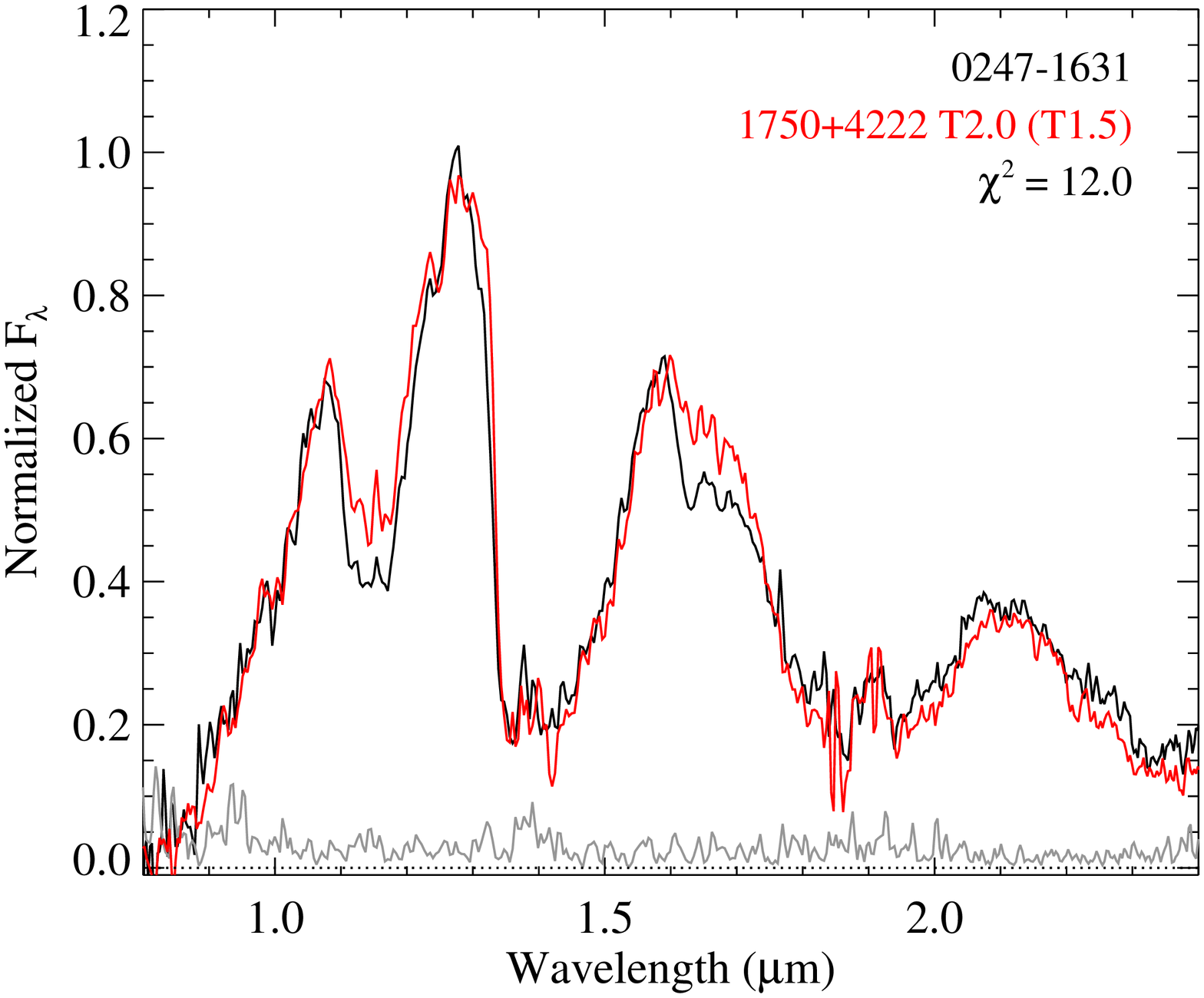}
\includegraphics[width=0.45\textwidth]{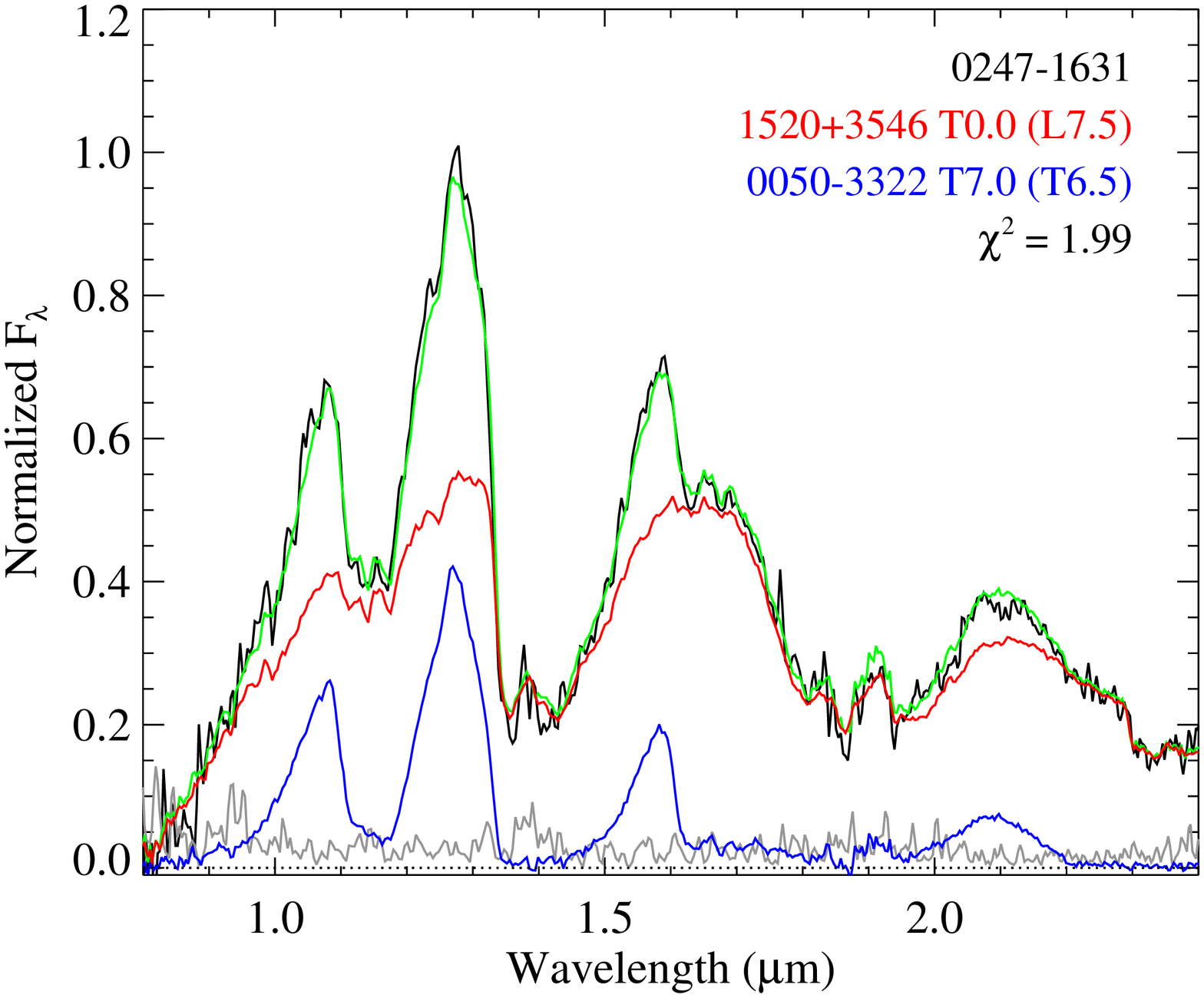}
\includegraphics[width=0.45\textwidth]{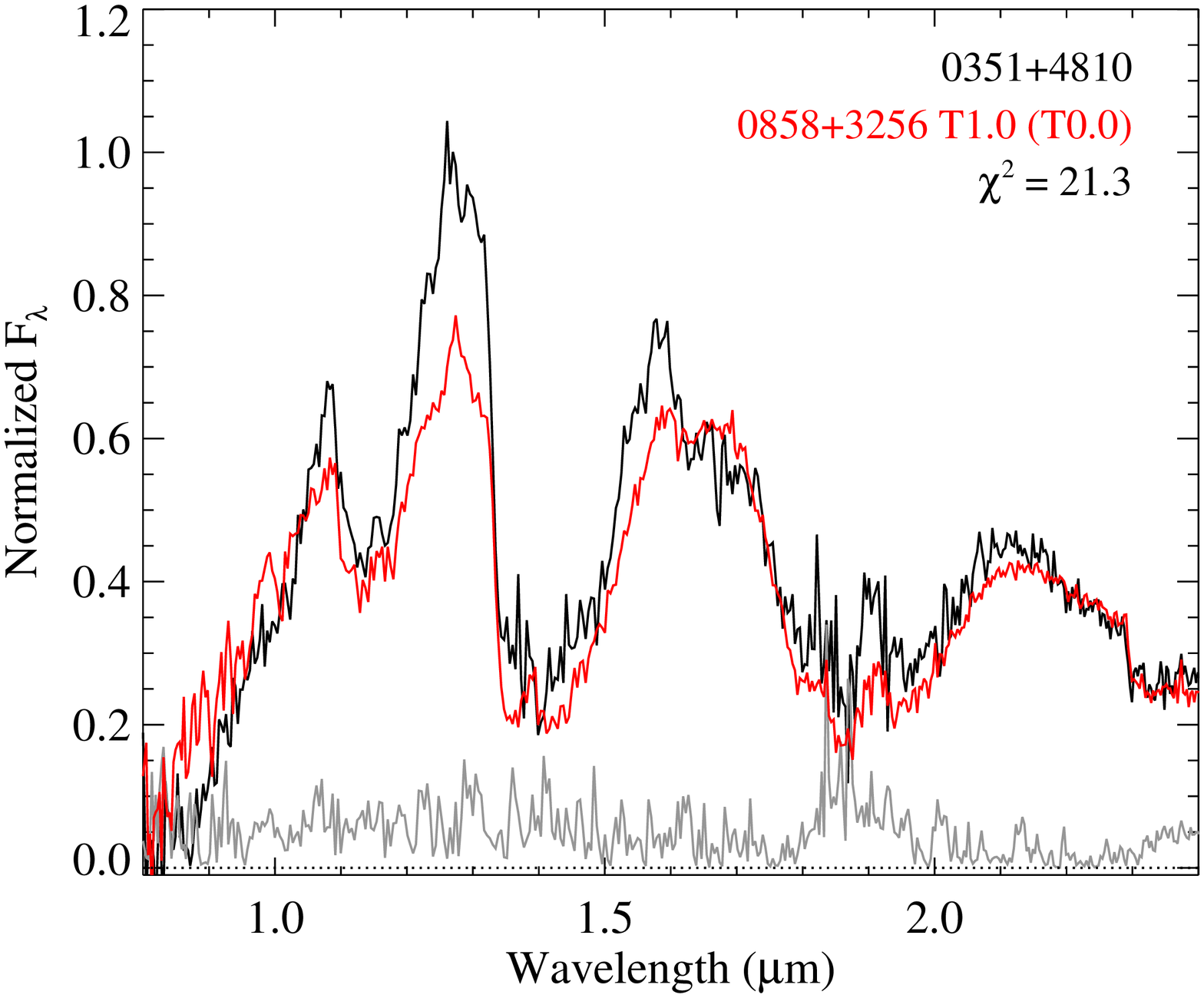}
\includegraphics[width=0.45\textwidth]{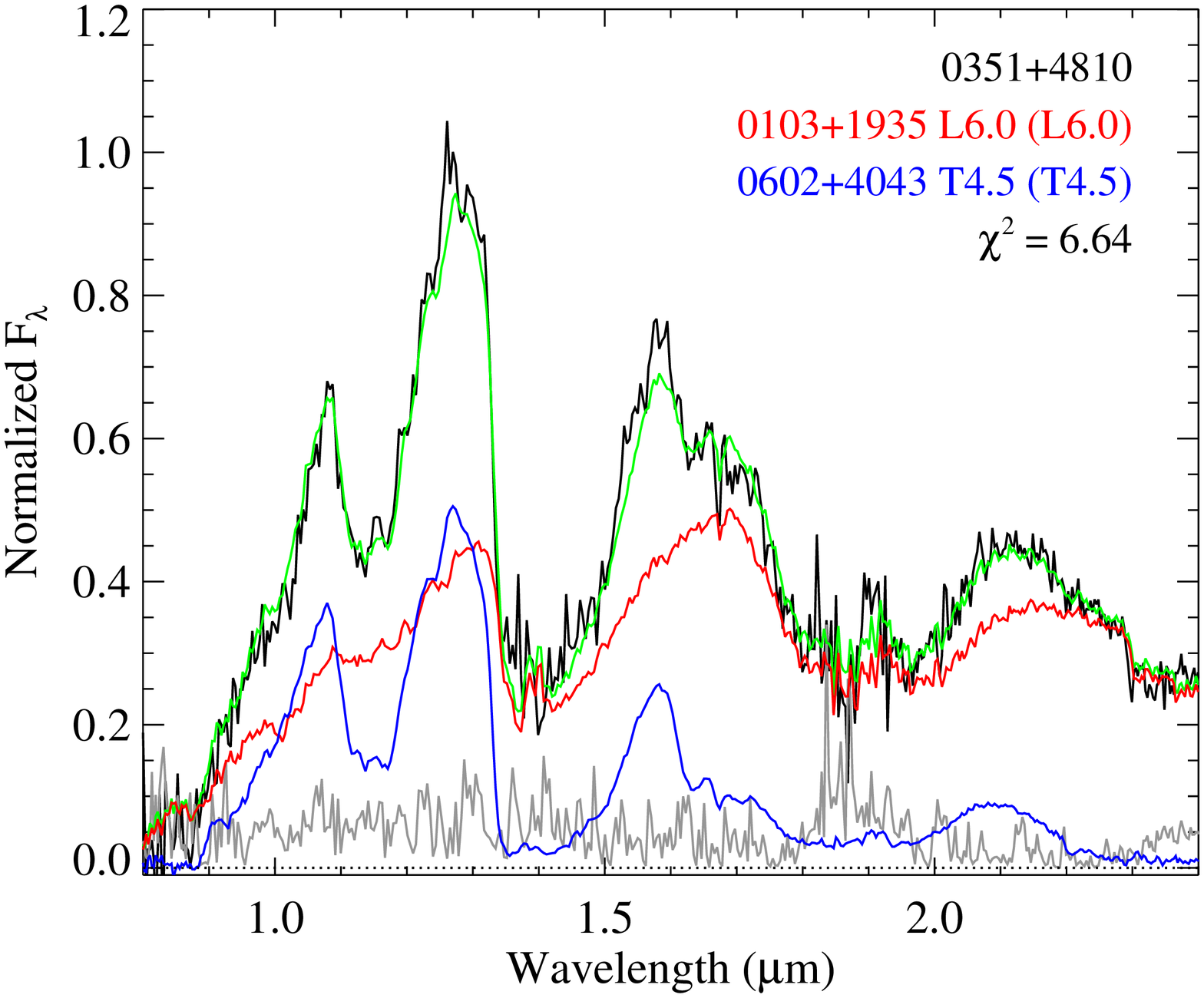}
\includegraphics[width=0.45\textwidth]{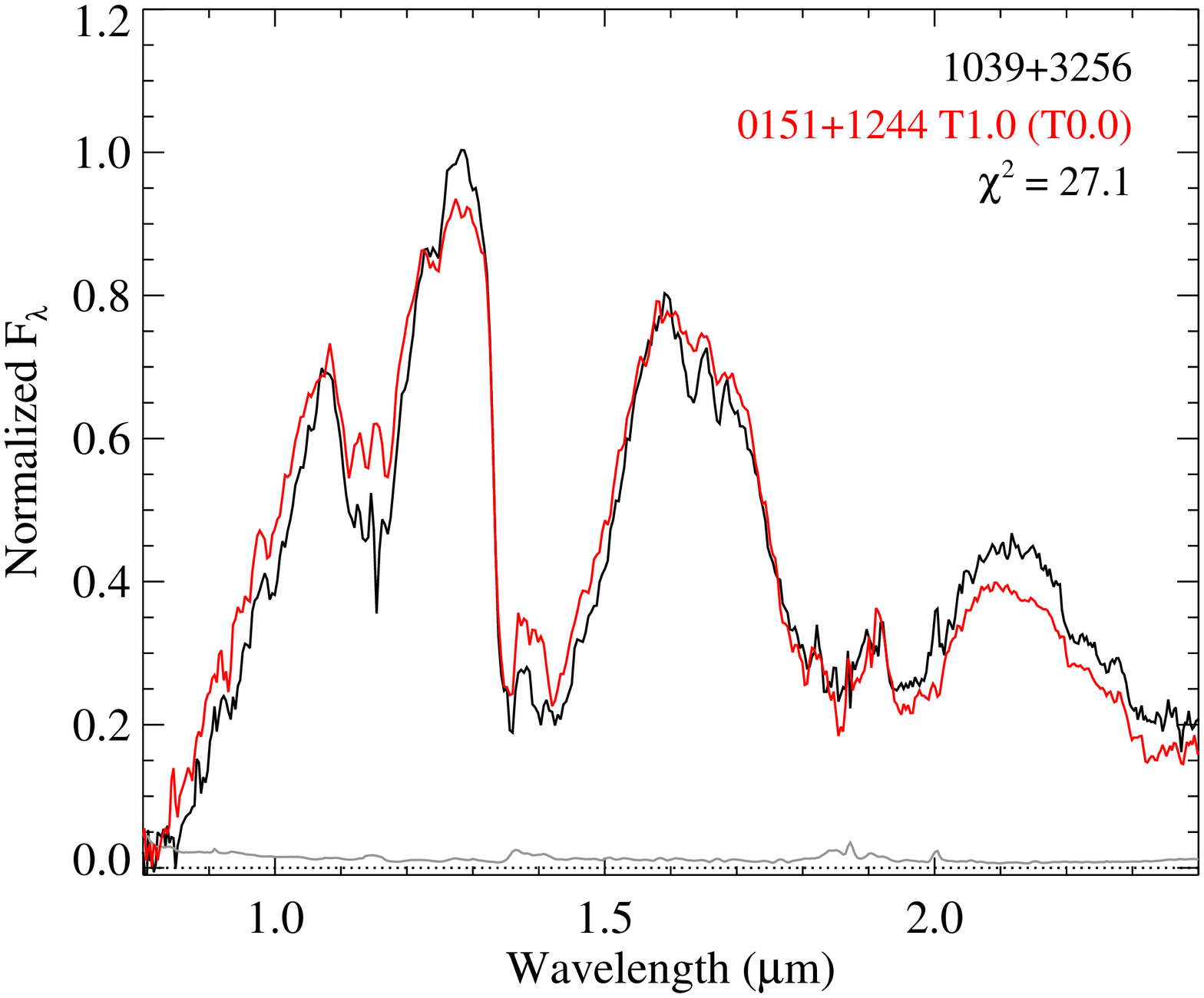}
\includegraphics[width=0.45\textwidth]{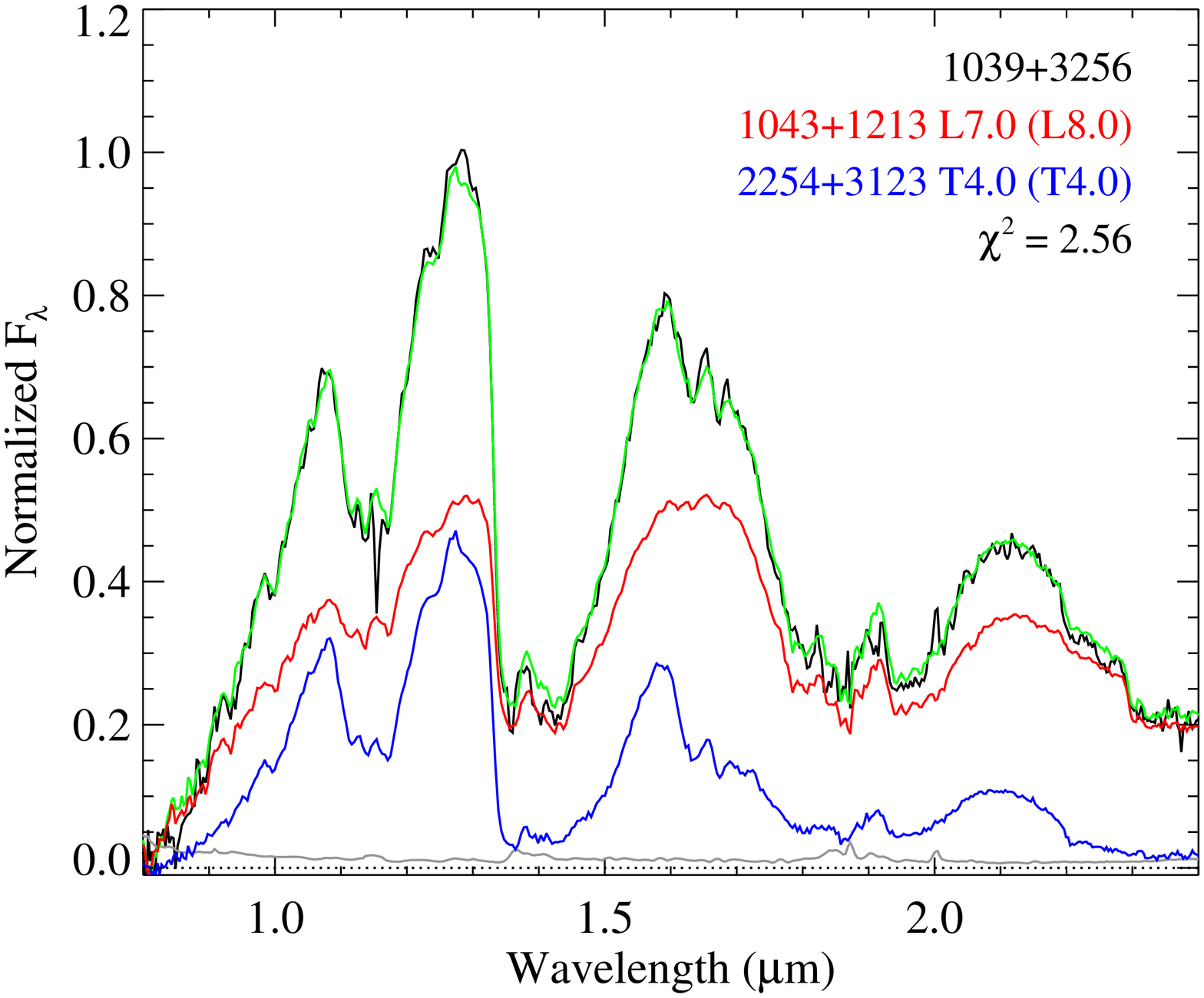}
\caption{\small Spectral fits to the sources identified as ``strong'' binary candidates (at least three selection criteria satisfied).  For each row, the left panel shows the best-fit single template (red lines) compared to the 
source spectrum (black lines, noise spectrum in grey), while the right panel shows the best-fit composite
(green lines), primary (red lines) and secondary spectra (blue lines).
Source spectra are normalized to the peak flux in the 1.0--1.3~$\micron$ range, while single and composites
are normalized to minimize {\chisq} deviations.  Component spectra in the right panels are 
normalized to their respective contribution to the composite, according to the faint \citet{2006ApJ...647.1393L}
relation (see Section~4.1). Template source names, literature and index-based spectral types (the latter in parentheses) and corresponding {\chisq} deviations are labeled in the top right corner of each panel.
\label{fig_fitsstrong}}
\end{figure}

\clearpage

\addtocounter{figure}{-1}

\begin{figure}
\centering
\epsscale{0.85}
\includegraphics[width=0.45\textwidth]{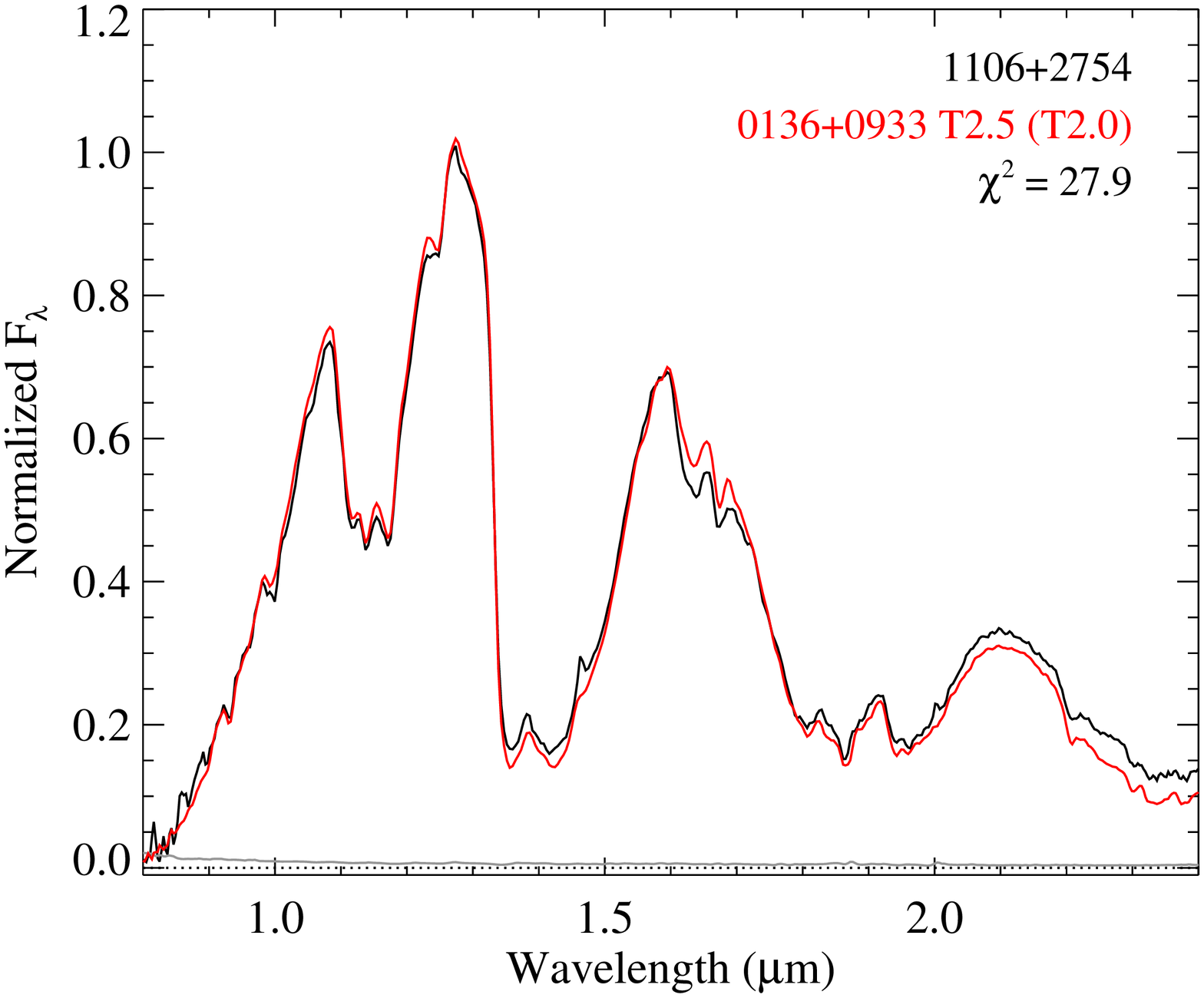}
\includegraphics[width=0.45\textwidth]{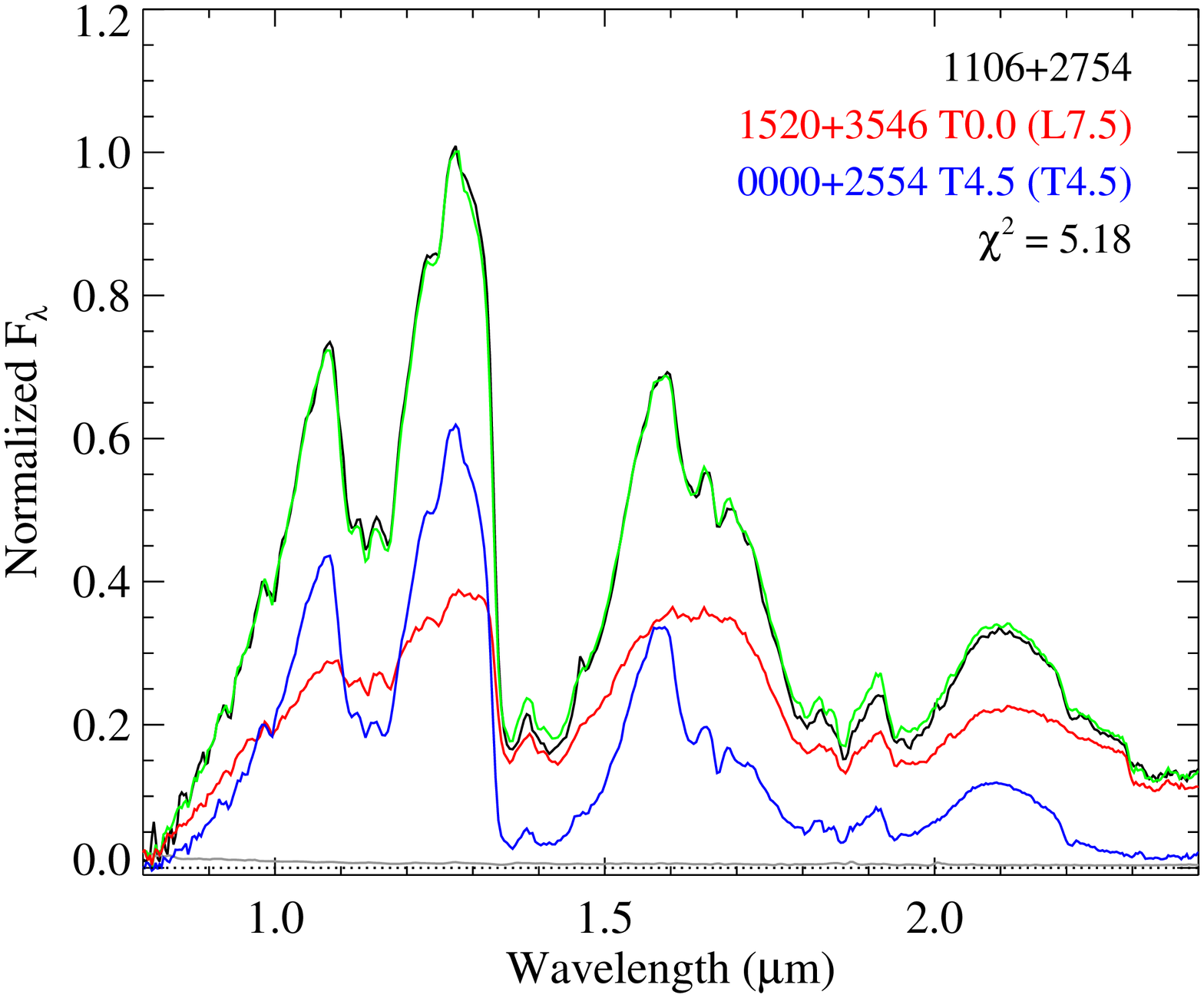}
\includegraphics[width=0.45\textwidth]{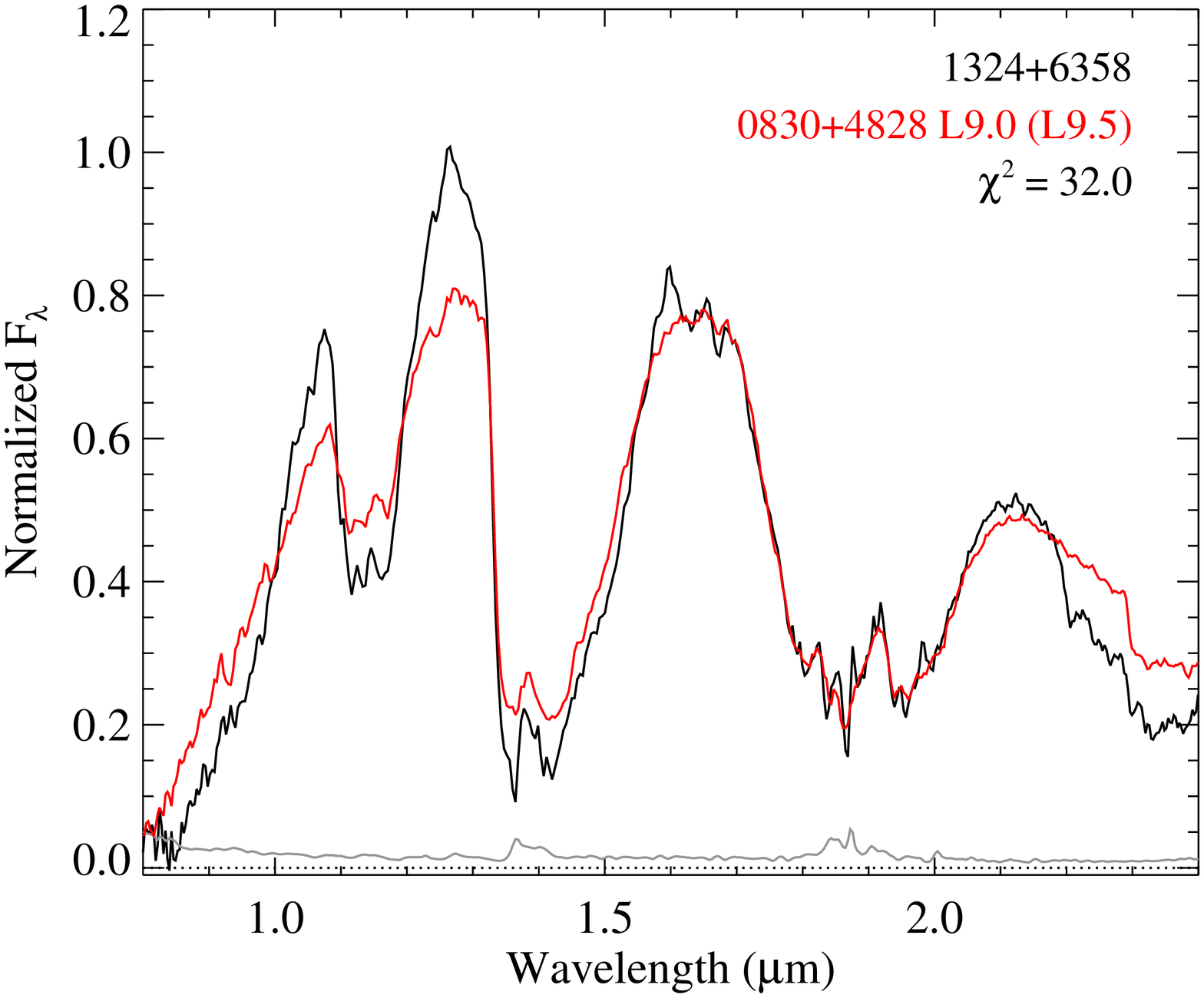}
\includegraphics[width=0.45\textwidth]{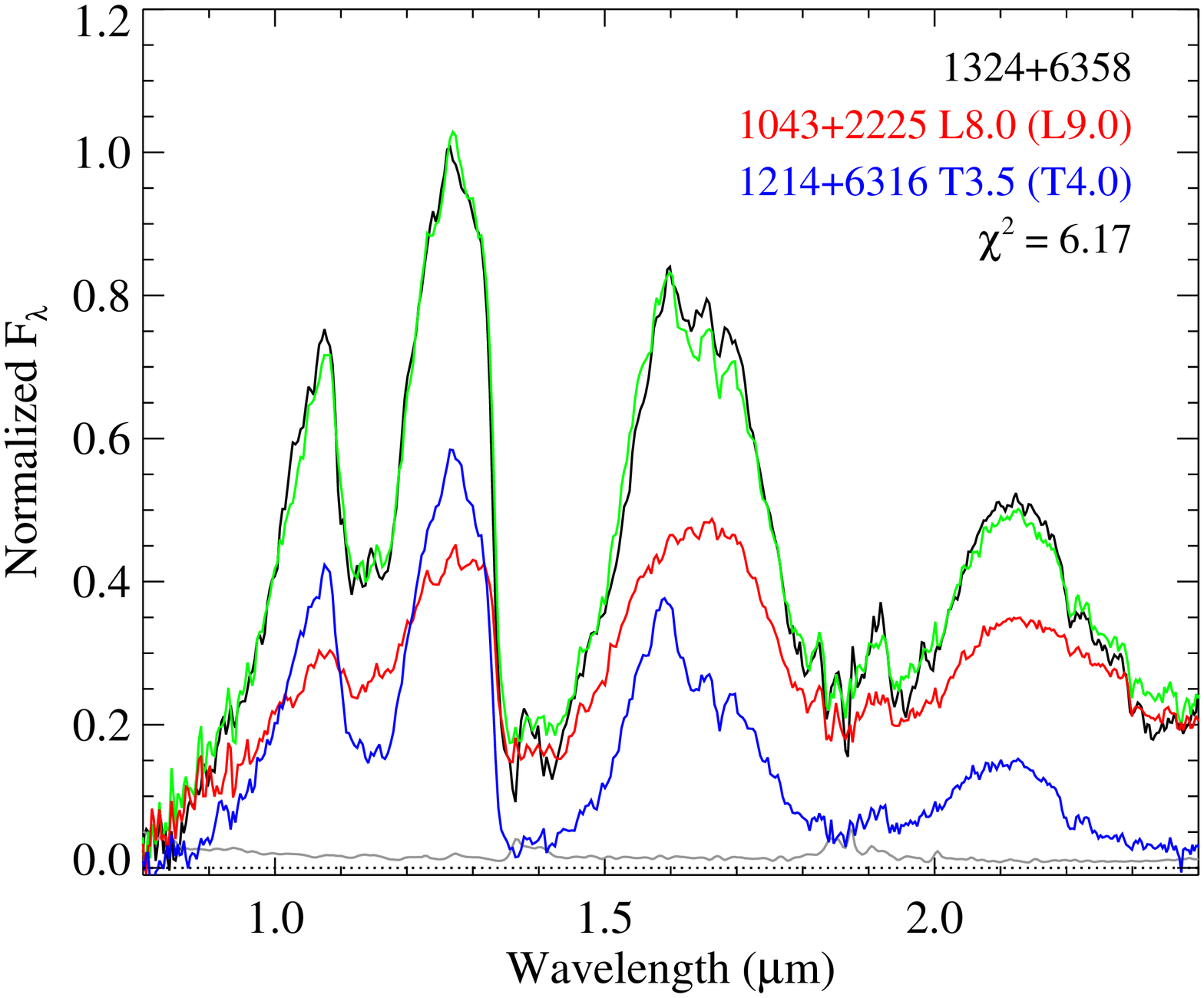}
\includegraphics[width=0.45\textwidth]{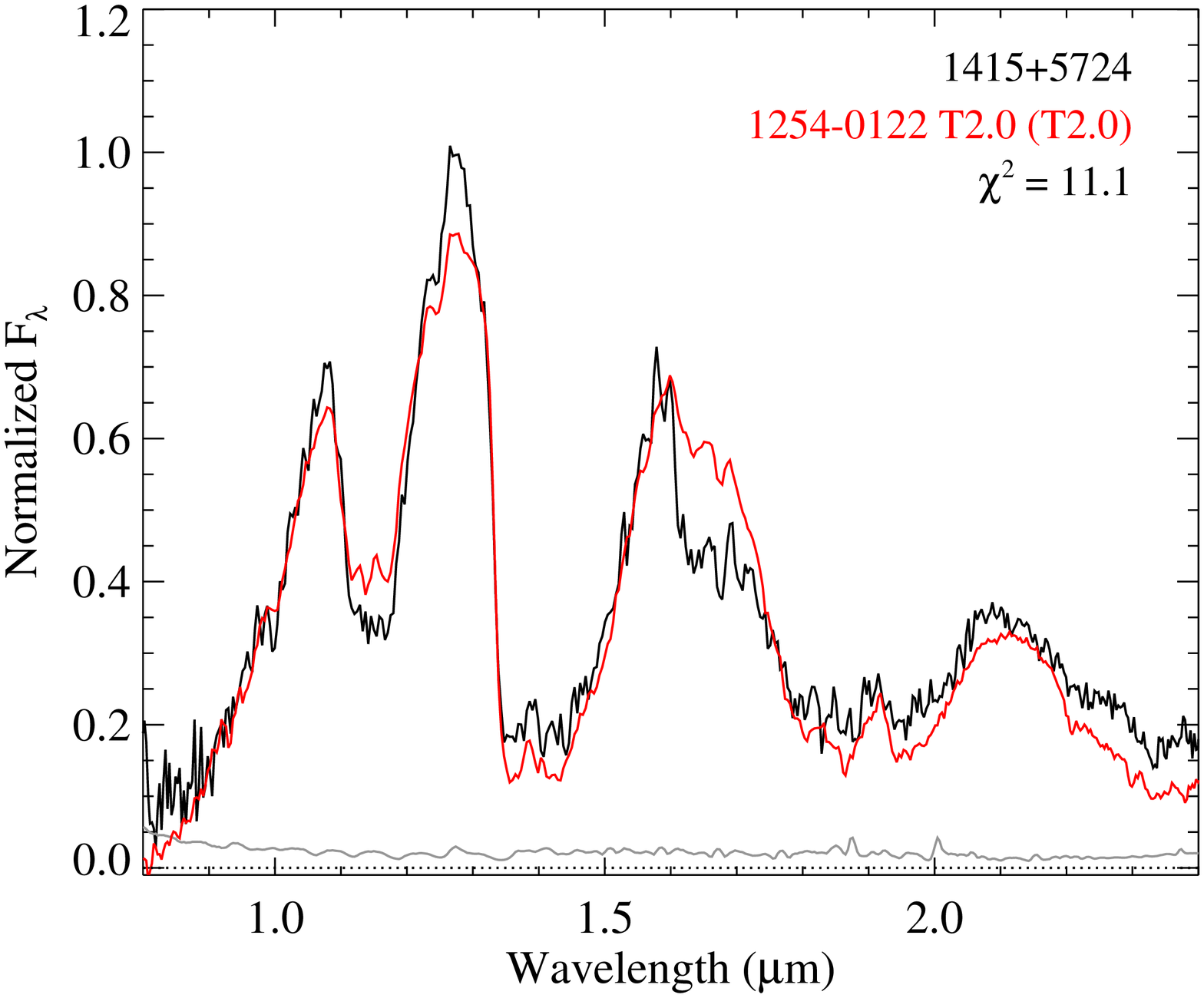}
\includegraphics[width=0.45\textwidth]{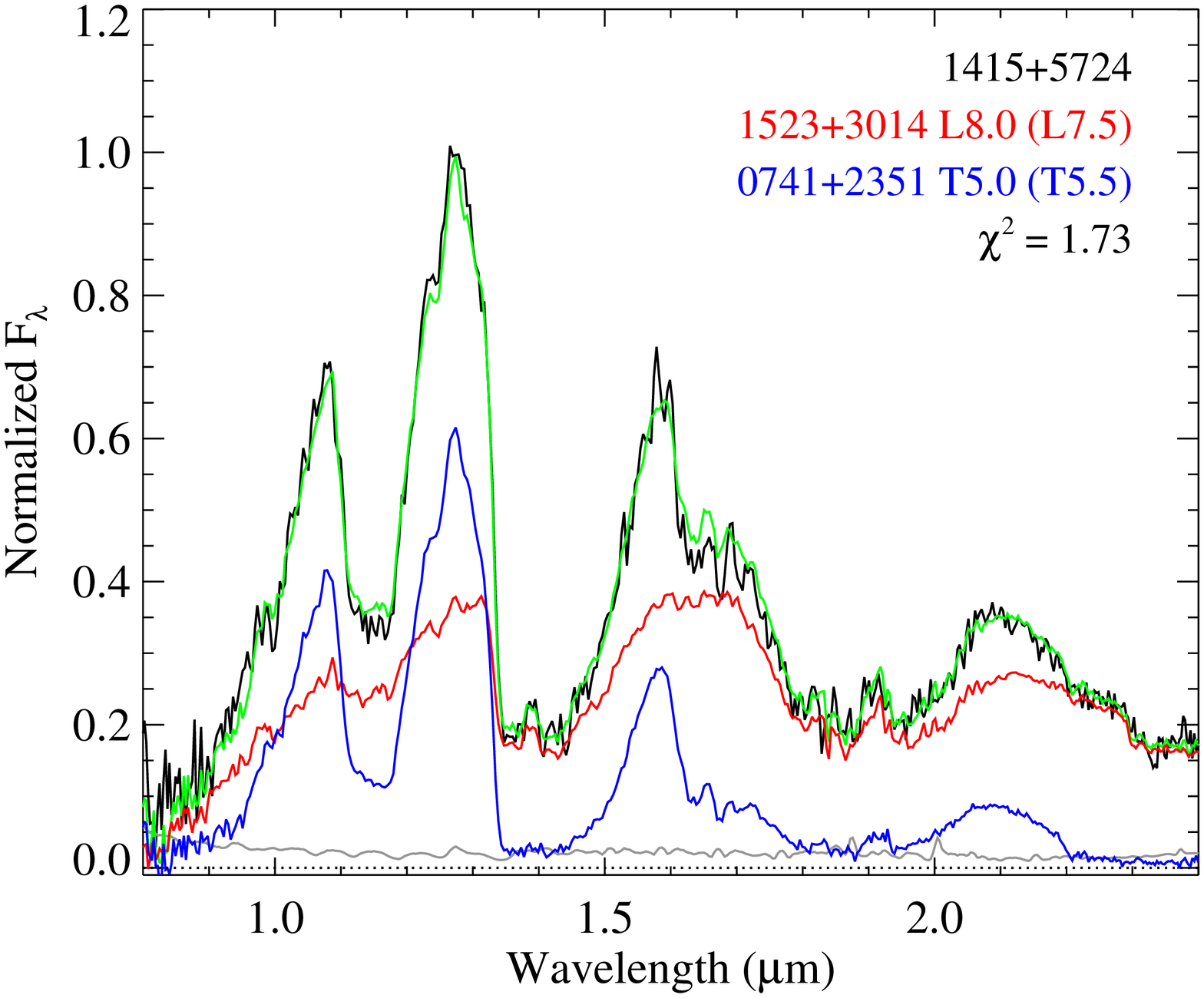}
\caption{Continued.}
\end{figure}

\clearpage

\addtocounter{figure}{-1}

\begin{figure}
\centering
\epsscale{0.85}
\includegraphics[width=0.45\textwidth]{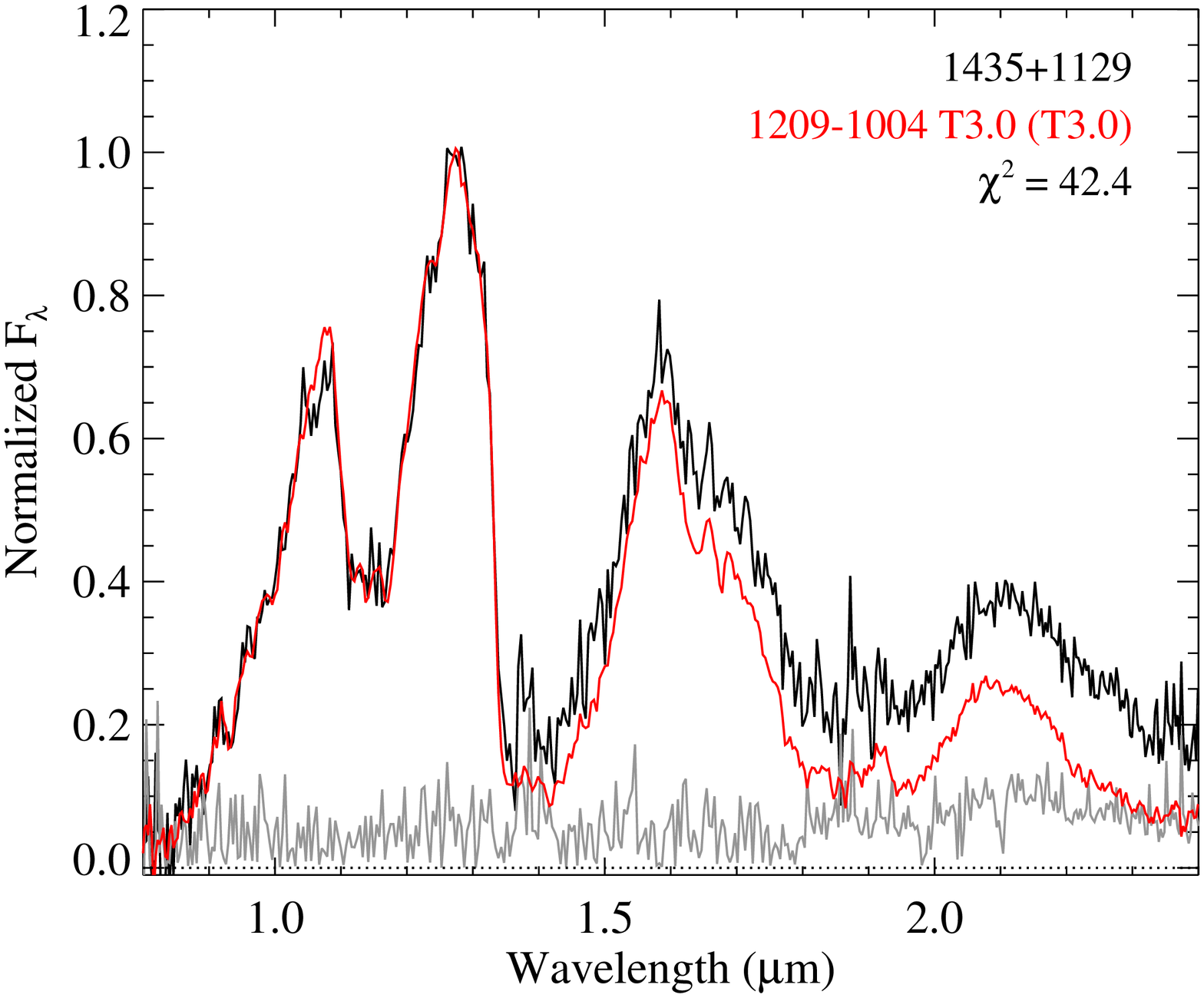}
\includegraphics[width=0.45\textwidth]{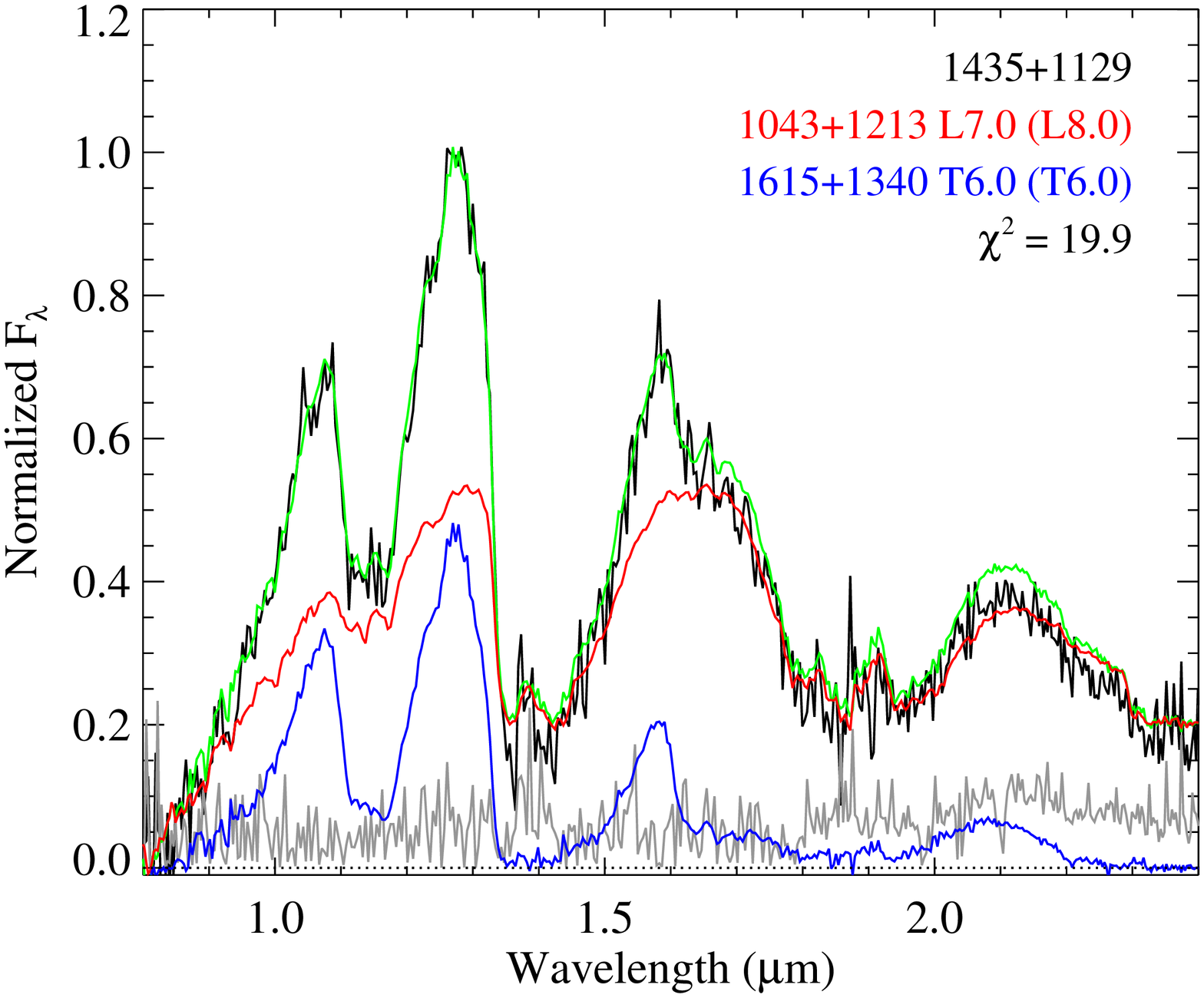}
\includegraphics[width=0.45\textwidth]{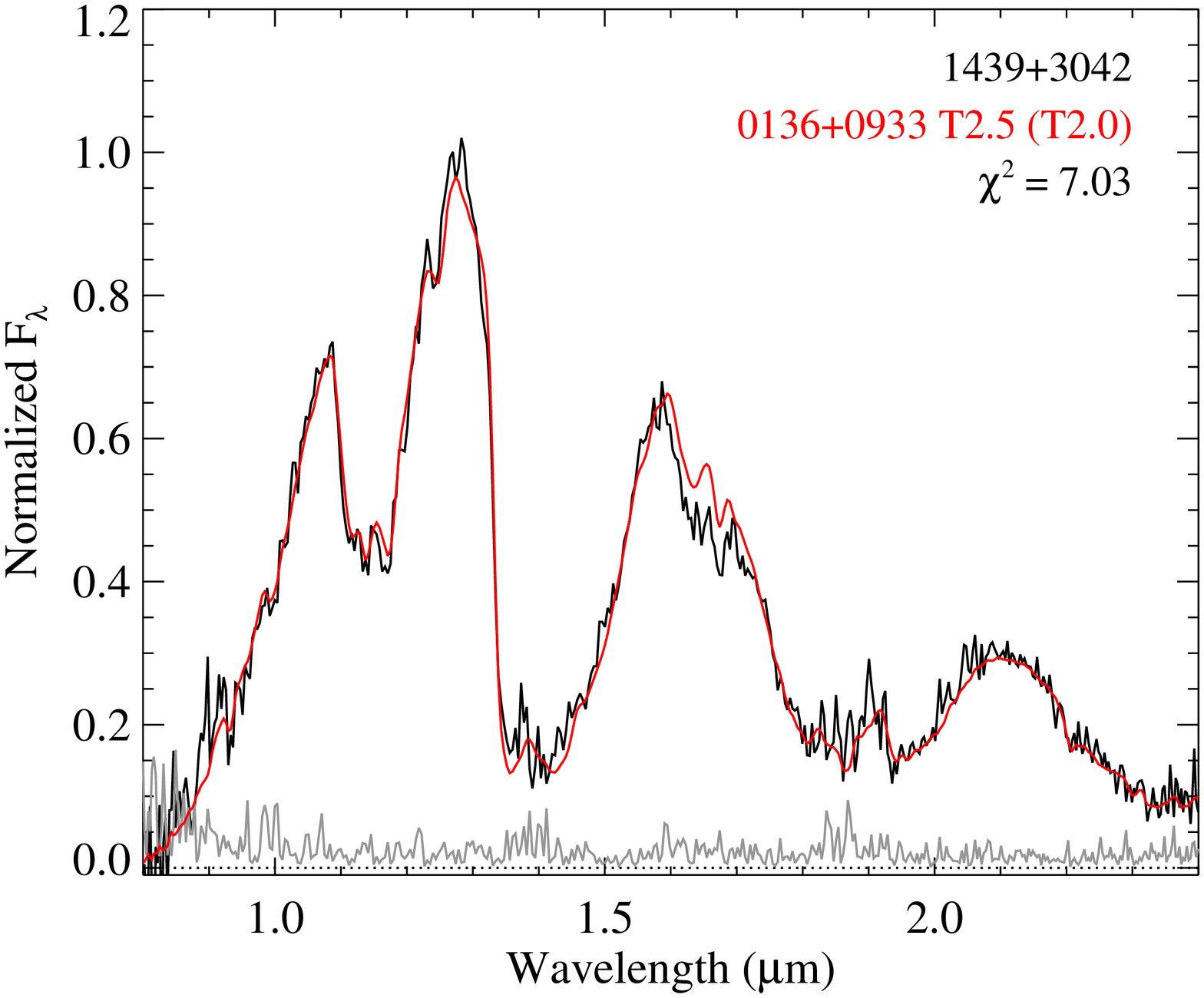}
\includegraphics[width=0.45\textwidth]{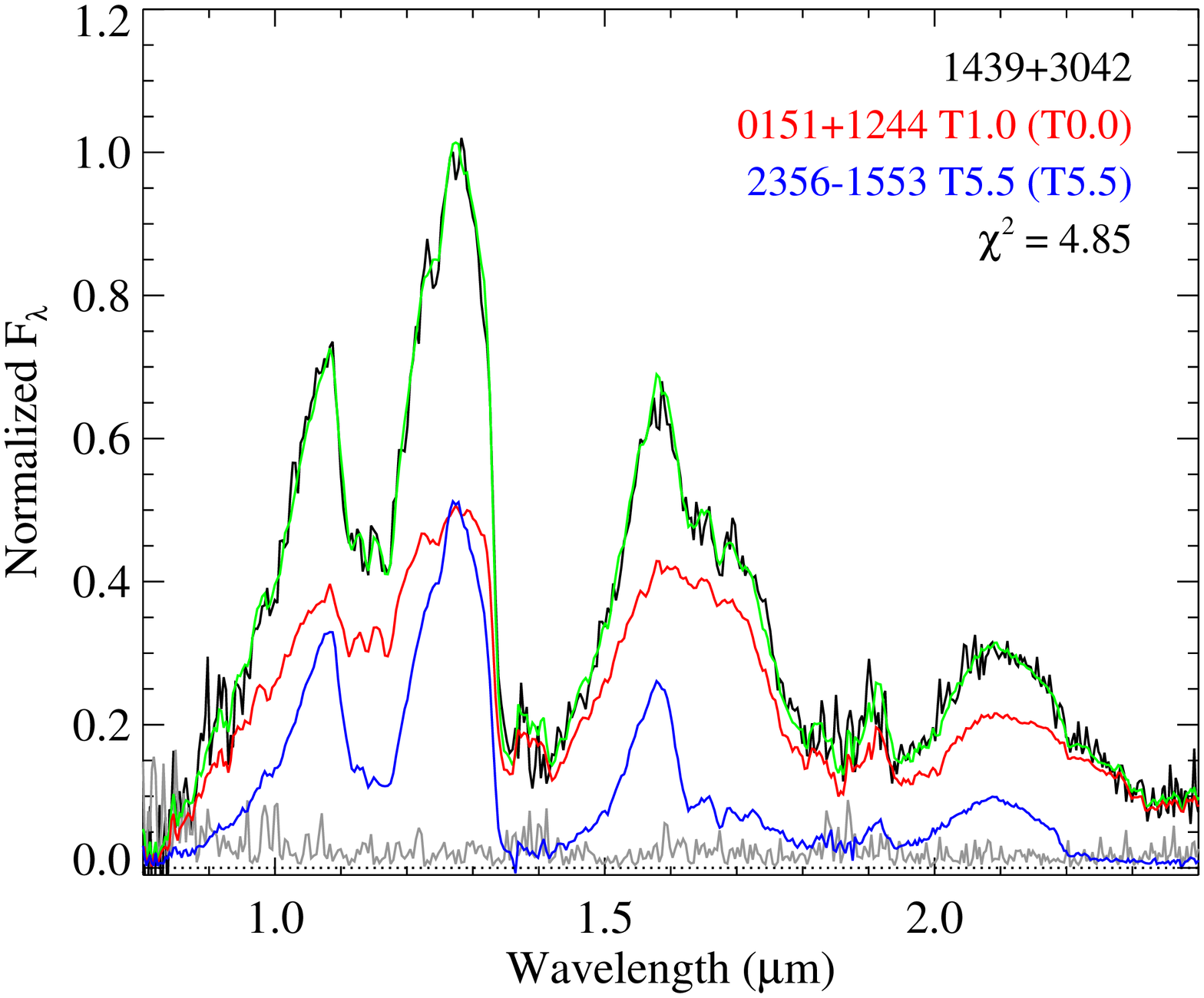}
\includegraphics[width=0.45\textwidth]{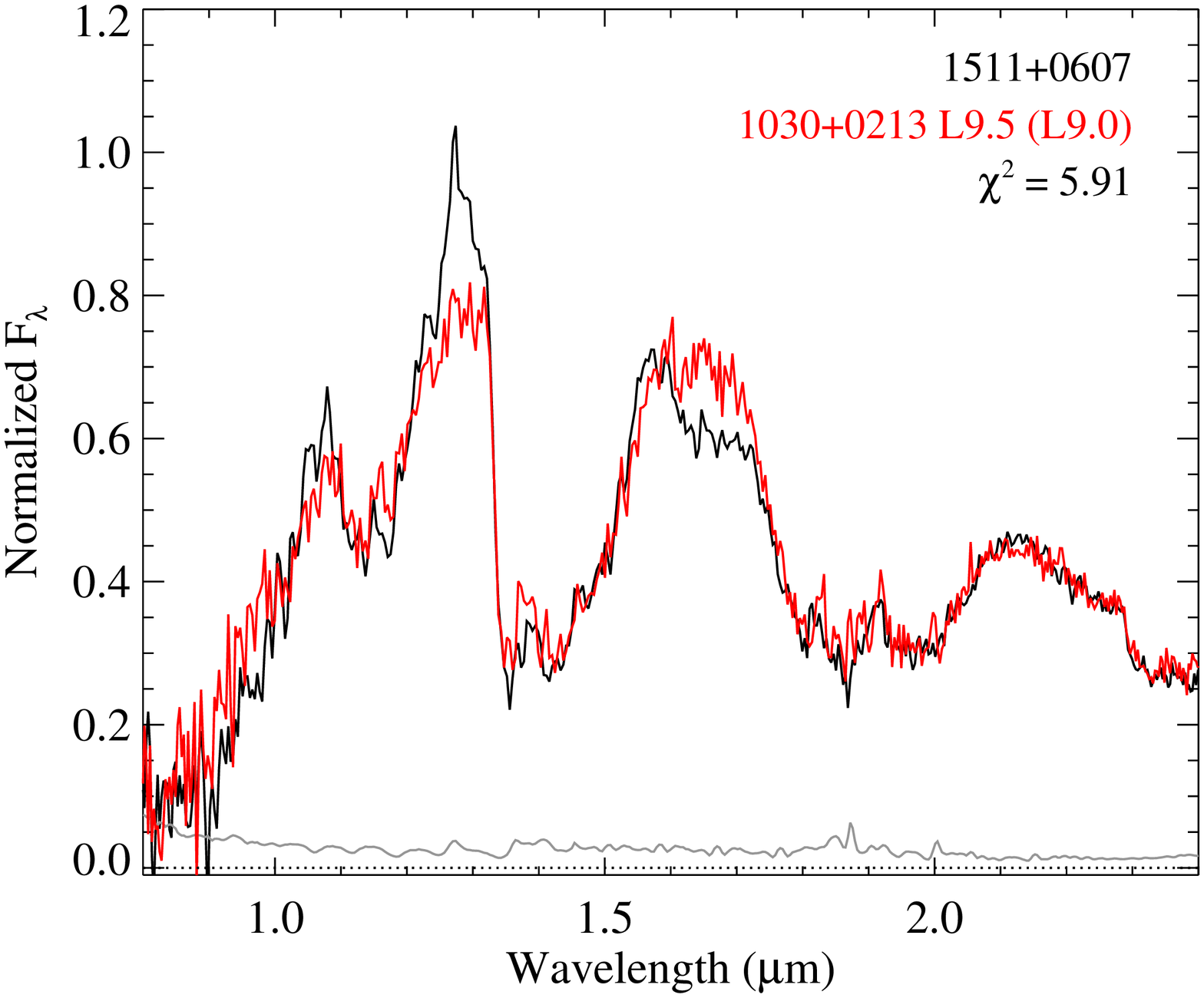}
\includegraphics[width=0.45\textwidth]{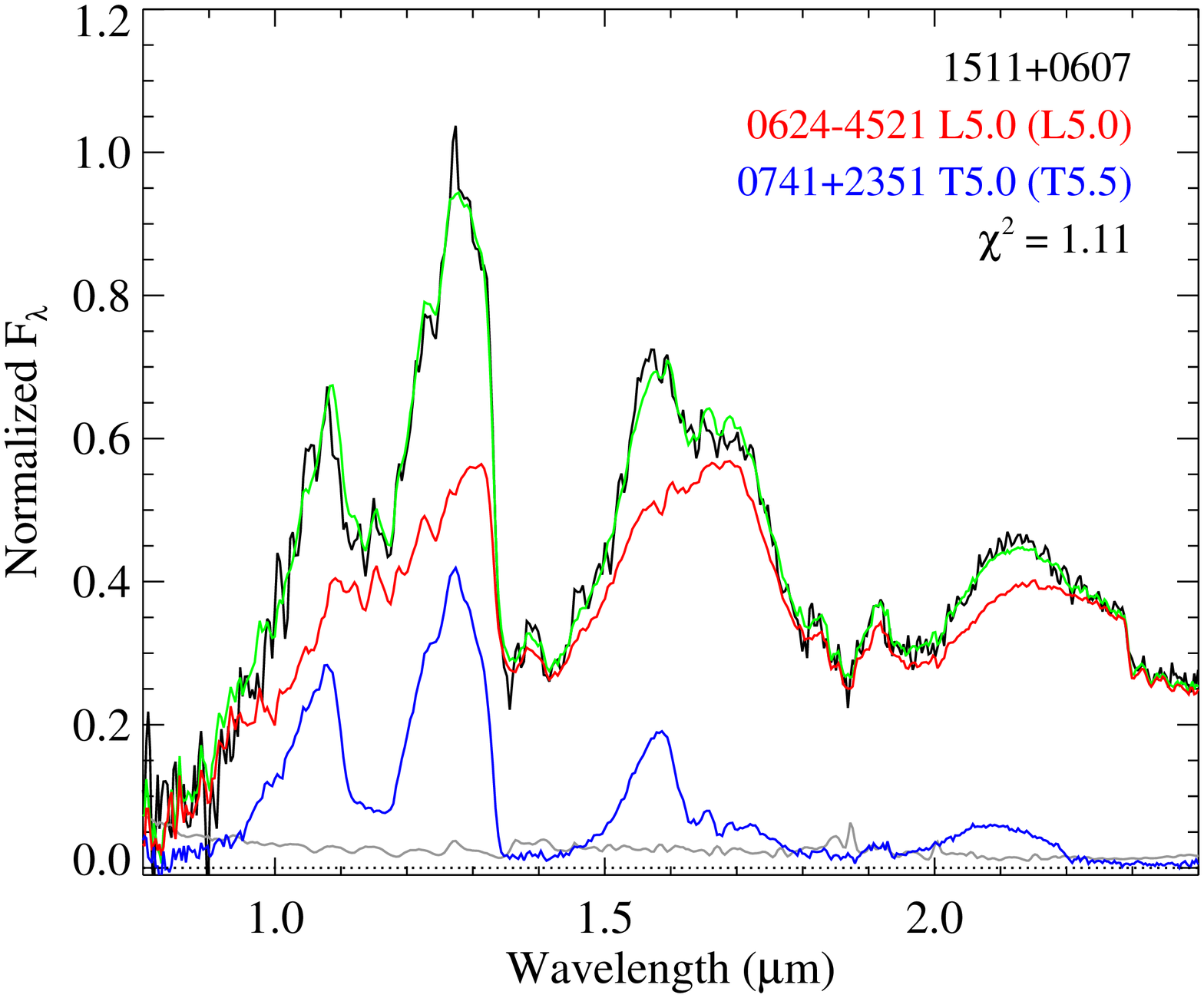}
\caption{Continued.}
\end{figure}

\clearpage

\addtocounter{figure}{-1}

\begin{figure}
\centering
\epsscale{0.85}
\includegraphics[width=0.45\textwidth]{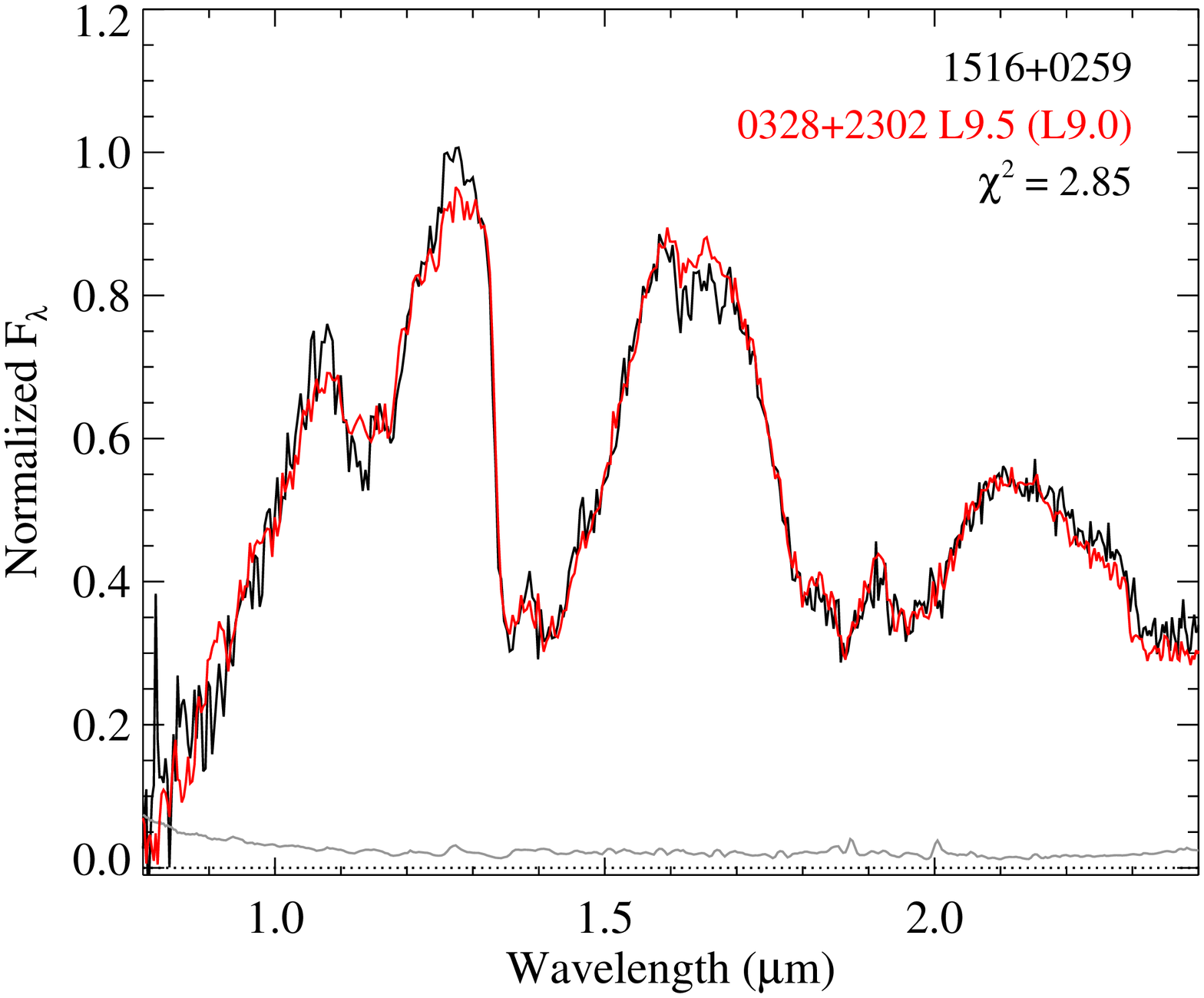}
\includegraphics[width=0.45\textwidth]{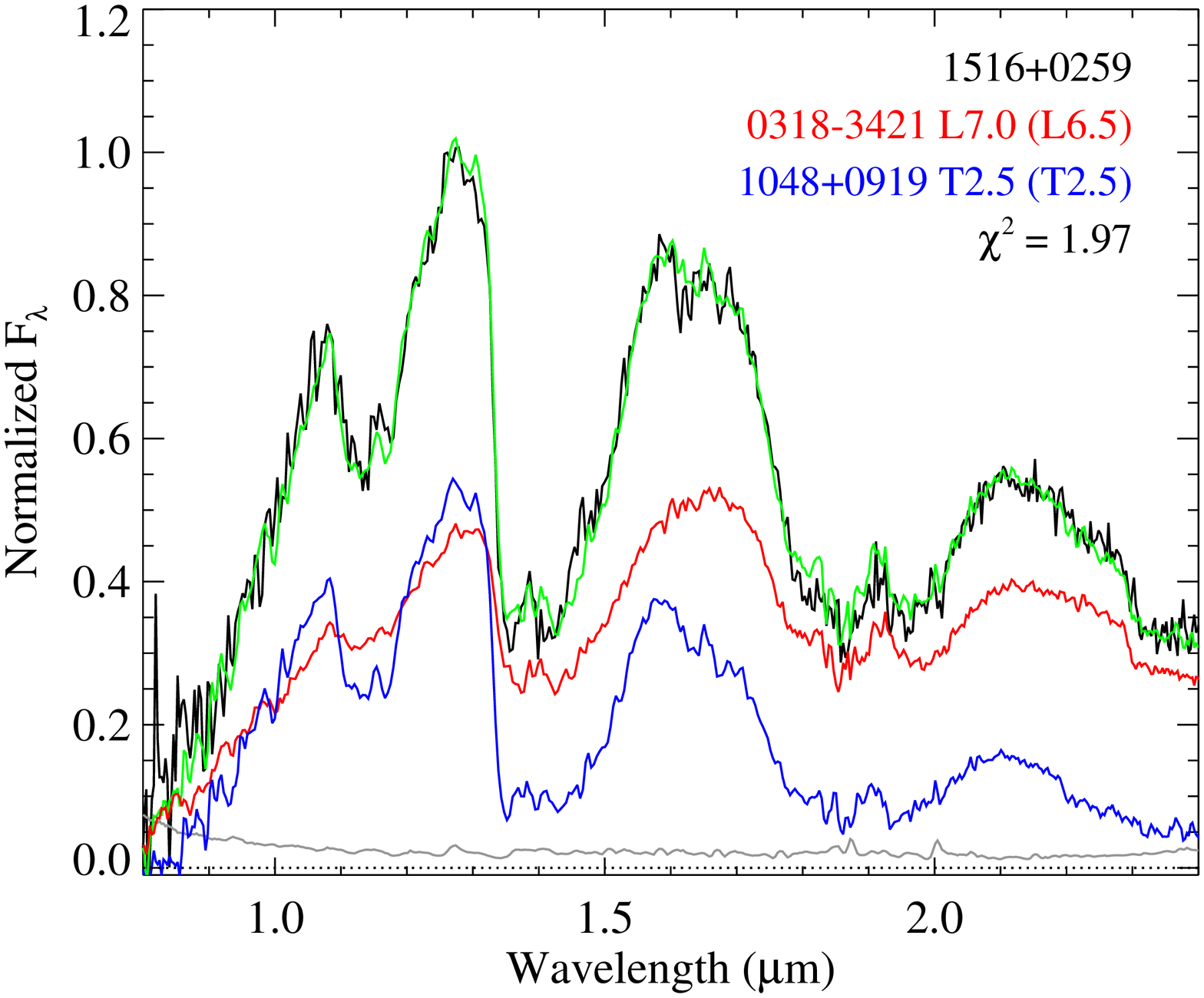}
\includegraphics[width=0.45\textwidth]{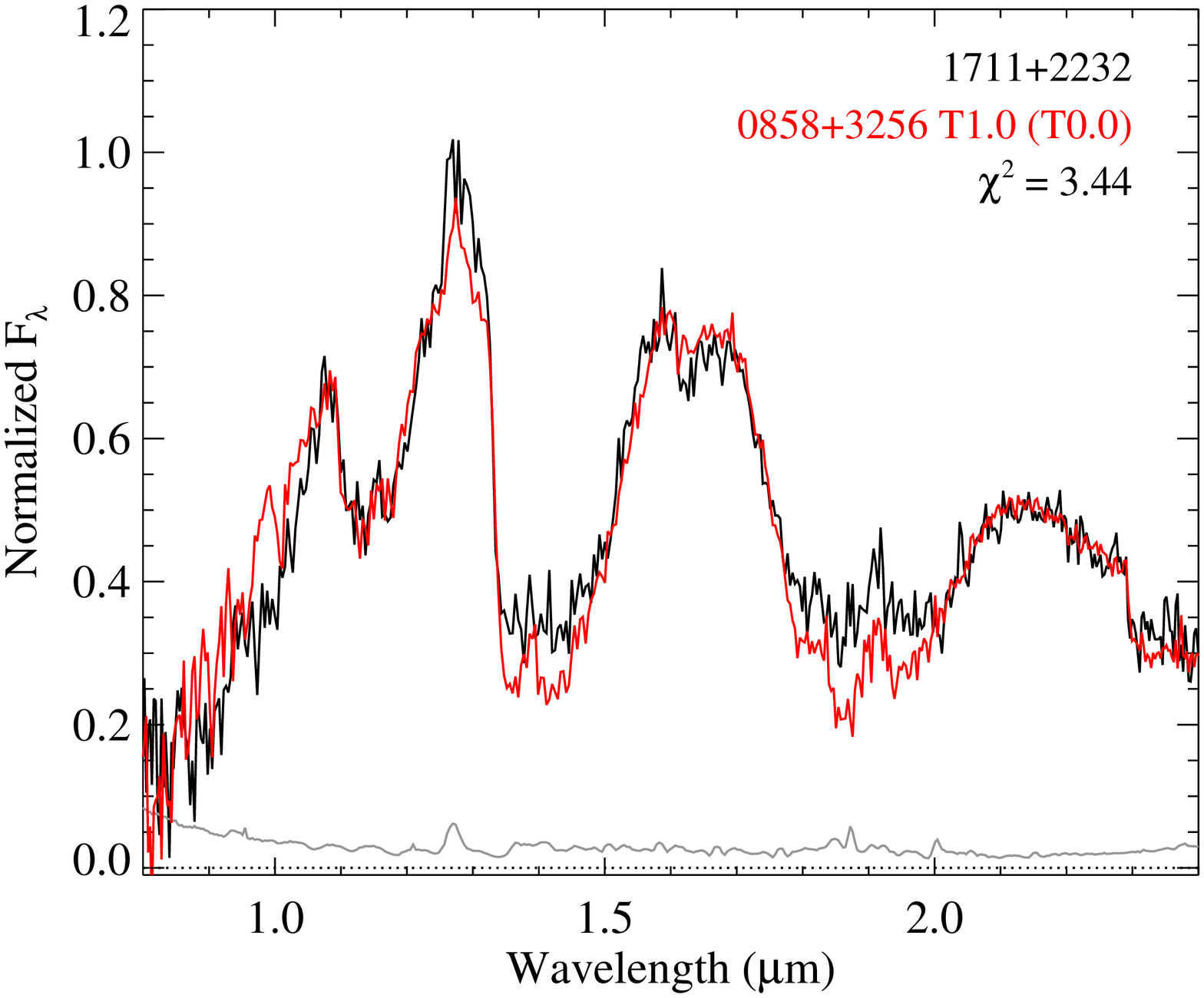}
\includegraphics[width=0.45\textwidth]{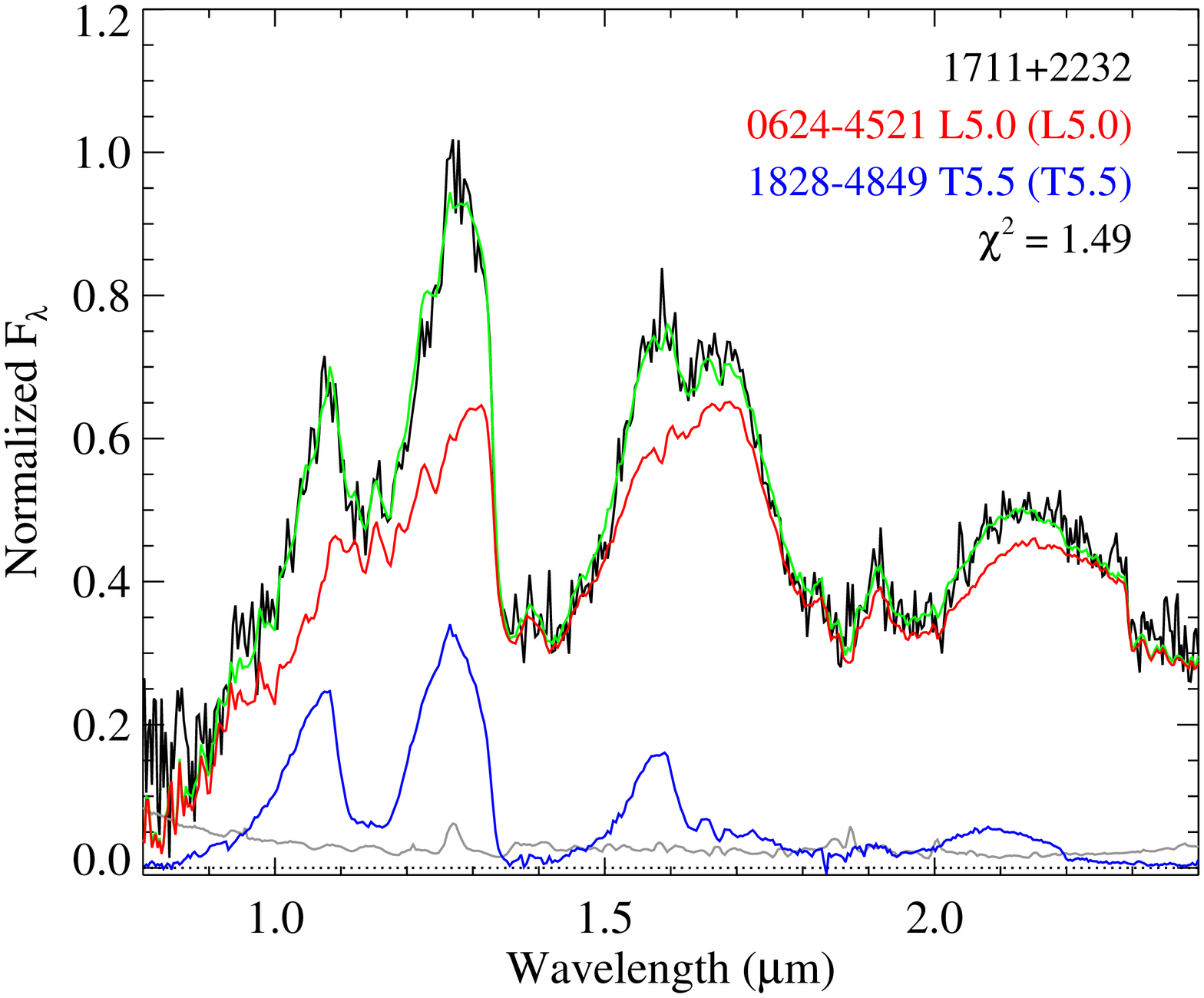}
\includegraphics[width=0.45\textwidth]{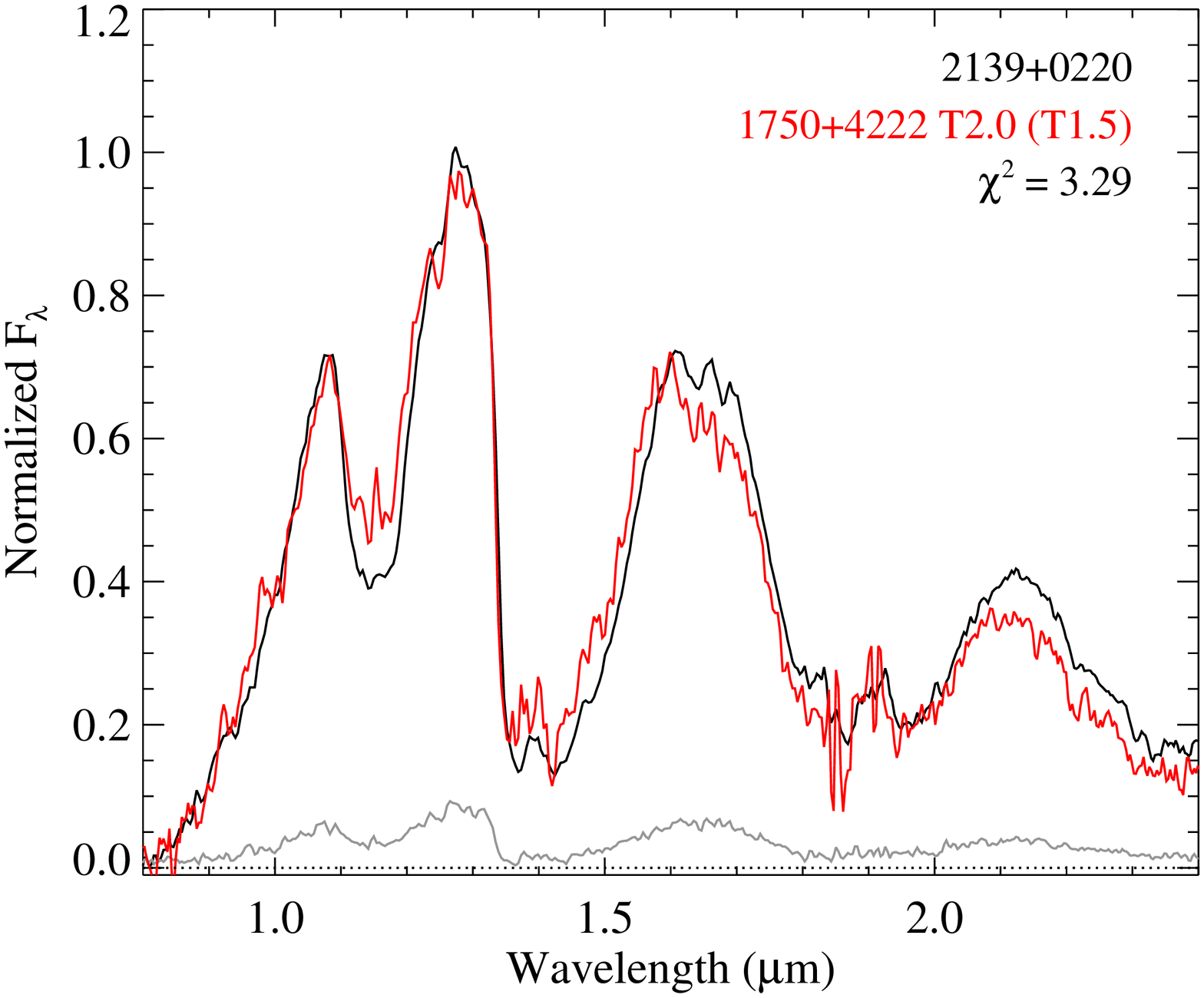}
\includegraphics[width=0.45\textwidth]{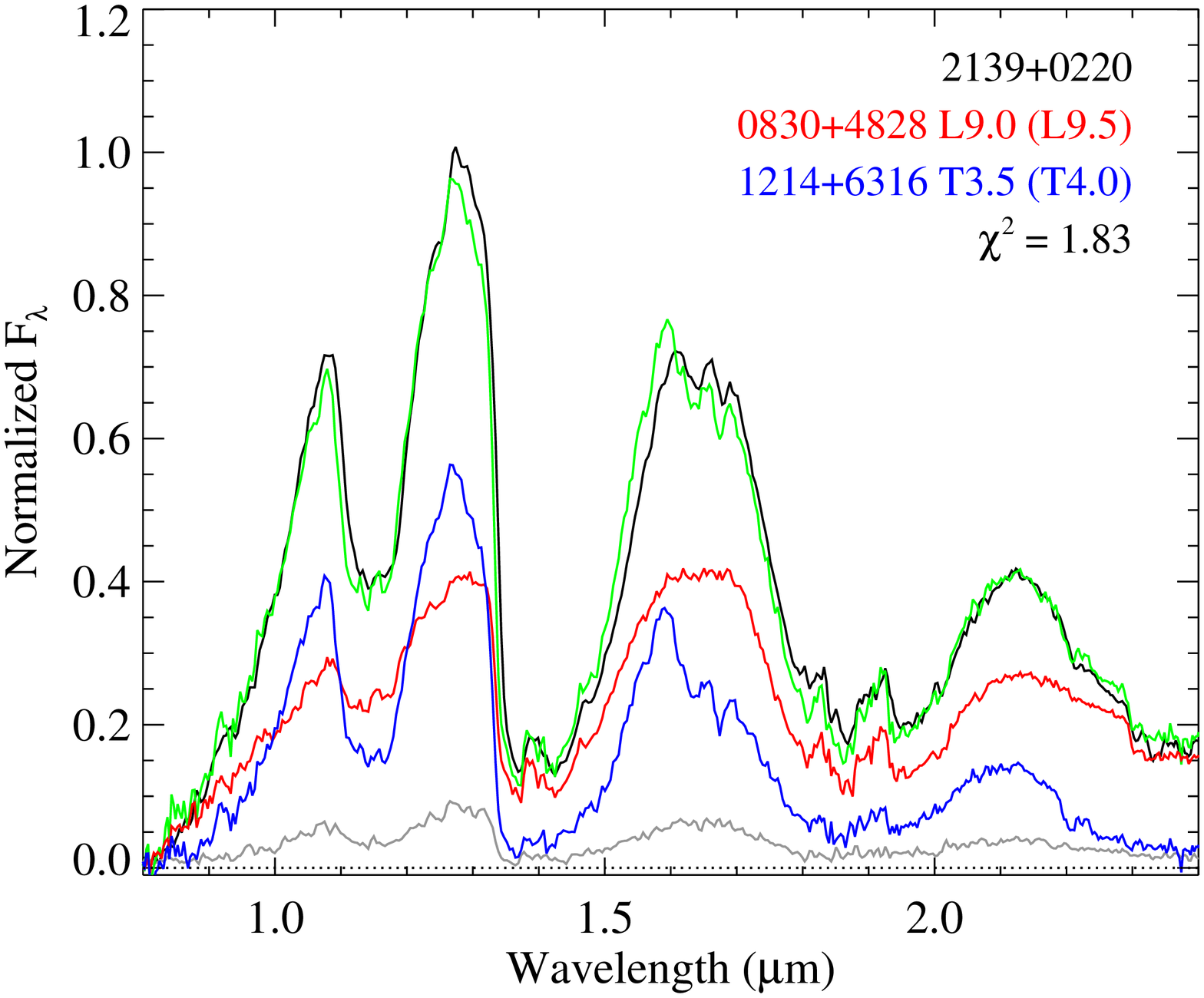}
\caption{Continued.}
\end{figure}

\clearpage

\begin{figure}
\centering
\epsscale{0.85}
\includegraphics[width=0.45\textwidth]{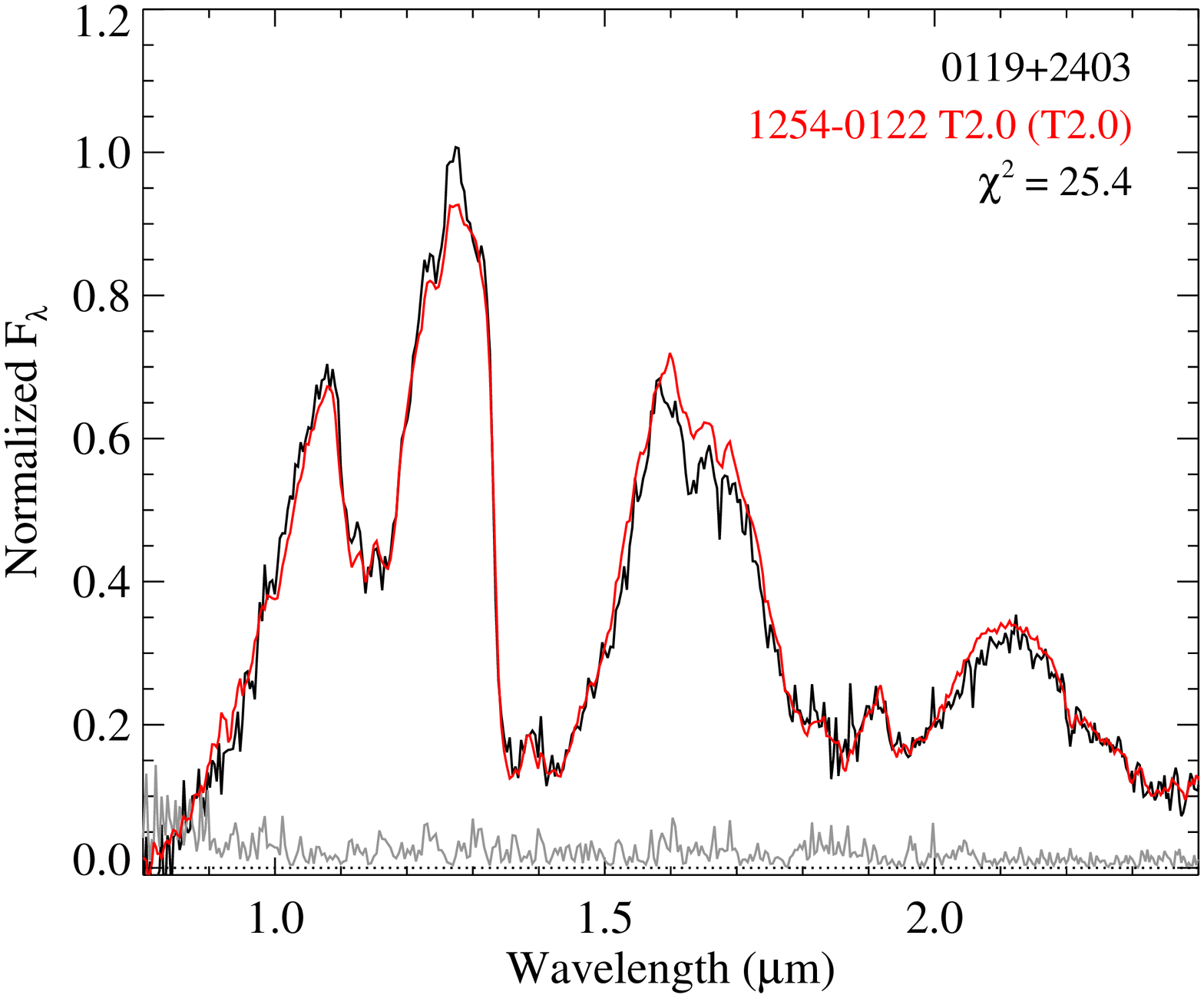}
\includegraphics[width=0.45\textwidth]{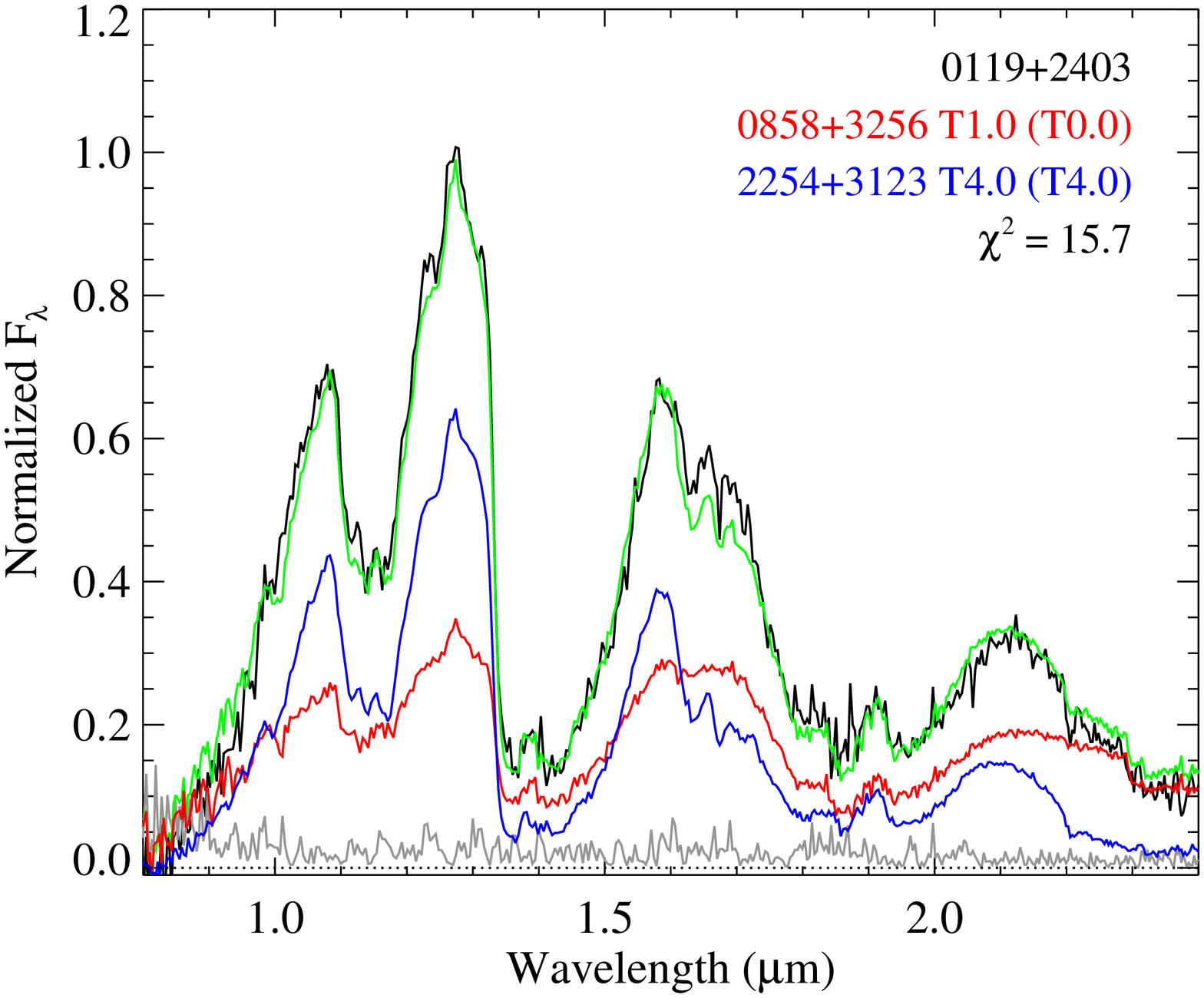}
\includegraphics[width=0.45\textwidth]{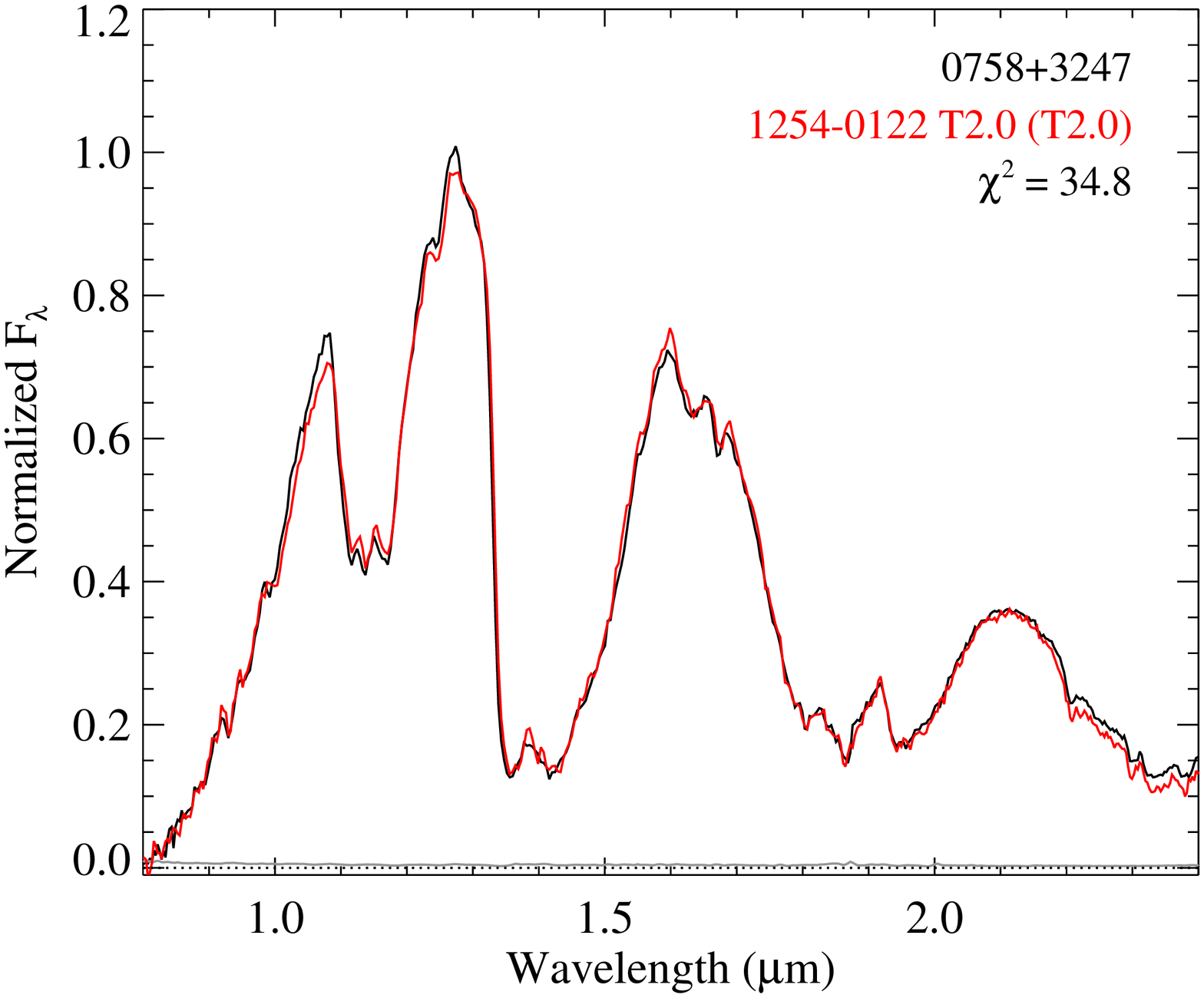}
\includegraphics[width=0.45\textwidth]{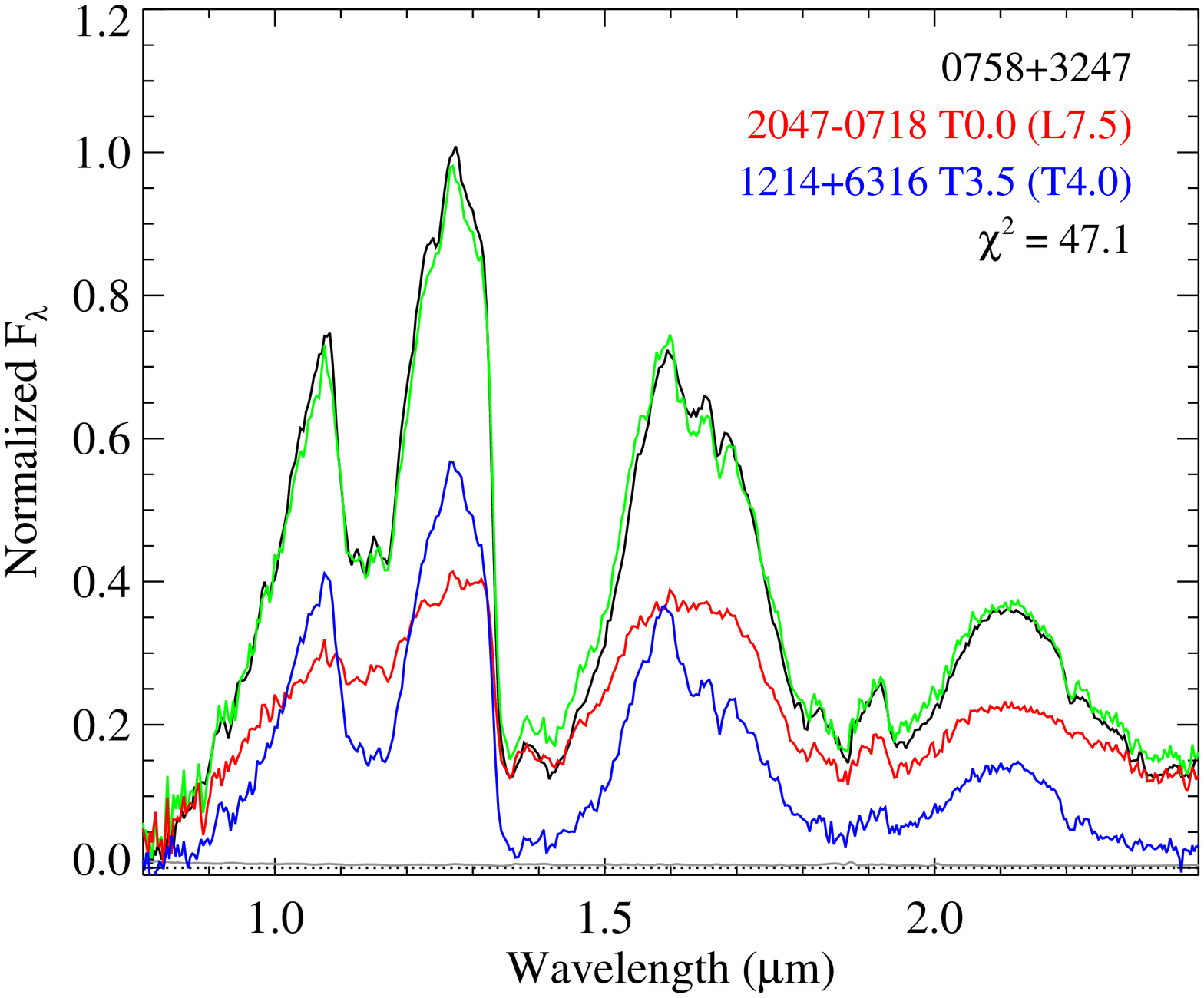}
\includegraphics[width=0.45\textwidth]{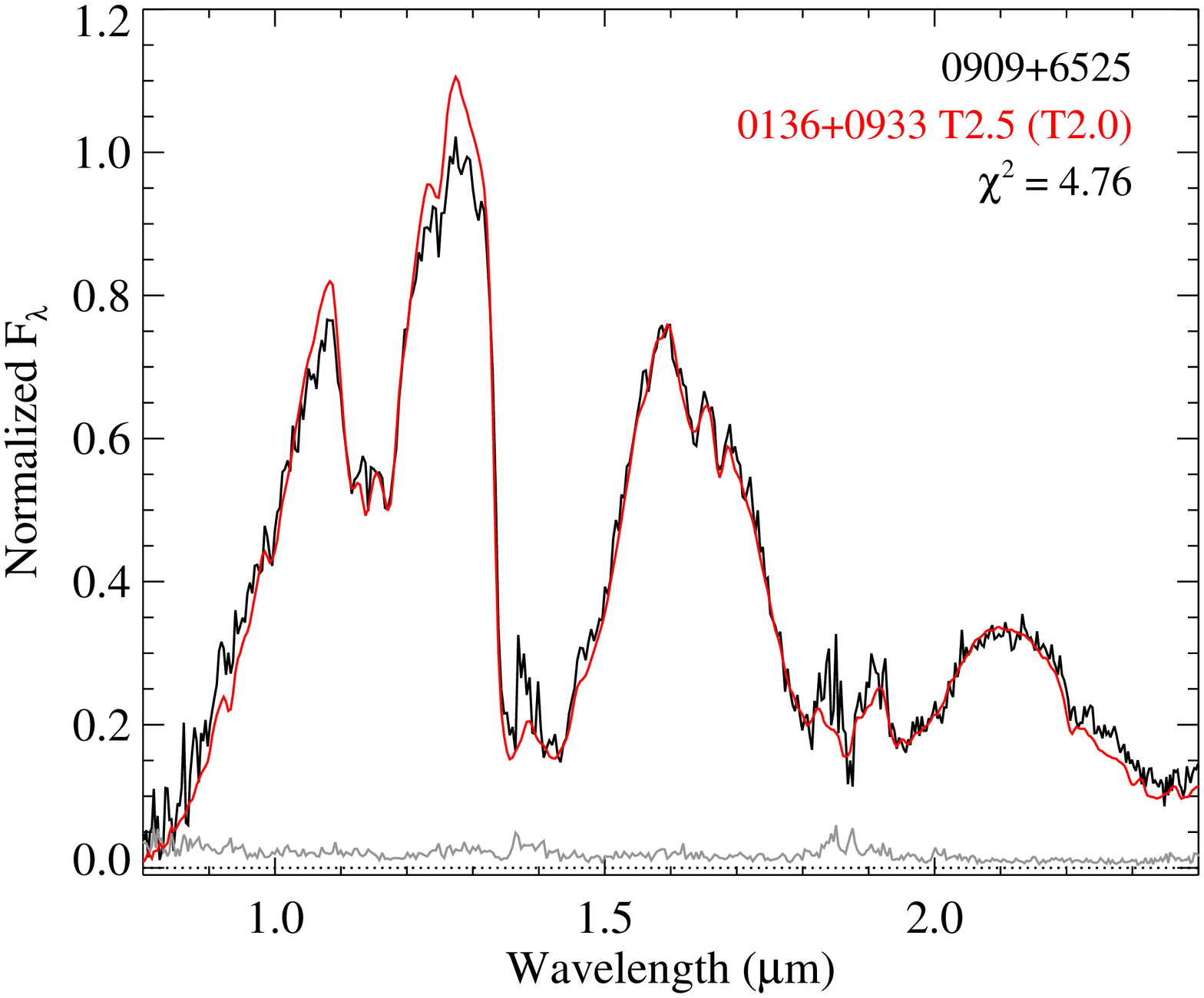}
\includegraphics[width=0.45\textwidth]{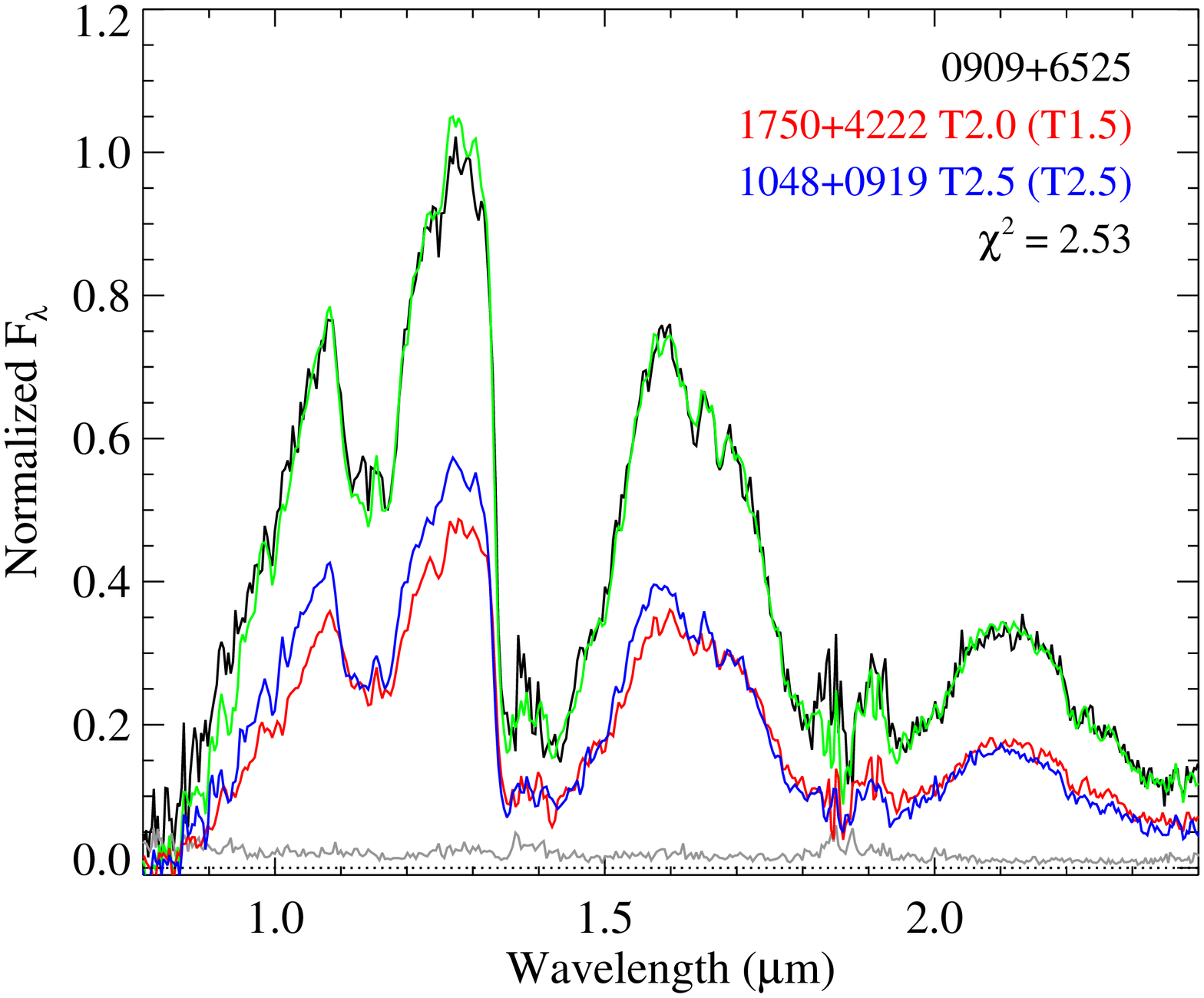}
\caption{Same as Figure~\ref{fig_fitsstrong} but for ``weak'' binary candidates (satisfying only 2 selection criteria).
\label{fig_fitsweak}}
\end{figure}

\clearpage

\addtocounter{figure}{-1}

\begin{figure}
\centering
\epsscale{0.85}
\includegraphics[width=0.45\textwidth]{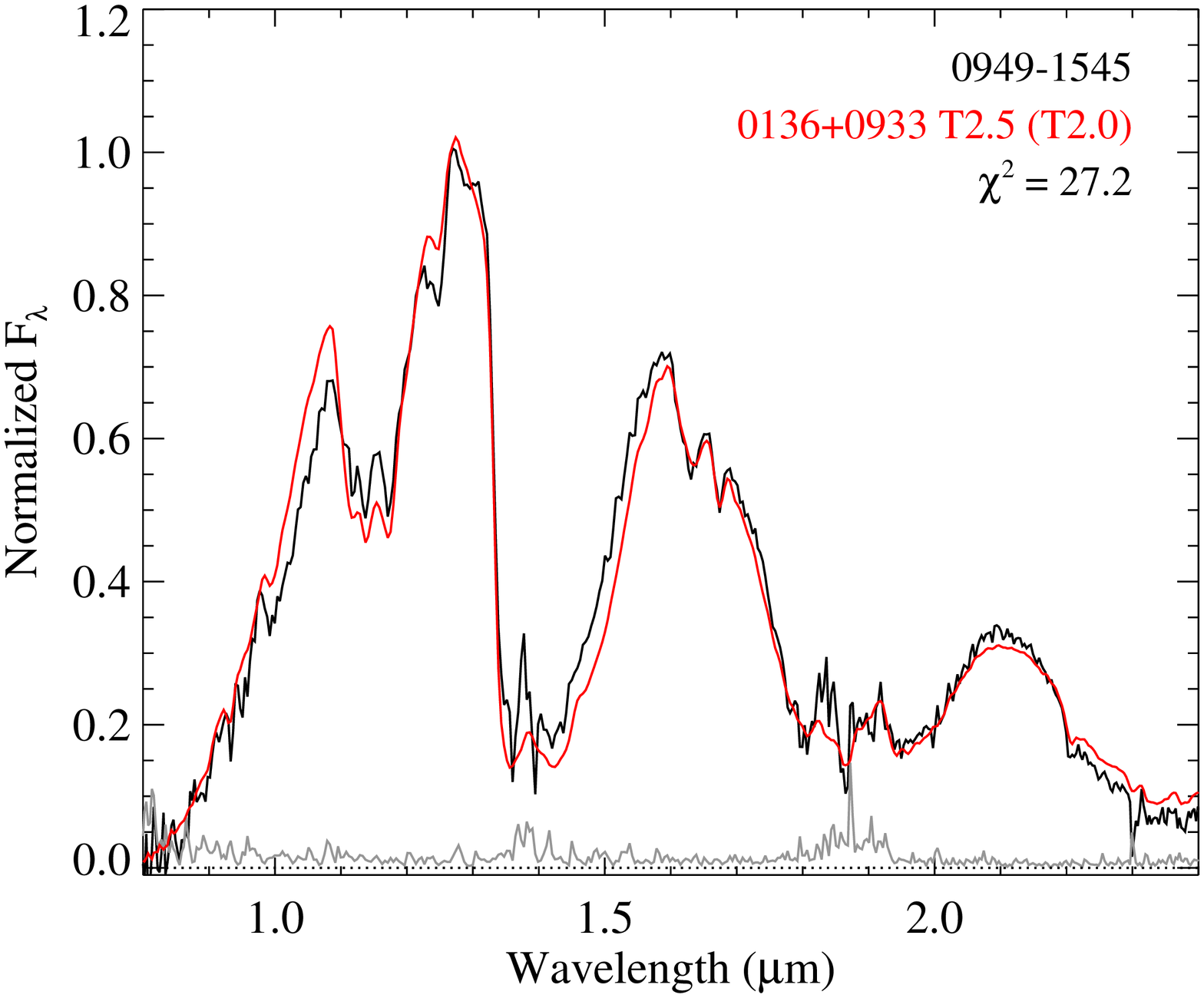}
\includegraphics[width=0.45\textwidth]{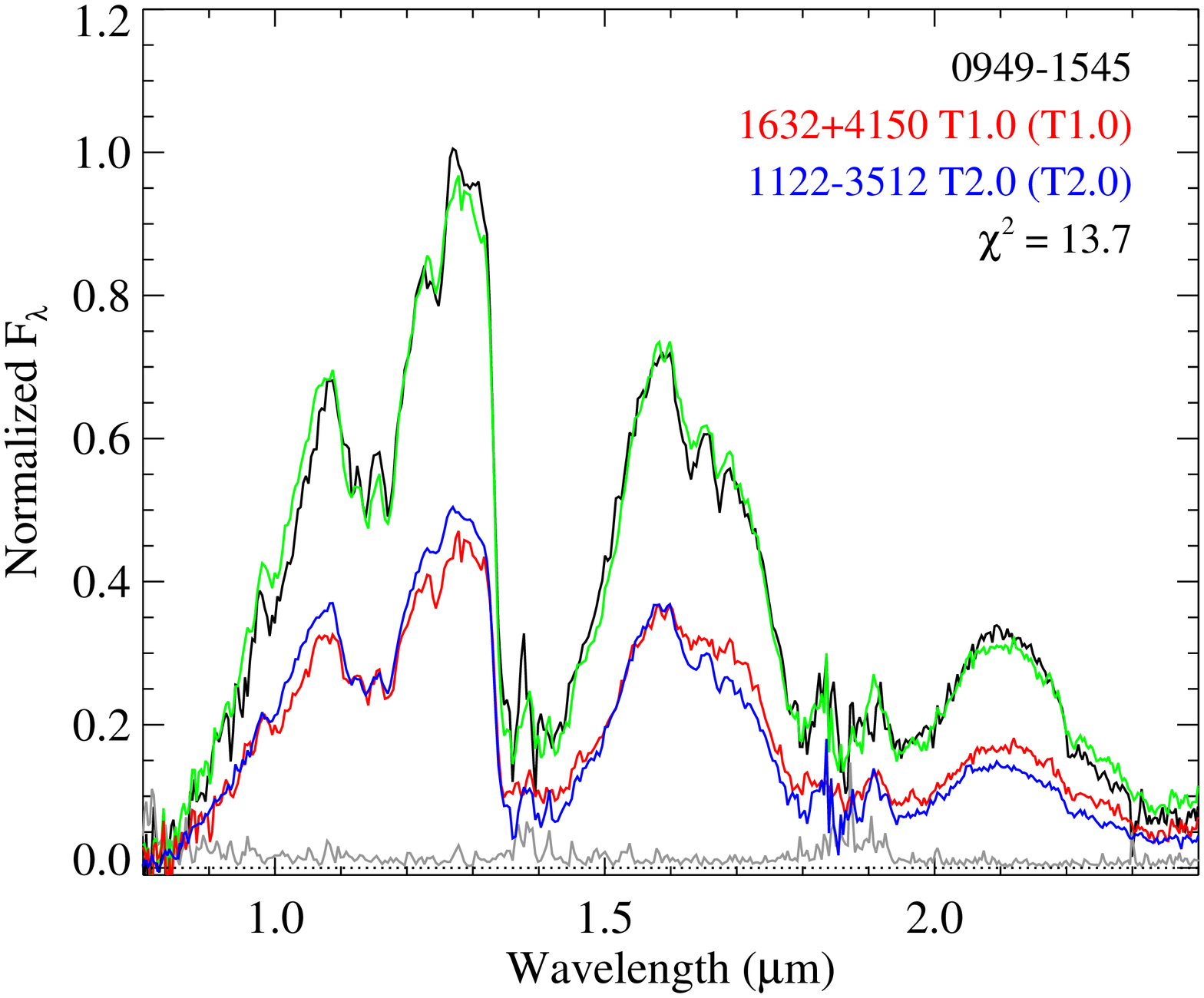}
\includegraphics[width=0.45\textwidth]{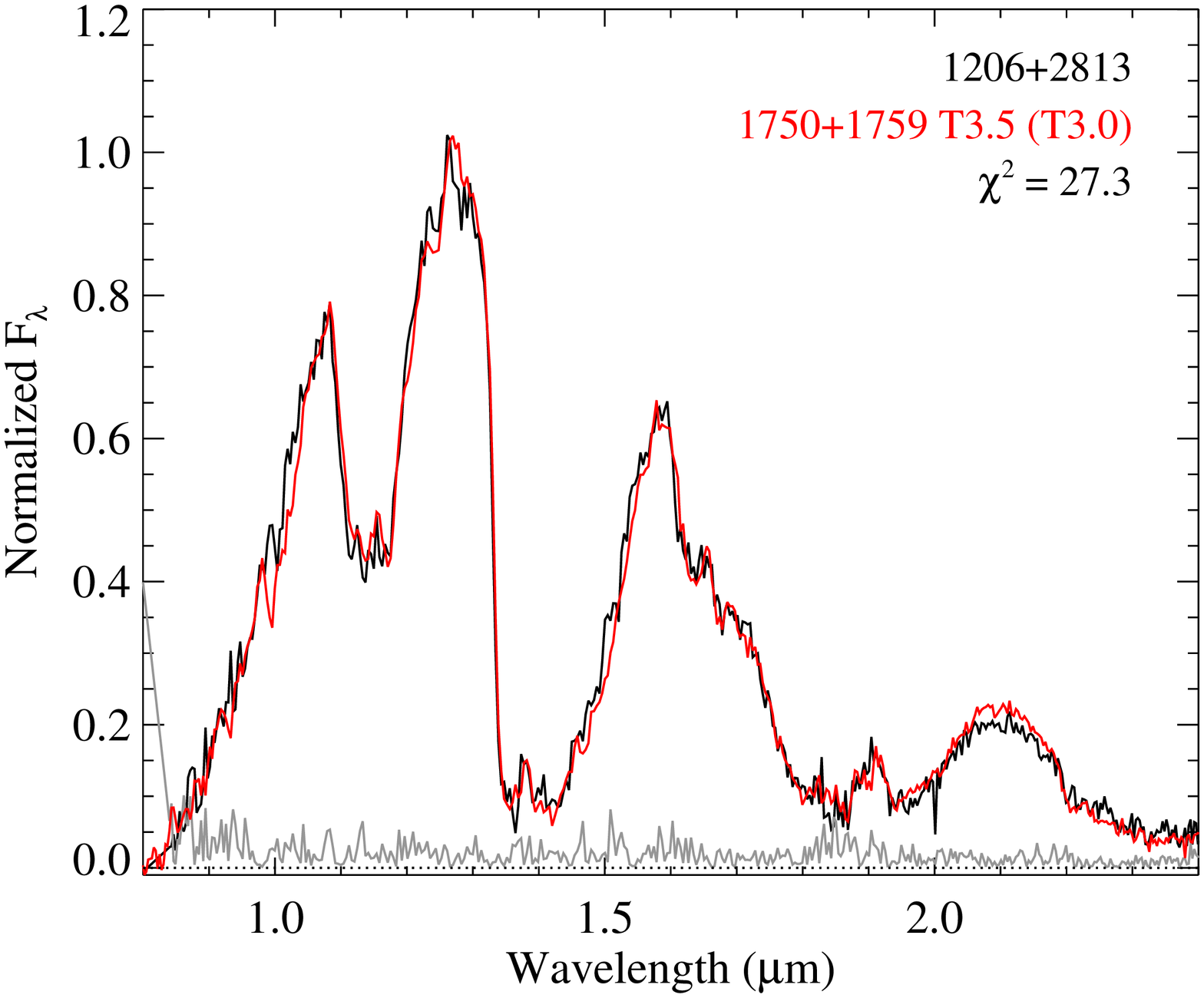}
\includegraphics[width=0.45\textwidth]{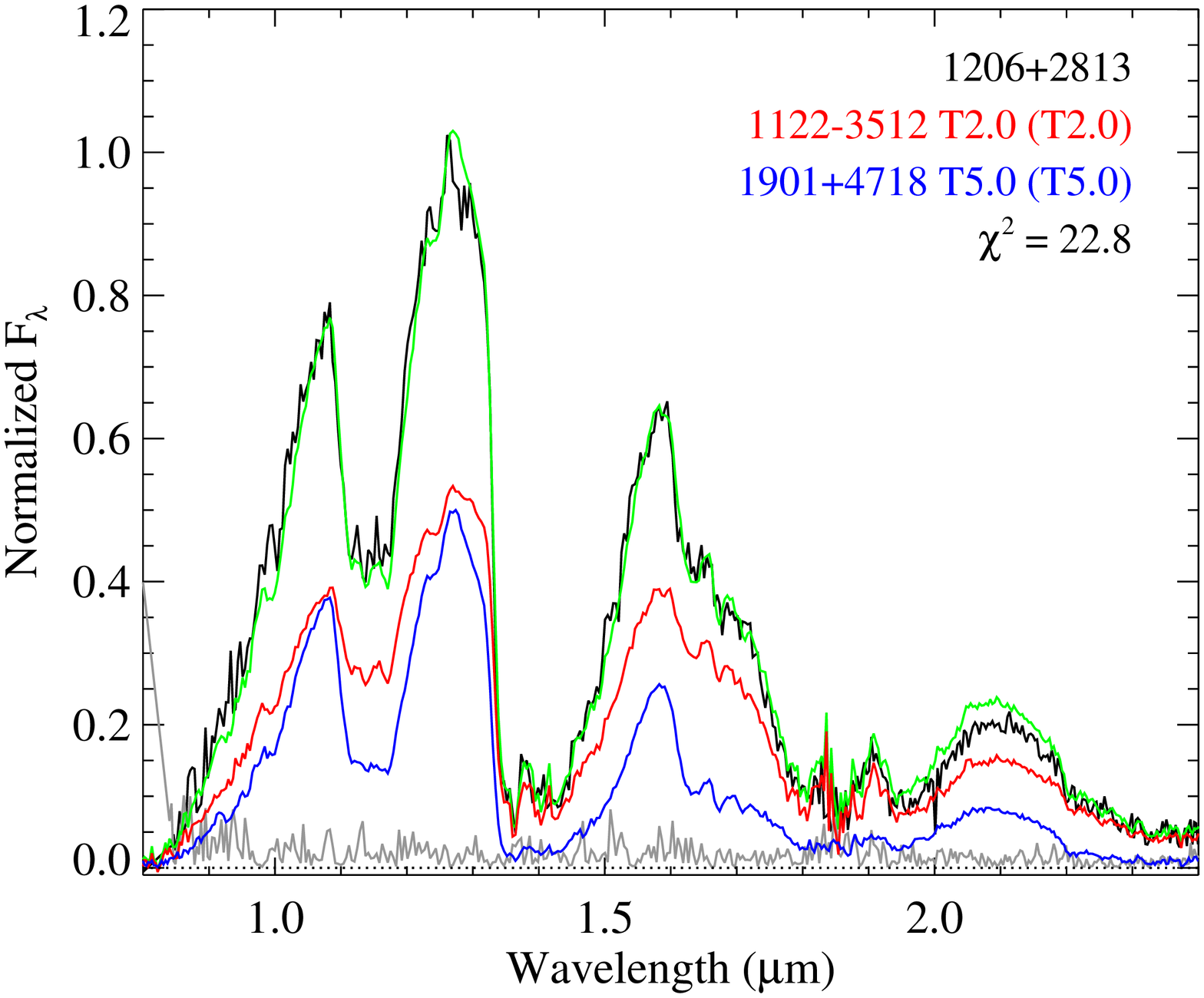}
\includegraphics[width=0.45\textwidth]{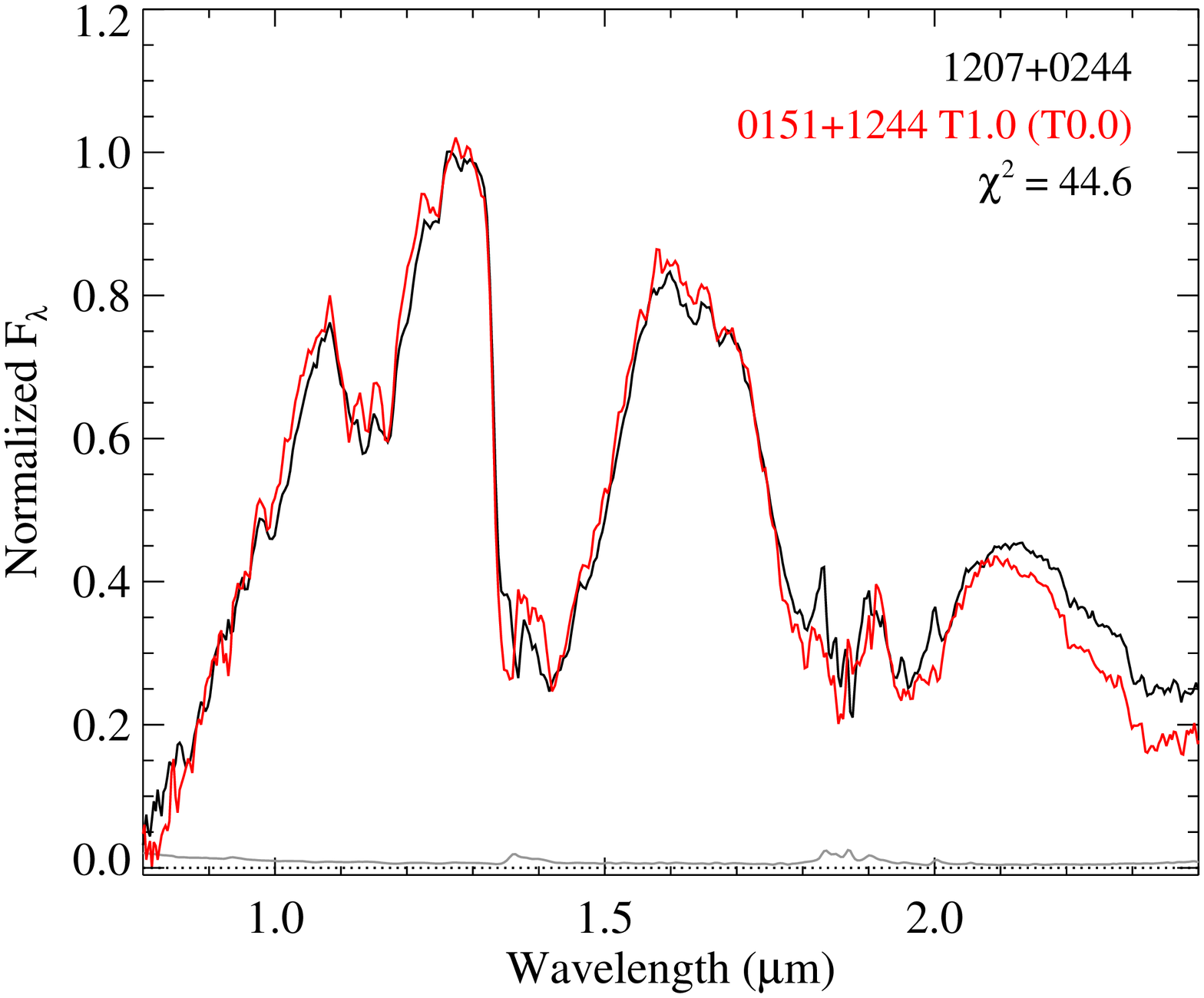}
\includegraphics[width=0.45\textwidth]{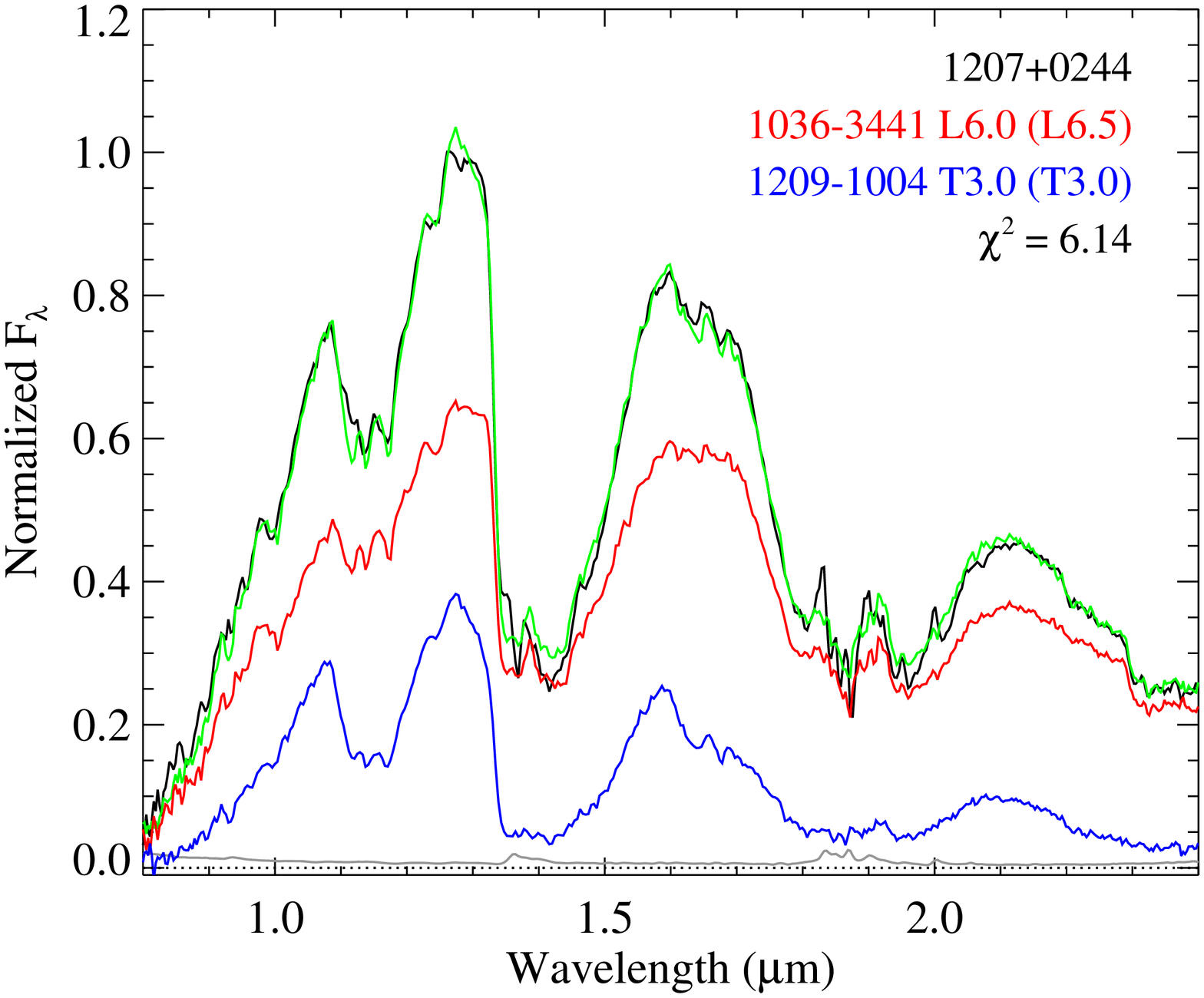}
\caption{Continued.}
\end{figure}

\clearpage

\addtocounter{figure}{-1}

\begin{figure}
\centering
\epsscale{0.85}
\includegraphics[width=0.45\textwidth]{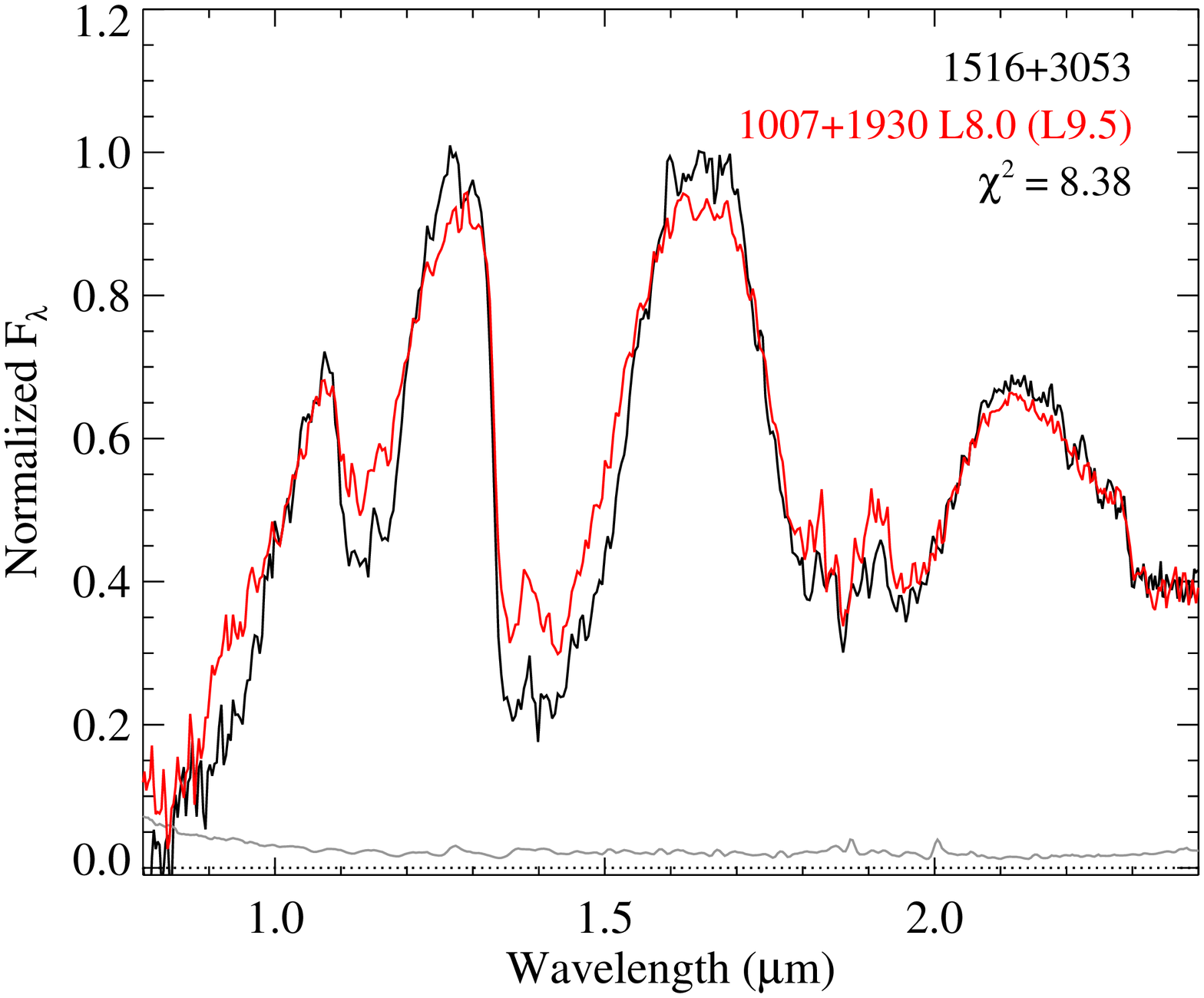}
\includegraphics[width=0.45\textwidth]{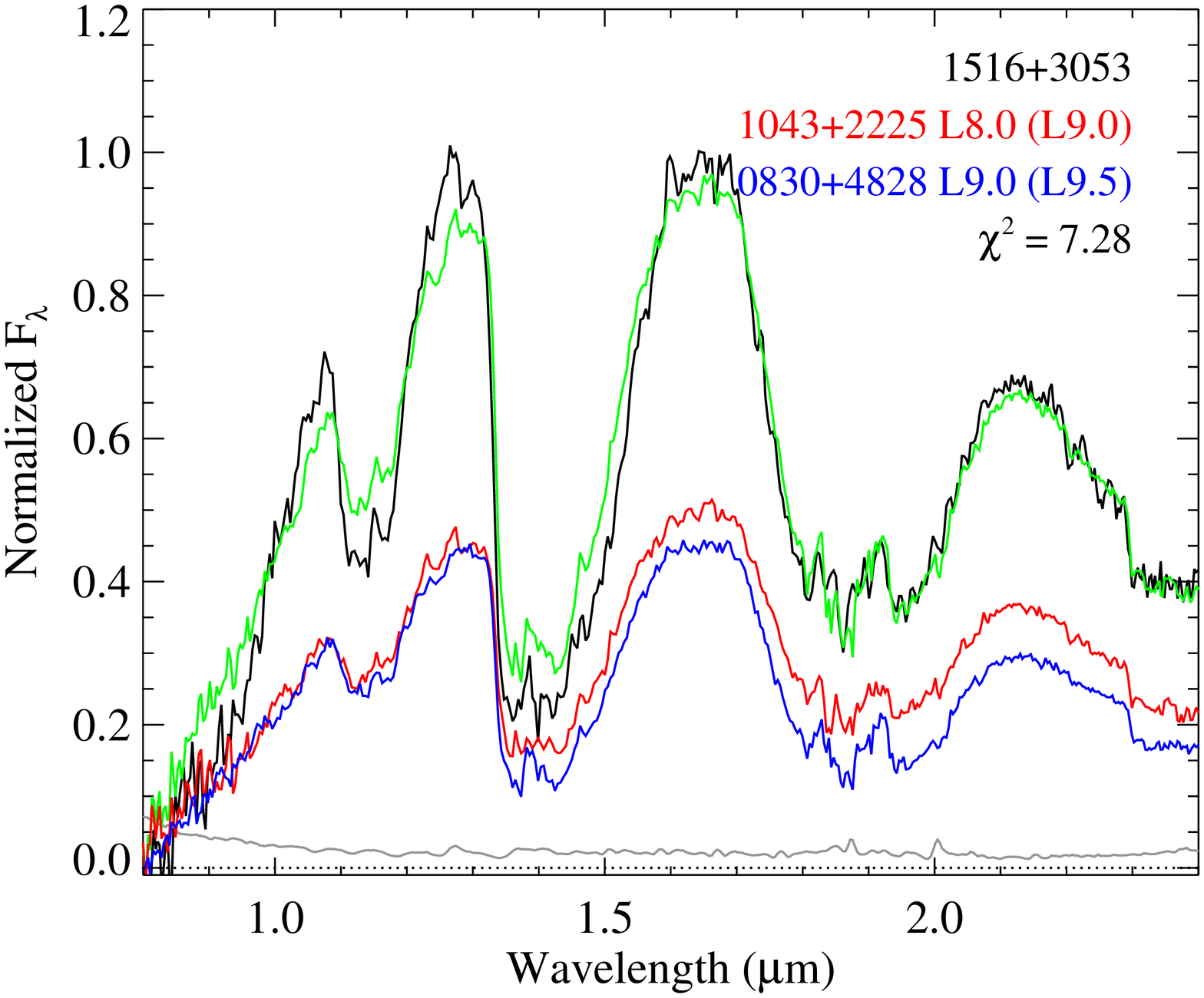}
\includegraphics[width=0.45\textwidth]{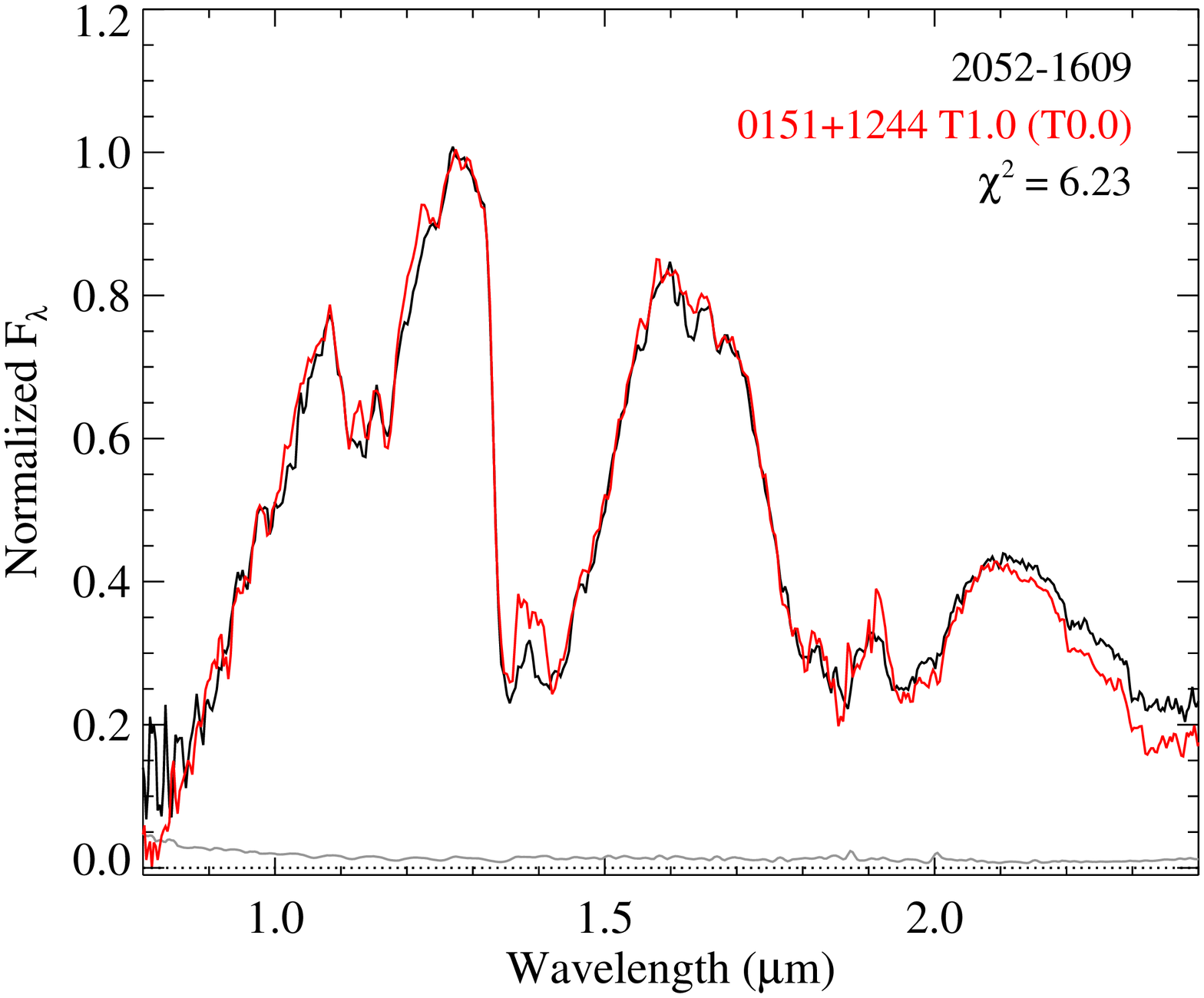}
\includegraphics[width=0.45\textwidth]{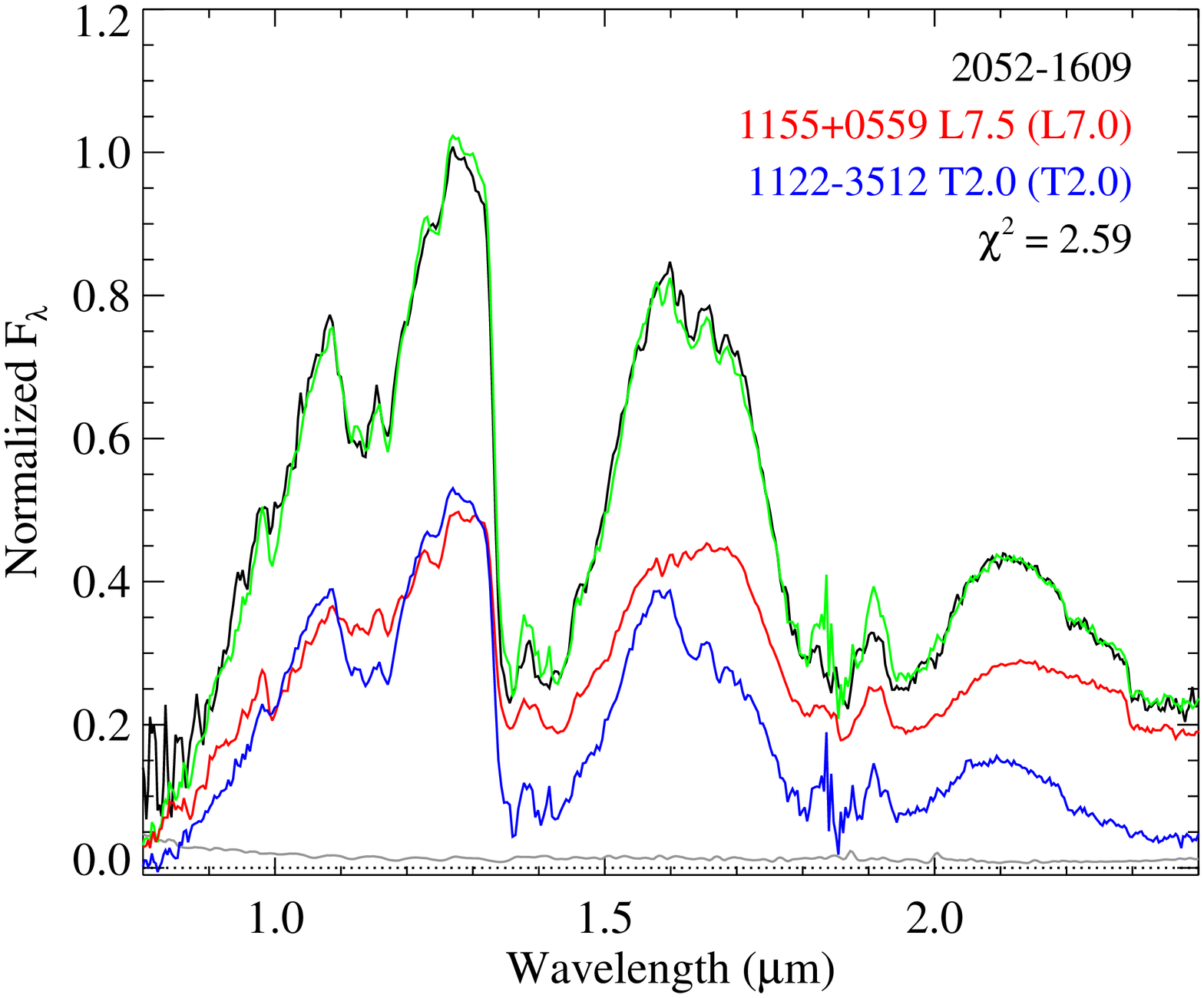}
\caption{Continued.}
\end{figure}

\clearpage

\begin{figure}
\centering
\epsscale{0.85}
\includegraphics[width=0.45\textwidth]{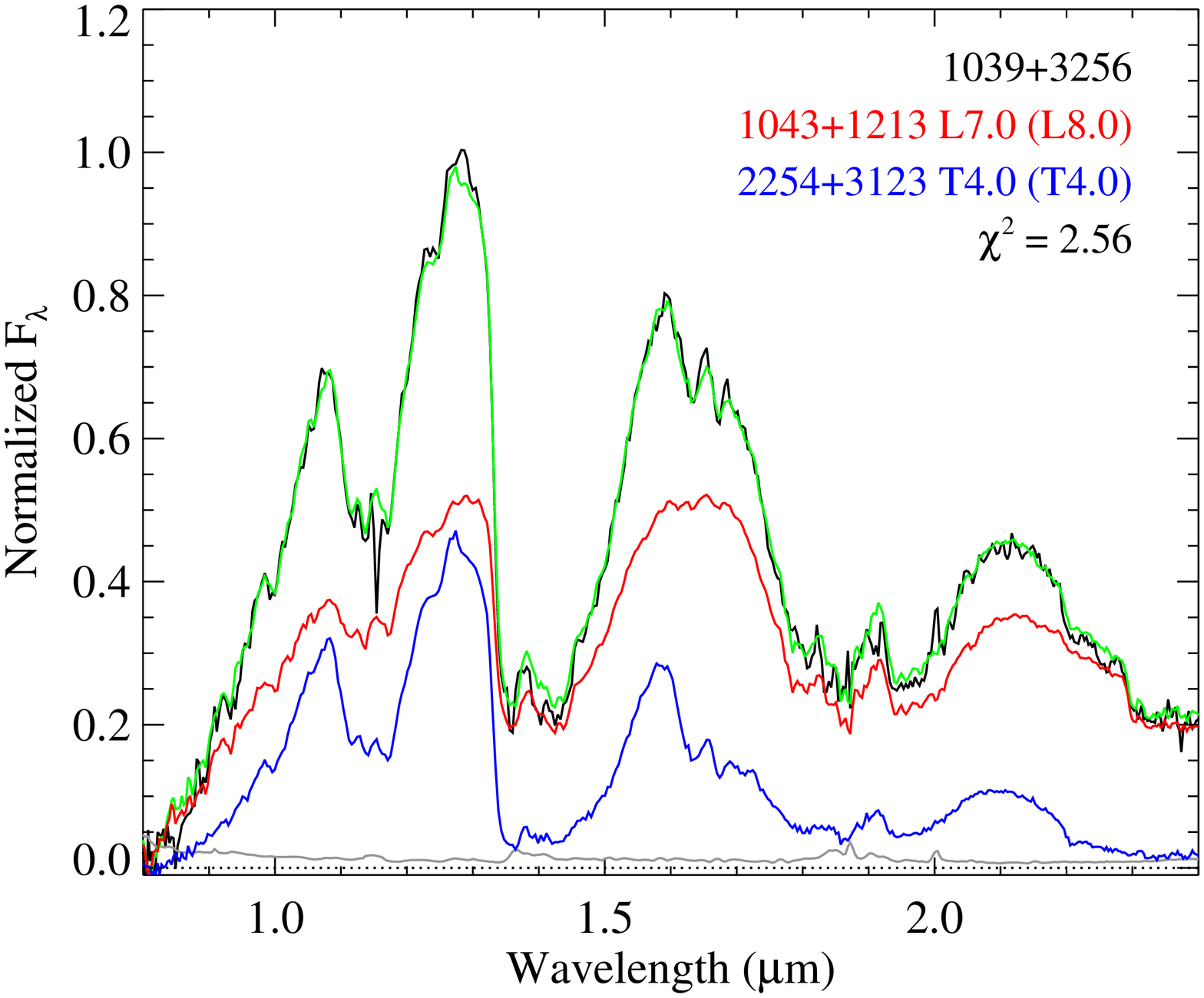}
\includegraphics[width=0.45\textwidth]{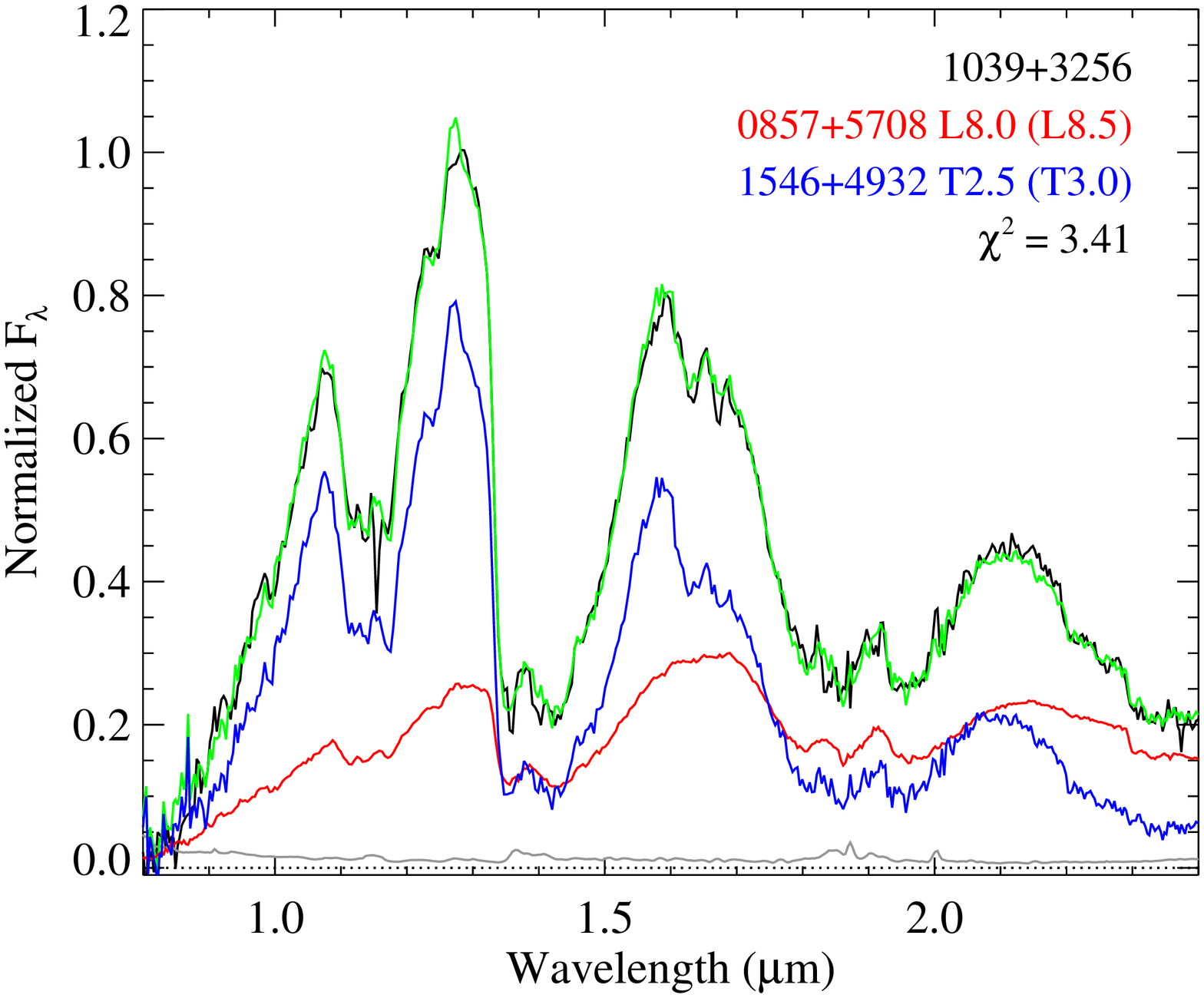}
\includegraphics[width=0.45\textwidth]{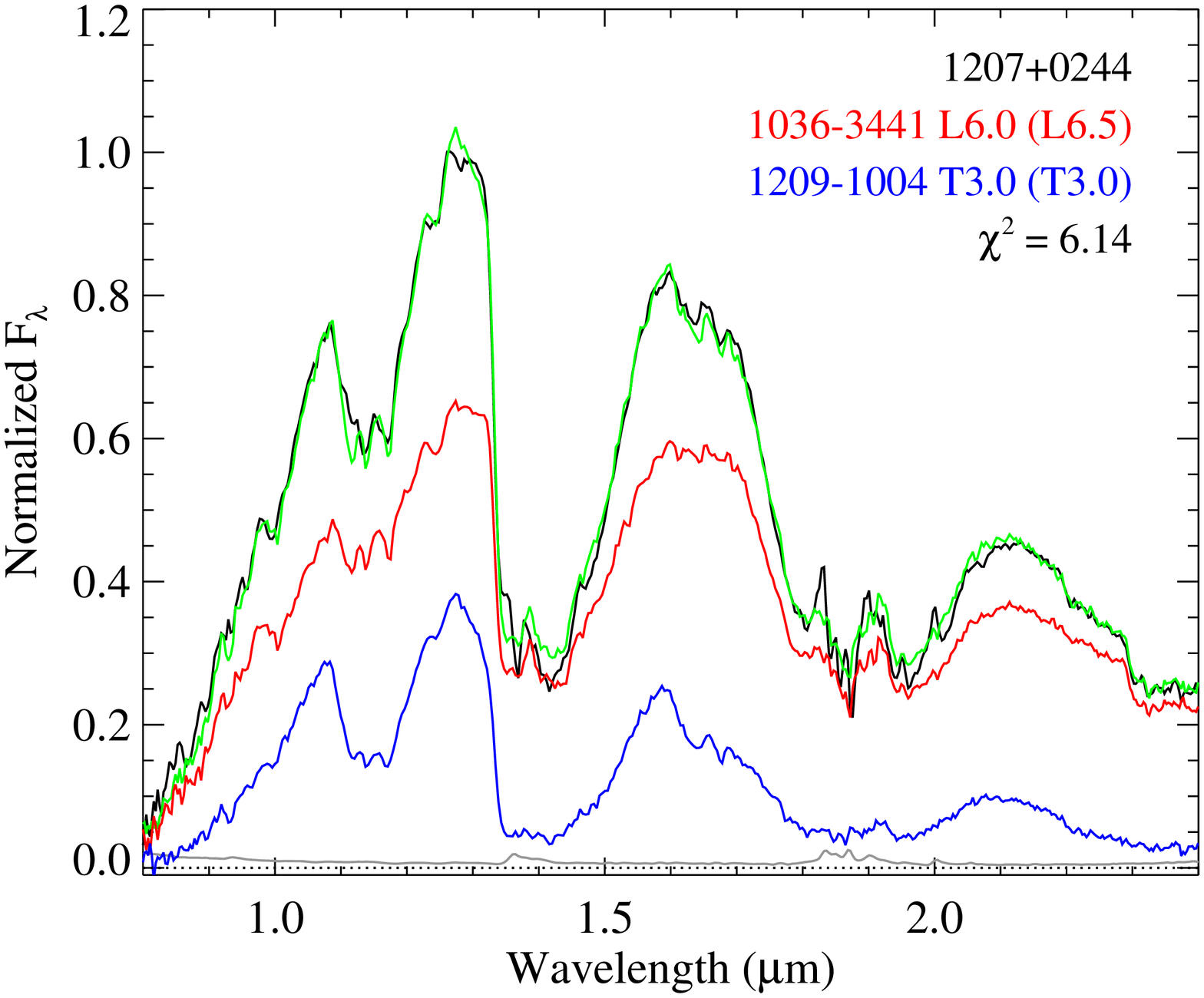}
\includegraphics[width=0.45\textwidth]{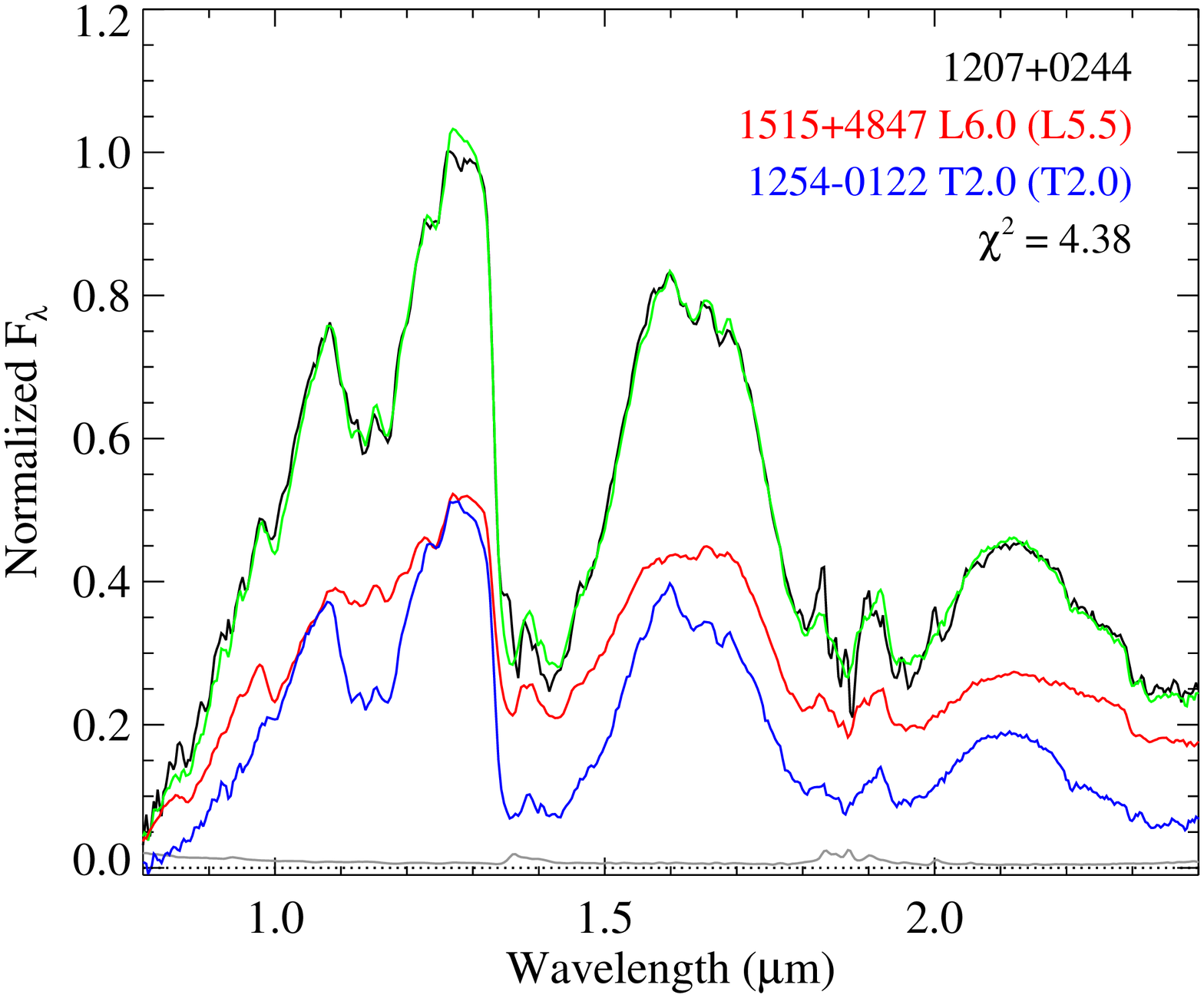}
\includegraphics[width=0.45\textwidth]{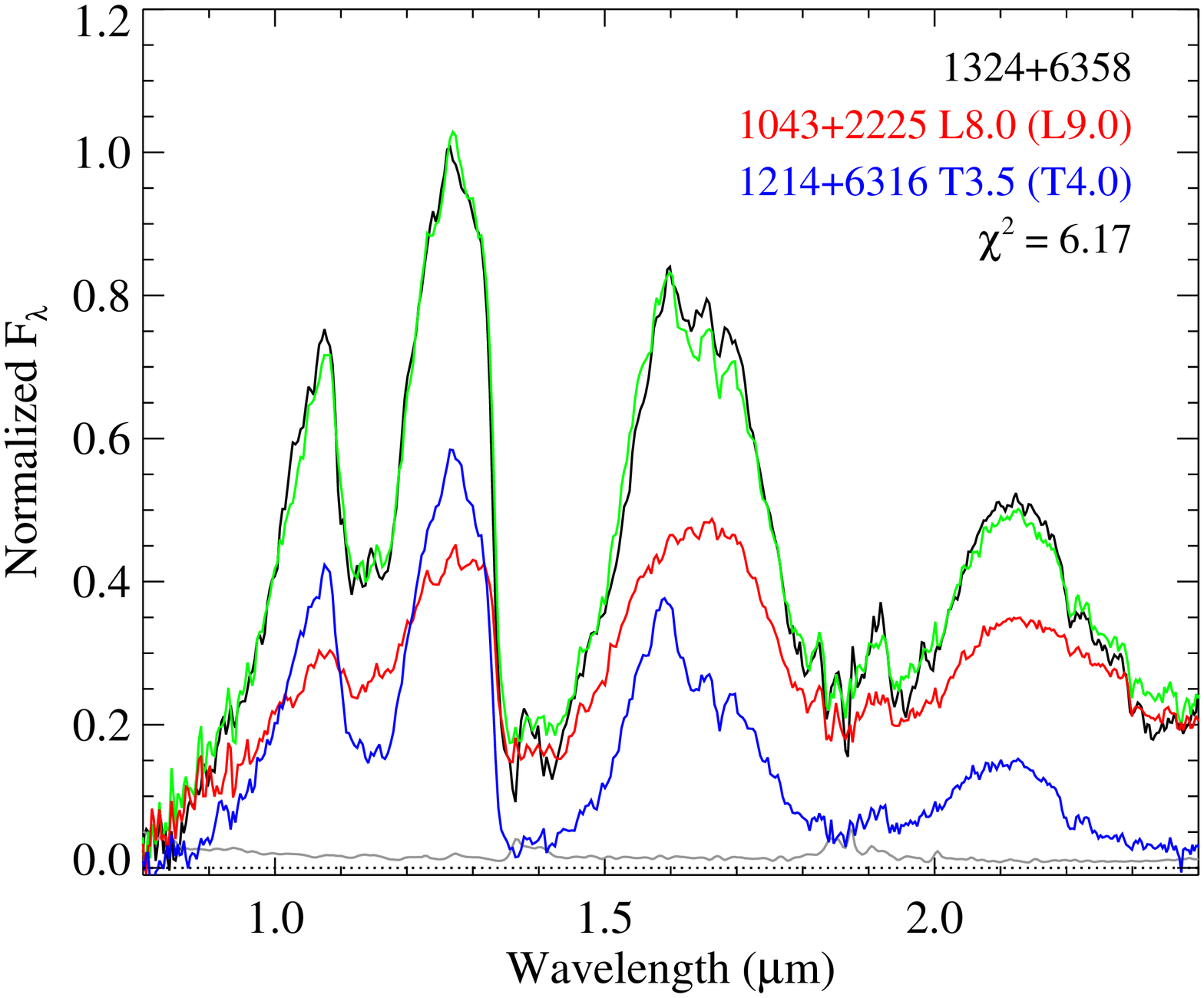}
\includegraphics[width=0.45\textwidth]{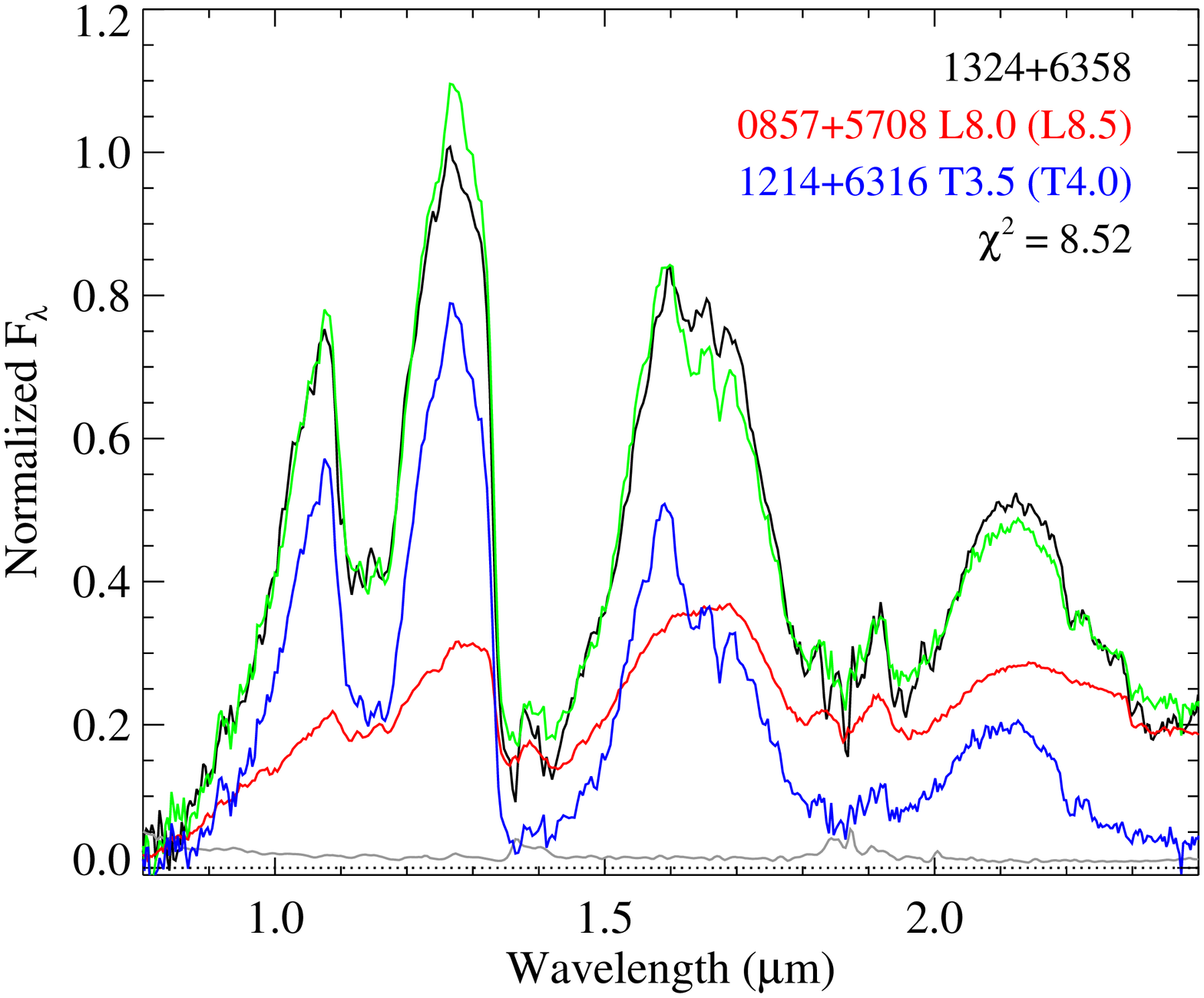}
\caption{Spectral fits for SDSS~J1039+3256 (top), SDSS~J1207+0244 (middle) 
and 2MASS~J1324+6358 (bottom) based on flux calibrations with the faint (left) or bright (right) absolute magnitude relations of \citet{2006ApJ...647.1393L}.  These were the only three sources for which the two relations gave statistically distinct {\chisq} residuals.  Format is the same as
the right panels in Figures~\ref{fig_fitsstrong} and~\ref{fig_fitsweak}.
\label{fig_fntvbrt}}
\end{figure}

\clearpage

\begin{figure}
\centering
\epsscale{0.8}
\includegraphics[width=0.49\textwidth]{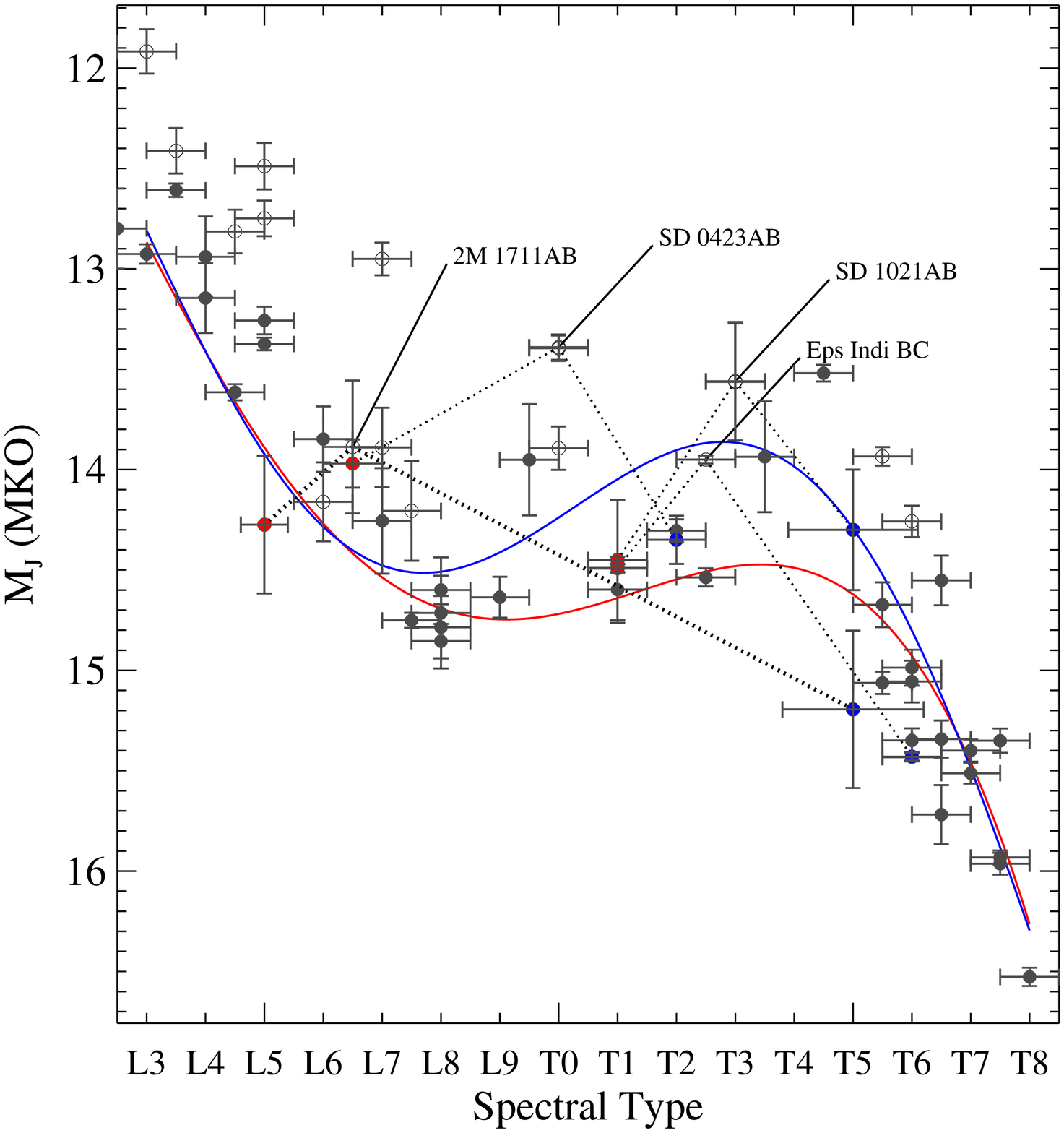}
\caption{Absolute MKO $J$ magnitudes versus spectral type for L and T dwarfs with parallax measurements.  Spectral types are based on optical classifications for 
L2-L8 dwarfs \citep{1999ApJ...519..802K} and near-infrared classifications for  L9-T8 dwarfs \citep{2006ApJ...637.1067B}.  Known multiples (including the candidate 2MASS~J1711+2232) and unresolved sources with absolute magnitude uncertainties $\le$0.3~mag are indicated as open and filled circles, respectively.  Absolute magnitudes for the primary (red) and secondary (blue) components of SDSS~J0423-0414, SDSS~J1021-0304, 2MASS~J1711+2232 and $\epsilon$ Indi BC (\citealt{2004A&A...413.1029M,2006ApJS..166..585B}; this paper)
are also indicated by filled circles and connected to their combined-light magnitude and spectral types (dotted lines).  The bright (top) and faint (bottom) absolute magnitude relations of \citet{2006ApJ...647.1393L} are delineated by solid blue and red lines, respectively.  
\label{fig_absmag}}
\end{figure}

\clearpage

\begin{deluxetable}{lllcclccccl}
\tabletypesize{\scriptsize}
\rotate
\tablecaption{New SpeX Spectral Observations. \label{tab_new}}
\tablewidth{0pt}
\tablehead{
\colhead{Source Name} &
\colhead{Opt SpT} &
\colhead{NIR SpT} &
\colhead{2MASS $J$} &
\colhead{UT Date} &
\colhead{Observer} &
\colhead{t$_{int}$ (s)} &
\colhead{Slit ($\arcsec$)} &
\colhead{Airmass} &
\colhead{Cal Star}  \\
\colhead{(1)} &
\colhead{(2)} &
\colhead{(3)} &
\colhead{(4)} &
\colhead{(5)} &
\colhead{(6)} &
\colhead{(7)} &
\colhead{(8)} &
\colhead{(9)} &
\colhead{(10)} \\
}
\startdata
2MASS J00165953-4056541 & L3.5 & \nodata & 15.32$\pm$0.06 & 2008 Sep 7 & AJB & 900 & 0.5 & 2.04 & HD 2339 \\ 
2MASS J00193927-3724392 & L3: & \nodata & 15.52$\pm$0.06 & 2003 Sep 4 & KLC & 720 & 0.8 & 1.26 & HD 9485 \\ 
2MASS J0028208+224905 & \nodata & L7: & 15.61$\pm$0.07 & 2003 Sep 3 & KLC & 600 & 0.8 & 1.09 & HD 198070 \\ 
2MASSW J0030300-145033 & L7 & \nodata & 16.28$\pm$0.11 & 2007 Jul 28 & DLL & 960 & 0.5 & 1.22 & HD 1154 \\ 
SDSSp J003259.36+141036.6 & \nodata & L8 & 16.83$\pm$0.17 & 2008 Jul 13 & AJB & 900 & 0.5 & 1.02 & HD 7215 \\ 
2MASSW J0051107-154417 & L3.5 & \nodata & 15.28$\pm$0.05 & 2008 Sep 8 & AJB & 960 & 0.5 & 1.24 & HD 1154 \\ 
2MASS J00531899-3631102 & L3.5 & \nodata & 14.45$\pm$0.03 & 2008 Sep 7 & AJB & 360 & 0.5 & 1.83 & HD 5061 \\ 
SDSSp J010752.33+004156.1 & L8 & L5.5 & 15.82$\pm$0.06 & 2007 Oct 12 & DLL &1200  & 0.5 & 1.16 & HD 9485 \\
\enddata
\end{deluxetable}


\begin{deluxetable}{llllccl}
\tabletypesize{\scriptsize}
\tablecaption{Spectral Templates. \label{tab_templates}}
\tablewidth{0pt}
\tablehead{
\colhead{Source Name} &
\colhead{Opt SpT} &
\colhead{NIR SpT} &
\colhead{SpeX SpT} &
\colhead{2MASS $J$} &
\colhead{$J-K_s$} &
\colhead{Ref\tablenotemark{a}} \\
\colhead{(1)} &
\colhead{(2)} &
\colhead{(3)} &
\colhead{(4)} &
\colhead{(5)} &
\colhead{(6)} &
\colhead{(7)} \\
}
\startdata
SDSS J000013.54+255418.6 & \nodata & T4.5 & T4.5 & 15.06$\pm$0.04 & 0.23$\pm$0.13 & 1; 2,1 \\
2MASSI J0013578-223520 & L4 & \nodata & L3.5 & 15.78$\pm$0.07 & 1.74$\pm$0.08 & 3; 4 \\
2MASS J00165953-4056541 & L3.5 & \nodata & L4.5$\pm$1 & 15.32$\pm$0.06 & 1.88$\pm$0.07 & 3; 5 \\
2MASS J00193927-3724392 & L3: & \nodata & L3.5$\pm$1 & 15.52$\pm$0.06 & 1.83$\pm$0.08 & 3; 6 \\
2MASS J0028208+224905\tablenotemark{d} & \nodata & L7: & L5.0 & 15.61$\pm$0.07 & 1.83$\pm$0.10 & 3; 6 \\
2MASSW J0030300-145033\tablenotemark{d} & L7 & \nodata & L4.5$\pm$2 & 16.28$\pm$0.11 & 1.80$\pm$0.15 & 3; 7 \\
SDSSp J003259.36+141036.6 & \nodata & L8 & L7.5 & 16.83$\pm$0.17 & 1.88$\pm$0.20 & 3; 8 \\
2MASS J00345157+0523050 & \nodata & T6.5 & T6.5 & 15.54$\pm$0.05 & $<$-0.71 & 9; 9,1 \\
\enddata
\end{deluxetable}

\clearpage

\begin{deluxetable}{lllllll}
\tabletypesize{\scriptsize}
\tablecaption{Reference L Dwarf/T Dwarf Transition Binaries. \label{tab_known}}
\tablewidth{0pt}
\tablehead{
\colhead{Source Name} &
\colhead{Opt SpT} &
\colhead{NIR SpT} &
\colhead{Component SpTs} &
\multicolumn{2}{c}{Component Relative Magnitudes\tablenotemark{a}} &
\colhead{Ref} \\
\colhead{(1)} &
\colhead{(2)} &
\colhead{(3)} &
\colhead{(4)} &
\multicolumn{2}{c}{(5)} &
\colhead{(6)} \\
}
\startdata
SDSSp J042348.57-041403.5 & L7.5 & T0 & L6.5+T2 & $\Delta$F110W = 0.526$\pm$0.015 & $\Delta$F170M = 0.820$\pm$0.013 & 1 \\
2MASS J05185995-2828372 & L7.5p & T1p & L6:+T4: & $\Delta$F110W = 0.8$\pm$0.5 & $\Delta$F170M = 0.9$\pm$0.6 & 2,3 \\
SDSS J102109.69-030420.1 & T4$\pm$2 & T3 & T1+T5 & $\Delta$F110W = 0.06$\pm$0.04 & $\Delta$F170M = 1.030$\pm$0.019 & 3 \\
2MASS J14044941-3159329 & T0 & T2.5 & T1+T5 & $\Delta$J = -0.53$\pm$0.16 & $\Delta$K = 1.20$\pm$0.21 & 4 \\
SDSS J153417.05+161546.1AB & \nodata & T3.5 & T1.5+T5.5 & $\Delta$J = -0.17$\pm$0.04 & $\Delta$K = 1.07$\pm$0.05  & 5 \\
DENIS-P J225210.73-173013.4 & \nodata & L7.5 & L6+T2 & $\Delta$F110W = 0.526$\pm$0.015 & $\Delta$F170M = 0.820$\pm$0.013 & 6 \\
\enddata
\tablenotetext{a}{Relative photometry defined as $M_B$-$M_A$; negative magnitudes indicate a secondary that is brighter than the primary.  Relative $JK$ photometry listed in on the MKO system.}
\tablerefs{(1) \citet{2005ApJ...634L.177B}; (2) \citet{2004ApJ...604L..61C}; (3) \citet{2006ApJS..166..585B};
(4) \citet{2008ApJ...685.1183L}; (5) \citet{2006ApJ...647.1393L}; (6) \citet{2006ApJ...639.1114R}}
\end{deluxetable}

\begin{deluxetable}{lccll}
\tabletypesize{\scriptsize}
\tablecaption{Spectral Indices. \label{tab_indices}}
\tablewidth{0pt}
\tablehead{
\colhead{} &
\colhead{Numerator} &
\colhead{Denominator} \\
\colhead{Index} &
\colhead{Range\tablenotemark{a}} &
\colhead{Range\tablenotemark{a}} &
\colhead{Feature} &
\colhead{Ref} \\
\colhead{(1)} &
\colhead{(2)} &
\colhead{(3)} &
\colhead{(4)} &
\colhead{(5)} \\
}
\startdata
H$_2$O-J & 1.14--1.165 & 1.26--1.285 & 1.15 $\micron$ H$_2$O & 1 \\
CH$_4$-J & 1.315--1.34 & 1.26--1.285 & 1.32 $\micron$ CH$_4$ & 1  \\
H$_2$O-H & 1.48--1.52 & 1.56--1.60 & 1.4 $\micron$ H$_2$O & 1  \\
CH$_4$-H & 1.635--1.675 & 1.56--1.60 & 1.65 $\micron$ CH$_4$ & 1  \\
H$_2$O-K & 1.975--1.995 & 2.08--2.10 & 1.9 $\micron$ H$_2$O & 1  \\
CH$_4$-K & 2.215--2.255 & 2.08--2.12 & 2.2 $\micron$ CH$_4$ & 1  \\
K/J & 2.060--2.10 & 1.25--1.29 & $J-K$ color & 1  \\
H-dip & 1.61--1.64 & 1.56--1.59 + 1.66--1.69\tablenotemark{b} & 1.65 $\micron$ CH$_4$\tablenotemark{c}  & 2 \\
\enddata
\tablenotetext{a}{Wavelength range (in microns) over which flux density ($f_{\lambda}$) is integrated.}
\tablenotetext{b}{Denominator is sum of these two wavelength ranges.}
\tablenotetext{c}{Specifically, this index samples the sharp {\meth} feature present in the near-infrared spectra of the suspected L dwarf plus T dwarf  binaries 2MASS~J0805+4812 \citep{2007AJ....134.1330B} and Kelu-1A \citep{2008arXiv0811.0556S}.}
\tablerefs{(1) \citet{2006ApJ...637.1067B}; (2) This paper.}
\end{deluxetable}

\begin{deluxetable}{lllc}
\tabletypesize{\scriptsize}
\tablecaption{Binary Index Selection Criteria.\tablenotemark{a} \label{tab_criteria}}
\tablewidth{0pt}
\tablehead{
\colhead{Ordinate} &
\colhead{Abcissa} &
\colhead{Inflection Points} &
\colhead{\#\tablenotemark{a}} \\
\colhead{(1)} &
\colhead{(2)} &
\colhead{(3)} &
\colhead{(4)} \\
}
\startdata
{\wat}-J & {\wat}-K & (0.325,0.5), (0.65,0.7) & 27 \\
{\meth}-H & {\meth}-K & (0.6,0.35), (1,0.775) & 24 \\
{\meth}-H & K/J & (0.65,0.25), (1,0.375) & 28 \\
{\wat}-H & H-dip & (0.5,0.49), (0.875,0.49) & 16 \\
Spex SpT & {\wat}-J/{\wat}-H & (L8.5,0.925), (T1.5,0.925), (T3.5,0.85) & 25 \\
Spex SpT & {\wat}-J/{\meth}-K & (L8.5,0.625), (T4.5,0.825) & 21 \\
\enddata
\tablenotetext{a}{See Figure~\ref{fig_indices}.}
\tablenotetext{b}{Number of sources satisfying selection criteria.}
\end{deluxetable}

\begin{deluxetable}{lllccccc}
\tabletypesize{\scriptsize}
\tablecaption{Summary of Fits to Reference Binaries. \label{tab_knownbfit}}
\tablewidth{0pt}
\tablehead{
 & \multicolumn{2}{c}{Spectral Types} \\
\cline{2-3}
\colhead{Source} &
\colhead{Primary} &
\colhead{Secondary} &
\colhead{$\Delta{J}$}  &
\colhead{$\Delta{K}$}  &
\colhead{$\Delta{F110W}$} &
\colhead{$\Delta{F170M}$} &
\colhead{CL\tablenotemark{a}} \\
\colhead{(1)} &
\colhead{(2)} &
\colhead{(3)} &
\colhead{(4)} &
\colhead{(5)} &
\colhead{(6)} &
\colhead{(7)} &
\colhead{(8)}  \\
}
\startdata
SDSSp J042348.57-041403.5 &  L7.5$\pm$0.3  &  T2.0$\pm$0.2  &  0.44$\pm$0.09  &  1.13$\pm$0.07  &  0.59$\pm$0.10  &  0.77$\pm$0.08  & $>$99\%   \\
2MASS J05185995-2828372 &  L7.5$\pm$0.2  &  T5.0$\pm$0.5  &  0.13$\pm$0.19  &  2.18$\pm$0.31  &  0.39$\pm$0.20  &  1.71$\pm$0.36  & $>$99\%   \\
SDSS J102109.69-030420.1 &  T1.0$\pm$0.4  &  T5.5$\pm$0.7  &  0.16$\pm$0.41  &  1.46$\pm$0.29  &  0.37$\pm$0.40  &  1.49$\pm$0.51  & $>$99\%   \\
2MASS J14044941-3159329 &  T0.0$\pm$0.2  &  T5.0$\pm$0.2  &  -0.27$\pm$0.07  &  1.41$\pm$0.08  &  -0.07$\pm$0.06  &  1.07$\pm$0.09  & $>$99\%   \\
SDSS J153417.05+161546.1AB &  T1.5$\pm$0.5  &  T5.0$\pm$1.0  &  0.06$\pm$0.67  &  1.16$\pm$0.47  &  0.23$\pm$0.66  &  1.11$\pm$0.76  & $>$99\%   \\
DENIS-P J225210.73-173013.4 &  L4.5$\pm$0.7  &  T4.5$\pm$0.4  &  1.29$\pm$0.25  &  2.88$\pm$0.32  &  1.54$\pm$0.25  &  2.50$\pm$0.33  & $>$99\%   \\
\enddata
\tablenotetext{a}{Confidence limit; see Section~4.2.}
\end{deluxetable}

\begin{deluxetable}{llllccc}
\tabletypesize{\scriptsize}
\tablecaption{Binary Candidates. \label{tab_candidates}}
\tablewidth{0pt}
\tablehead{
\colhead{Source} &
\colhead{Opt SpT} &
\colhead{NIR SpT} &
\colhead{SpeX SpT} &
\colhead{2MASS $J$} &
\colhead{$J-K_s$} &
\colhead{\# Criteria} \\
\colhead{(1)} &
\colhead{(2)} &
\colhead{(3)} &
\colhead{(4)} &
\colhead{(5)} &
\colhead{(6)} &
\colhead{(7)} \\
}
\startdata
\cline{1-7}
\multicolumn{7}{c}{Strong Candidates} \\
\cline{1-7}
SDSS J024749.90-163112.6 & \nodata & T2$\pm$1.5 & T2.5$\pm$1 & 17.19$\pm$0.18 & 1.57$\pm$0.27 & 6 \\
SDSS J035104.37+481046.8 & \nodata & T1$\pm$1.5 & T1.5$\pm$1 & 16.47$\pm$0.13 & 1.47$\pm$0.18 & 6 \\
SDSS J103931.35+325625.5 & \nodata & T1 & T1.5 & 16.41$\pm$0.15 & 1.25$\pm$0.22 & 6 \\
2MASS J11061197+2754225 & \nodata & T2.5 & T2.0 & 14.82$\pm$0.04 & 1.02$\pm$0.07 & 3 \\
2MASS J13243559+6358284 & \nodata & T2: & T2.0$\pm$1 & 15.60$\pm$0.07 & 1.54$\pm$0.09 & 3 \\
SDSS J141530.05+572428.7 & \nodata & T3$\pm$1 & T3.0$\pm$1 & 16.73$\pm$0.16 & $<$1.19 & 6 \\
SDSS J143553.25+112948.6 & \nodata & T2$\pm$1 & T2.5$\pm$1 & 17.14$\pm$0.23 & $<$0.23 & 6 \\
\enddata
\end{deluxetable}

\begin{deluxetable}{lllllllcccccc}
\tabletypesize{\scriptsize}
\tablecaption{Template Fits to Strong Binary Candidates. \label{tab_fitsstrong}\tablenotemark{a}}
\tablewidth{0pt}
\rotate
\tablehead{
\colhead{Mode\tablenotemark{b}} &
\colhead{Primary} &
\colhead{Pub SpT} &
\colhead{SpeX SpT} &
\colhead{Secondary} &
\colhead{Pub SpT} &
\colhead{SpeX SpT} &
\colhead{$\Delta{J}$} &
\colhead{$\Delta{H}$} &
\colhead{$\Delta{K}$} &
\colhead{$\Delta{F110W}$} &
\colhead{$\Delta{F170M}$} &
\colhead{$\chi^2$}  \\
\colhead{(1)} &
\colhead{(2)} &
\colhead{(3)} &
\colhead{(4)} &
\colhead{(5)} &
\colhead{(6)} &
\colhead{(7)} &
\colhead{(8)} &
\colhead{(9)} &
\colhead{(10)} &
\colhead{(11)} &
\colhead{(12)} &
\colhead{(13)} \\
}
\startdata
\cline{1-13}
\multicolumn{13}{c}{SDSS J024749.90-163112.6} \\
\cline{1-13}
Single & 1750+4222 & T2.0 & T1.5 & \nodata & \nodata & \nodata & \nodata & \nodata & \nodata & \nodata & \nodata & 12.0 \\
L06-faint & 1520+3546 & T0.0 & L7.5 & 0050-3322 & T7.0 & T6.5 & 0.75 & 2.07 & 2.52 & 1.06 & 2.75 & 1.99 \\
 & $<$Primary$>$ & T0.0$\pm$0.2 & L7.5$\pm$0.2 & $<$Secondary$>$ & T7.0$\pm$0.3 & T6.5$\pm$0.3 & 0.68$\pm$0.10 & 1.96$\pm$0.14 & 2.42$\pm$0.17 & 0.98$\pm$0.11 & 2.58$\pm$0.24 &  \nodata \\
L06-faint (IC) & 1520+3546 & T0.0 & L7.5 & 1624+0029 & T6.0 & T6.0 & 0.50 & 1.75 & 2.44 & 0.78 & 2.23 & 1.70 \\
 & $<$Primary$>$ & T0.0$\pm$0.2 & L7.5$\pm$0.2 & $<$Secondary$>$ & T6.0$\pm$0.2 & T6.0$\pm$0.2 & 0.50$\pm$0.04 & 1.74$\pm$0.05 & 2.42$\pm$0.07 & 0.77$\pm$0.04 & 2.21$\pm$0.08 &  \nodata \\
L06-bright & 1520+3546 & T0.0 & L7.5 & 2356-1553 & T5.5 & T5.5 & 0.36 & 1.47 & 1.97 & 0.61 & 1.87 & 1.81 \\
 & $<$Primary$>$ & L9.5$\pm$1.1 & L7.5$\pm$0.9 & $<$Secondary$>$ & T5.5$\pm$0.4 & T5.5$\pm$0.4 & 0.30$\pm$0.21 & 1.42$\pm$0.27 & 2.00$\pm$0.23 & 0.55$\pm$0.22 & 1.82$\pm$0.33 &  \nodata \\
\enddata
\tablenotetext{a}{For each source, we list the parameters of the best-fitting template (minimum {\chisq}) for each of the fitting modes; for the composite fits, we also give mean parameters for the
primary ($<$Primary$>$) and secondary ($<$Secondary$>$), as well as the predicted relative magnitudes in the MKO $JHK$ and HST/NICMOS $F110W$ and $F170M$ bands.}
\tablenotetext{b}{Spectral fitting mode: Single: fits to single templates; L06-faint and L06-bright: fits to composite templates using the \citet{2006ApJ...647.1393L} faint and bright $M_K$/spectral type relations, respectively; L06-faint (IC) and L06-bright (IC): fits to composite templates using the \citet{2006ApJ...647.1393L} faint and bright relations, respectively, and the index-based spectral classifications.}
\end{deluxetable}

\begin{deluxetable}{lllllllcccccc}
\tabletypesize{\scriptsize}
\tablecaption{Template Fits to Weak Binary Candidates. \label{tab_fitsweak}\tablenotemark{a}}
\tablewidth{0pt}
\rotate
\tablehead{
\colhead{Mode\tablenotemark{b}} &
\colhead{Primary} &
\colhead{Pub SpT} &
\colhead{SpeX SpT} &
\colhead{Secondary} &
\colhead{Pub SpT} &
\colhead{SpeX SpT} &
\colhead{$\Delta{J}$} &
\colhead{$\Delta{H}$} &
\colhead{$\Delta{K}$} &
\colhead{$\Delta{F110W}$} &
\colhead{$\Delta{F170M}$} &
\colhead{$\chi^2$}  \\
\colhead{(1)} &
\colhead{(2)} &
\colhead{(3)} &
\colhead{(4)} &
\colhead{(5)} &
\colhead{(6)} &
\colhead{(7)} &
\colhead{(8)} &
\colhead{(9)} &
\colhead{(10)} &
\colhead{(11)} &
\colhead{(12)} &
\colhead{(13)} \\
}
\startdata
\cline{1-13}
\multicolumn{13}{c}{SDSS J011912.22+240331.6} \\
\cline{1-13}
Single & 1254-0122 & T2.0 & T2.0 & \nodata & \nodata & \nodata & \nodata & \nodata & \nodata & \nodata & \nodata & 25.4 \\
L06-faint & 0858+3256 & T1.0 & T0.0 & 2254+3123 & T4.0 & T4.0 & -0.60 & 0.08 & 0.83 & -0.38 & 0.29 & 15.7 \\
 & $<$Primary$>$ & T0.0$\pm$0.7 & L9.0$\pm$0.9 & $<$Secondary$>$ & T4.0$\pm$0.4 & T4.5$\pm$0.5 & -0.42$\pm$0.19 & 0.43$\pm$0.33 & 1.12$\pm$0.28 & -0.21$\pm$0.20 & 0.66$\pm$0.37 &  \nodata \\
L06-faint (IC) & 0858+3256 & T1.0 & T0.0 & 1750+1759 & T3.5 & T3.0 & -0.69 & -0.06 & 0.79 & -0.55 & 0.14 & 14.9 \\
 & $<$Primary$>$ & T0.0$\pm$1.0 & T0.0$\pm$0.4 & $<$Secondary$>$ & T4.0$\pm$0.5 & T3.5$\pm$0.7 & -0.54$\pm$0.23 & 0.20$\pm$0.39 & 0.99$\pm$0.33 & -0.39$\pm$0.27 & 0.42$\pm$0.45 &  \nodata \\
L06-bright & 0328+2302 & L9.5 & L9.0 & 2151-4853 & T4.0 & T4.0 & -0.47 & 0.38 & 1.04 & -0.26 & 0.58 & 14.2 \\
 & $<$Primary$>$ & L9.0$\pm$0.7 & L9.0$\pm$0.8 & $<$Secondary$>$ & T4.0$\pm$0.3 & T4.0$\pm$0.5 & -0.54$\pm$0.11 & 0.25$\pm$0.20 & 0.97$\pm$0.16 & -0.36$\pm$0.16 & 0.46$\pm$0.21 &  \nodata \\
\enddata
\tablenotetext{a}{For each source, we list the parameters of the best-fitting template (minimum {\chisq}) for each of the fitting modes; for the composite fits, we also give mean parameters for the
primary ($<$Primary$>$) and secondary ($<$Secondary$>$), as well as the predicted relative magnitudes in the MKO $JHK$ and HST/NICMOS $F110W$ and $F170M$ bands.}
\tablenotetext{b}{Spectral fitting mode: Single: fits to single templates; L06-faint and L06-bright: fits to composite templates using the \citet{2006ApJ...647.1393L} faint and bright $M_K$/spectral type relations, respectively; L06-faint (IC) and L06-bright (IC): fits to composite templates using the \citet{2006ApJ...647.1393L} faint and bright relations, respectively, and the index-based spectral classifications.}
\end{deluxetable}

\begin{deluxetable}{lllllcccl}
\tabletypesize{\scriptsize}
\tablecaption{Summary of Binary Candidates. \label{tab_summary}}
\tablewidth{0pt}
\tablehead{
 & \multicolumn{4}{c}{Spectral Types} &  &  & \colhead{AO/} & \\
\cline{2-5}
\colhead{Source} &
\colhead{Pub} &
\colhead{SpeX} &
\colhead{Primary} &
\colhead{Secondary} &
\colhead{$\Delta{J}$} &
\colhead{CL\tablenotemark{a}} &
\colhead{HST?} &
\colhead{Notes}  \\
\colhead{(1)} &
\colhead{(2)} &
\colhead{(3)} &
\colhead{(4)} &
\colhead{(5)} &
\colhead{(6)} &
\colhead{(7)} &
\colhead{(8)} &
\colhead{(9)}  \\
}
\startdata
\cline{1-9}
\multicolumn{9}{c}{Probable Binaries} \\
\cline{1-9}
SDSS J011912.22+240331.6 & T2 & T2.5 &  T0.0$\pm$0.7  &  T4.0$\pm$0.4  &  -0.42$\pm$0.19  & $>$99\% & No & Large $J$ flux reversal \\
\multicolumn{8}{c}{} & Cloudy primary? \\
SDSS J024749.90-163112.6 & T2$\pm$1.5 & T2.5$\pm$1 &  T0.0$\pm$0.2  &  T7.0$\pm$0.3  &  0.68$\pm$0.10  & $>$99\% & No & Old/metal-poor? \\
SDSS J035104.37+481046.8 & T1$\pm$1.5 & T1.5$\pm$1 &  L6.5$\pm$0.7  &  T5.0$\pm$0.7  &  0.31$\pm$0.31  & $>$99\% & No & Small $J$ flux reversal \\
\multicolumn{8}{c}{} & Cloudy primary? \\
SDSS J090900.73+652527.2 & T1.5 & T1.5 &  T1.5$\pm$0.5  &  T2.5$\pm$0.3  &  -0.12$\pm$0.10  & $>$99\% & No & Small $J$ flux reversal \\
2MASS J09490860-1545485 & T2 & T1.5$\pm$1 &  T1.0$\pm$0.2  &  T2.0$\pm$0.2  &  -0.07$\pm$0.05  & $>$99\% & No & Small $J$ flux reversal  \\
\enddata
\tablenotetext{a}{Confidence limit; see Section~4.2.}
\tablenotetext{b}{HST/WFPC2 observations in the F1042M band by \citet{2003AJ....125.3302G} did not have sufficient sensitivity to detect any putative T dwarf secondary in this system.}
\end{deluxetable}

\begin{deluxetable}{lcccccc}
\tabletypesize{\scriptsize}
\tablecaption{Estimated Binary Fractions for SpeX Sample. \label{tab_bf}}
\tablewidth{0pt}
\tablehead{
 & \multicolumn{4}{c}{Number of Sources} & \multicolumn{2}{c}{Binary Fraction\tablenotemark{a}}  \\
\cline{2-5}
\cline{6-7}
\colhead{SpT} &
\colhead{Total} &
\colhead{Resolved} &
\colhead{Candidates} &
\colhead{Non-candidates} &
\colhead{Observed} &
\colhead{Predicted\tablenotemark{b}} \\
\colhead{(1)} &
\colhead{(2)} &
\colhead{(3)} &
\colhead{(4)} &
\colhead{(5)} &
\colhead{(6)} &
\colhead{(7)}  \\
}
\startdata
L8--L9 & 12 & 0 & 0 & 12 & $<$13\% & 33\% \\
L9--T0 & 9 & 1 & 0 & 8 & 11$^{+18}_{-4}$\% & 49\% \\
T0--T1 & 9 & 3 & 3 & 3 & 67$^{+11}_{-17}$\% & 70\% \\
T1--T2 & 10 & 1 & 5 & 4 & 60$^{+13}_{-16}$\% & 68\% \\
T2--T3 & 17 & 1 & 7 & 9 & 47$^{+12}_{-11}$\% & 45\% \\
T3--T4 & 7 & 2 & 1 & 4 & 43$^{+18}_{-15}$\% & 23\% \\
T4--T5 & 10 & 1 & 0 & 9 & 10$^{+17}_{-3}$\% & 26\% \\
T5--T6 & 20 & 1 & 0 & 19 & 5$^{+10}_{-2}$\% & 30\% \\
\cline{1-7}
T0--T4 & 43 & 7 & 16 & 20 & 53$\pm$7\% & 56\% \\
\enddata
\tablenotetext{a}{Including resolved binaries plus binary candidates.}
\tablenotetext{b}{Based on the simulations of \citet{2007ApJ...659..655B}, assuming a mass function $\Psi$(M) $\propto$ M$^{-0.5}$, a constant birthrate over 0.01 $< t <$ 10~Gyr, an intrinsic binary fraction $\epsilon_b$ = 15\%, a mass ratio distribution $P(q)$ $\propto$ $q^4$, evolutionary models from \citet{2003A&A...402..701B}, $K$-band absolute magnitude and bolometric magnitude relations as defined in that study, and a magnitude-limited sample.}
\end{deluxetable}


\clearpage

\begin{thebibliography}{129}
\expandafter\ifx\csname natexlab\endcsname\relax\def\natexlab#1{#1}\fi

\bibitem[{{Ackerman} \& {Marley}(2001)}]{2001ApJ...556..872A}
{Ackerman}, A.~S., \& {Marley}, M.~S. 2001, \apj, 556, 872

\bibitem[{{Allen}(2007)}]{2007ApJ...668..492A}
{Allen}, P.~R. 2007, \apj, 668, 492

\bibitem[{{Artigau} {et~al.}(2006){Artigau}, {Doyon}, {Lafreni{\`e}re},
  {Nadeau}, {Robert}, \& {Albert}}]{2006ApJ...651L..57A}
{Artigau}, {\'E}., {Doyon}, R., {Lafreni{\`e}re}, D., {Nadeau}, D., {Robert},
  J., \& {Albert}, L. 2006, \apjl, 651, L57

\bibitem[{{Baraffe} {et~al.}(2003){Baraffe}, {Chabrier}, {Barman}, {Allard}, \&
  {Hauschildt}}]{2003A&A...402..701B}
{Baraffe}, I., {Chabrier}, G., {Barman}, T.~S., {Allard}, F., \& {Hauschildt},
  P.~H. 2003, \aap, 402, 701

\bibitem[{{Basri} \& {Mart{\'{\i}}n}(1999)}]{1999AJ....118.2460B}
{Basri}, G., \& {Mart{\'{\i}}n}, E.~L. 1999, \aj, 118, 2460

\bibitem[{{Basri} \& {Reiners}(2006)}]{2006AJ....132..663B}
{Basri}, G., \& {Reiners}, A. 2006, \aj, 132, 663

\bibitem[{{Blake} {et~al.}(2008){Blake}, {Charbonneau}, {White}, {Torres},
  {Marley}, \& {Saumon}}]{2008ApJ...678L.125B}
{Blake}, C.~H., {Charbonneau}, D., {White}, R.~J., {Torres}, G., {Marley},
  M.~S., \& {Saumon}, D. 2008, \apjl, 678, L125

\bibitem[{{Bouy} {et~al.}(2005){Bouy}, {Mart{\'{\i}}n}, {Brandner}, \&
  {Bouvier}}]{2005AJ....129..511B}
{Bouy}, H., {Mart{\'{\i}}n}, E.~L., {Brandner}, W., \& {Bouvier}, J. 2005, \aj,
  129, 511

\bibitem[{{Burgasser}(2007{\natexlab{a}})}]{2007ApJ...659..655B}
{Burgasser}, A.~J. 2007{\natexlab{a}}, \apj, 659, 655

\bibitem[{{Burgasser}(2007{\natexlab{b}})}]{2007AJ....134.1330B}
---. 2007{\natexlab{b}}, \aj, 134, 1330

\bibitem[{{Burgasser}(2007{\natexlab{c}})}]{2007ApJ...658..617B}
---. 2007{\natexlab{c}}, \apj, 658, 617

\bibitem[{{Burgasser} {et~al.}(2006{\natexlab{a}}){Burgasser}, {Burrows}, \&
  {Kirkpatrick}}]{2006ApJ...639.1095B}
{Burgasser}, A.~J., {Burrows}, A., \& {Kirkpatrick}, J.~D. 2006{\natexlab{a}},
  \apj, 639, 1095

\bibitem[{{Burgasser} {et~al.}(2006{\natexlab{b}}){Burgasser}, {Geballe},
  {Leggett}, {Kirkpatrick}, \& {Golimowski}}]{2006ApJ...637.1067B}
{Burgasser}, A.~J., {Geballe}, T.~R., {Leggett}, S.~K., {Kirkpatrick}, J.~D.,
  \& {Golimowski}, D.~A. 2006{\natexlab{b}}, \apj, 637, 1067

\bibitem[{{Burgasser} {et~al.}(2003{\natexlab{a}}){Burgasser}, {Kirkpatrick},
  {Burrows}, {Liebert}, {Reid}, {Gizis}, {McGovern}, {Prato}, \&
  {McLean}}]{2003ApJ...592.1186B}
{Burgasser}, A.~J., {Kirkpatrick}, J.~D., {Burrows}, A., {Liebert}, J., {Reid},
  I.~N., {Gizis}, J.~E., {McGovern}, M.~R., {Prato}, L., \& {McLean}, I.~S.
  2003{\natexlab{a}}, \apj, 592, 1186

\bibitem[{{Burgasser} {et~al.}(2006{\natexlab{c}}){Burgasser}, {Kirkpatrick},
  {Cruz}, {Reid}, {Leggett}, {Liebert}, {Burrows}, \&
  {Brown}}]{2006ApJS..166..585B}
{Burgasser}, A.~J., {Kirkpatrick}, J.~D., {Cruz}, K.~L., {Reid}, I.~N.,
  {Leggett}, S.~K., {Liebert}, J., {Burrows}, A., \& {Brown}, M.~E.
  2006{\natexlab{c}}, \apjs, 166, 585

\bibitem[{{Burgasser} {et~al.}(2003{\natexlab{b}}){Burgasser}, {Kirkpatrick},
  {Liebert}, \& {Burrows}}]{2003ApJ...594..510B}
{Burgasser}, A.~J., {Kirkpatrick}, J.~D., {Liebert}, J., \& {Burrows}, A.
  2003{\natexlab{b}}, \apj, 594, 510

\bibitem[{{Burgasser} {et~al.}(2005{\natexlab{a}}){Burgasser}, {Kirkpatrick},
  \& {Lowrance}}]{2005AJ....129.2849B}
{Burgasser}, A.~J., {Kirkpatrick}, J.~D., \& {Lowrance}, P.~J.
  2005{\natexlab{a}}, \aj, 129, 2849

\bibitem[{{Burgasser} {et~al.}(2003{\natexlab{c}}){Burgasser}, {Kirkpatrick},
  {McElwain}, {Cutri}, {Burgasser}, \& {Skrutskie}}]{2003AJ....125..850B}
{Burgasser}, A.~J., {Kirkpatrick}, J.~D., {McElwain}, M.~W., {Cutri}, R.~M.,
  {Burgasser}, A.~J., \& {Skrutskie}, M.~F. 2003{\natexlab{c}}, \aj, 125, 850

\bibitem[{{Burgasser} {et~al.}(2008{\natexlab{a}}){Burgasser}, {Liu},
  {Ireland}, {Cruz}, \& {Dupuy}}]{2008ApJ...681..579B}
{Burgasser}, A.~J., {Liu}, M.~C., {Ireland}, M.~J., {Cruz}, K.~L., \& {Dupuy},
  T.~J. 2008{\natexlab{a}}, \apj, 681, 579

\bibitem[{{Burgasser} {et~al.}(2008{\natexlab{b}}){Burgasser}, {Looper},
  {Kirkpatrick}, {Cruz}, \& {Swift}}]{2008ApJ...674..451B}
{Burgasser}, A.~J., {Looper}, D.~L., {Kirkpatrick}, J.~D., {Cruz}, K.~L., \&
  {Swift}, B.~J. 2008{\natexlab{b}}, \apj, 674, 451

\bibitem[{{Burgasser} {et~al.}(2007{\natexlab{a}}){Burgasser}, {Looper},
  {Kirkpatrick}, \& {Liu}}]{2007ApJ...658..557B}
{Burgasser}, A.~J., {Looper}, D.~L., {Kirkpatrick}, J.~D., \& {Liu}, M.~C.
  2007{\natexlab{a}}, \apj, 658, 557

\bibitem[{{Burgasser} \& {McElwain}(2006)}]{2006AJ....131.1007B}
{Burgasser}, A.~J., \& {McElwain}, M.~W. 2006, \aj, 131, 1007

\bibitem[{{Burgasser} {et~al.}(2003{\natexlab{d}}){Burgasser}, {McElwain}, \&
  {Kirkpatrick}}]{2003AJ....126.2487B}
{Burgasser}, A.~J., {McElwain}, M.~W., \& {Kirkpatrick}, J.~D.
  2003{\natexlab{d}}, \aj, 126, 2487

\bibitem[{{Burgasser} {et~al.}(2004){Burgasser}, {McElwain}, {Kirkpatrick},
  {Cruz}, {Tinney}, \& {Reid}}]{2004AJ....127.2856B}
{Burgasser}, A.~J., {McElwain}, M.~W., {Kirkpatrick}, J.~D., {Cruz}, K.~L.,
  {Tinney}, C.~G., \& {Reid}, I.~N. 2004, \aj, 127, 2856

\bibitem[{{Burgasser} {et~al.}(2005{\natexlab{b}}){Burgasser}, {Reid},
  {Leggett}, {Kirkpatrick}, {Liebert}, \& {Burrows}}]{2005ApJ...634L.177B}
{Burgasser}, A.~J., {Reid}, I.~N., {Leggett}, S.~K., {Kirkpatrick}, J.~D.,
  {Liebert}, J., \& {Burrows}, A. 2005{\natexlab{b}}, \apjl, 634, L177

\bibitem[{{Burgasser} {et~al.}(2007{\natexlab{b}}){Burgasser}, {Reid},
  {Siegler}, {Close}, {Allen}, {Lowrance}, \& {Gizis}}]{2007prpl.conf..427B}
{Burgasser}, A.~J., {Reid}, I.~N., {Siegler}, N., {Close}, L., {Allen}, P.,
  {Lowrance}, P., \& {Gizis}, J. 2007{\natexlab{b}}, Protostars and Planets V,
  427

\bibitem[{{Burgasser} {et~al.}(1999)}]{1999ApJ...522L..65B}
{Burgasser}, A.~J., {et~al.} 1999, \apjl, 522, L65

\bibitem[{{Burgasser} {et~al.}(2000{\natexlab{a}})}]{2000AJ....120.1100B}
---. 2000{\natexlab{a}}, \aj, 120, 1100

\bibitem[{{Burgasser} {et~al.}(2000{\natexlab{b}})}]{2000ApJ...531L..57B}
---. 2000{\natexlab{b}}, \apjl, 531, L57

\bibitem[{{Burgasser} {et~al.}(2002{\natexlab{a}}){Burgasser}, {Kirkpatrick},
  {Brown}, {Reid}, {Burrows}, {Liebert}, {Matthews}, {Gizis}, {Dahn}, {Monet},
  {Cutri}, \& {Skrutskie}}]{2002ApJ...564..421B}
{Burgasser}, A.~J., {et~al.}  2002{\natexlab{a}},
  \apj, 564, 421

\bibitem[{{Burgasser} {et~al.}(2002{\natexlab{b}})}]{2002ApJ...571L.151B}
---. 2002{\natexlab{b}}, \apjl, 571, L151

\bibitem[{{Burrows} {et~al.}(2006){Burrows}, {Sudarsky}, \&
  {Hubeny}}]{2006ApJ...640.1063B}
{Burrows}, A., {Sudarsky}, D., \& {Hubeny}, I. 2006, \apj, 640, 1063

\bibitem[{{Chabrier} {et~al.}(2009){Chabrier}, {Baraffe}, {Leconte},
  {Gallardo}, \& {Barman}}]{2009AIPC.1094..102C}
{Chabrier}, G., {Baraffe}, I., {Leconte}, J., {Gallardo}, J., \& {Barman}, T.
  2009, in American Institute of Physics Conference Series, Vol. 1094, American
  Institute of Physics Conference Series, ed. E.~{Stempels}, 102--111

\bibitem[{{Chiu} {et~al.}(2006){Chiu}, {Fan}, {Leggett}, {Golimowski}, {Zheng},
  {Geballe}, {Schneider}, \& {Brinkmann}}]{2006AJ....131.2722C}
{Chiu}, K., {Fan}, X., {Leggett}, S.~K., {Golimowski}, D.~A., {Zheng}, W.,
  {Geballe}, T.~R., {Schneider}, D.~P., \& {Brinkmann}, J. 2006, \aj, 131, 2722

\bibitem[{{Cruz} {et~al.}(2004){Cruz}, {Burgasser}, {Reid}, \&
  {Liebert}}]{2004ApJ...604L..61C}
{Cruz}, K.~L., {Burgasser}, A.~J., {Reid}, I.~N., \& {Liebert}, J. 2004, \apjl,
  604, L61

\bibitem[{{Cruz} {et~al.}(2009){Cruz}, {Kirkpatrick}, \&
  {Burgasser}}]{2009AJ....137.3345C}
{Cruz}, K.~L., {Kirkpatrick}, J.~D., \& {Burgasser}, A.~J. 2009, \aj, 137, 3345

\bibitem[{{Cruz} {et~al.}(2007){Cruz}, {Reid}, {Kirkpatrick}, {Burgasser},
  {Liebert}, {Solomon}, {Schmidt}, {Allen}, {Hawley}, \&
  {Covey}}]{2007AJ....133..439C}
{Cruz}, K.~L., {et~al.} 2007, \aj, 133, 439

\bibitem[{{Cruz} {et~al.}(2003){Cruz}, {Reid}, {Liebert}, {Kirkpatrick}, \&
  {Lowrance}}]{2003AJ....126.2421C}
{Cruz}, K.~L., {Reid}, I.~N., {Liebert}, J., {Kirkpatrick}, J.~D., \&
  {Lowrance}, P.~J. 2003, \aj, 126, 2421

\bibitem[{{Cushing} {et~al.}(2008){Cushing}, {Marley}, {Saumon}, {Kelly},
  {Vacca}, {Rayner}, {Freedman}, {Lodders}, \& {Roellig}}]{2008ApJ...678.1372C}
{Cushing}, M.~C., {Marley}, M.~S., {Saumon}, D., {Kelly}, B.~C., {Vacca},
  W.~D., {Rayner}, J.~T., {Freedman}, R.~S., {Lodders}, K., \& {Roellig}, T.~L.
  2008, \apj, 678, 1372

\bibitem[{{Cushing} {et~al.}(2005){Cushing}, {Rayner}, \&
  {Vacca}}]{2005ApJ...623.1115C}
{Cushing}, M.~C., {Rayner}, J.~T., \& {Vacca}, W.~D. 2005, \apj, 623, 1115

\bibitem[{{Cushing} {et~al.}(2006){Cushing}, {Roellig}, {Marley}, {Saumon},
  {Leggett}, {Kirkpatrick}, {Wilson}, {Sloan}, {Mainzer}, {Van Cleve}, \&
  {Houck}}]{2006ApJ...648..614C}
{Cushing}, M.~C., {et~al.} 2006, \apj, 648, 614

\bibitem[{{Cushing} {et~al.}(2004){Cushing}, {Vacca}, \&
  {Rayner}}]{2004PASP..116..362C}
{Cushing}, M.~C., {Vacca}, W.~D., \& {Rayner}, J.~T. 2004, \pasp, 116, 362

\bibitem[{{Dahn} {et~al.}(2002)}]{2002AJ....124.1170D}
{Dahn}, C.~C., {et~al.} 2002, \aj, 124, 1170

\bibitem[{{Delfosse} {et~al.}(1997)}]{1997A&A...327L..25D}
{Delfosse}, X., {et~al.} 1997, \aap, 327, L25

\bibitem[{{Dupuy} {et~al.}(2009){Dupuy}, {Liu}, \&
  {Ireland}}]{2009ApJ...692..729D}
{Dupuy}, T.~J., {Liu}, M.~C., \& {Ireland}, M.~J. 2009, \apj, 692, 729

\bibitem[{{Duquennoy} \& {Mayor}(1991)}]{1991A&A...248..485D}
{Duquennoy}, A., \& {Mayor}, M. 1991, \aap, 248, 485

\bibitem[{{Ellis} {et~al.}(2005){Ellis}, {Tinney}, {Burgasser}, {Kirkpatrick},
  \& {McElwain}}]{2005AJ....130.2347E}
{Ellis}, S.~C., {Tinney}, C.~G., {Burgasser}, A.~J., {Kirkpatrick}, J.~D., \&
  {McElwain}, M.~W. 2005, \aj, 130, 2347

\bibitem[{{Epchtein} {et~al.}(1997)}]{1997Msngr..87...27E}
{Epchtein}, N., {et~al.} 1997, The Messenger, 87, 27

\bibitem[{{Faherty} {et~al.}(2009){Faherty}, {Burgasser}, {Cruz}, {Shara},
  {Walter}, \& {Gelino}}]{2009AJ....137....1F}
{Faherty}, J.~K., {Burgasser}, A.~J., {Cruz}, K.~L., {Shara}, M.~M., {Walter},
  F.~M., \& {Gelino}, C.~R. 2009, \aj, 137, 1

\bibitem[{{Fan} {et~al.}(2000)}]{2000AJ....119..928F}
{Fan}, X., {et~al.} 2000, \aj, 119, 928

\bibitem[{{Fischer} \& {Marcy}(1992)}]{1992ApJ...396..178F}
{Fischer}, D.~A., \& {Marcy}, G.~W. 1992, \apj, 396, 178

\bibitem[{{Geballe} {et~al.}(2002)}]{2002ApJ...564..466G}
{Geballe}, T.~R., {et~al.} 2002, \apj, 564, 466

\bibitem[{{Gelino} {et~al.}(2006){Gelino}, {Kulkarni}, \&
  {Stephens}}]{2006PASP..118..611G}
{Gelino}, C.~R., {Kulkarni}, S.~R., \& {Stephens}, D.~C. 2006, \pasp, 118, 611

\bibitem[{{Gizis}(2002)}]{2002ApJ...575..484G}
{Gizis}, J.~E. 2002, \apj, 575, 484

\bibitem[{{Gizis} {et~al.}(2000){Gizis}, {Monet}, {Reid}, {Kirkpatrick},
  {Liebert}, \& {Williams}}]{2000AJ....120.1085G}
{Gizis}, J.~E., {Monet}, D.~G., {Reid}, I.~N., {Kirkpatrick}, J.~D., {Liebert},
  J., \& {Williams}, R.~J. 2000, \aj, 120, 1085

\bibitem[{{Gizis} {et~al.}(2003){Gizis}, {Reid}, {Knapp}, {Liebert},
  {Kirkpatrick}, {Koerner}, \& {Burgasser}}]{2003AJ....125.3302G}
{Gizis}, J.~E., {Reid}, I.~N., {Knapp}, G.~R., {Liebert}, J., {Kirkpatrick},
  J.~D., {Koerner}, D.~W., \& {Burgasser}, A.~J. 2003, \aj, 125, 3302

\bibitem[{{Goldman} {et~al.}(2008){Goldman}, {Bouy}, {Zapatero Osorio},
  {Stumpf}, {Brandner}, \& {Henning}}]{2008A&A...490..763G}
{Goldman}, B., {Bouy}, H., {Zapatero Osorio}, M.~R., {Stumpf}, M.~B.,
  {Brandner}, W., \& {Henning}, T. 2008, \aap, 490, 763

\bibitem[{{Golimowski} {et~al.}(2004)}]{2004AJ....127.3516G}
{Golimowski}, D.~A., {et~al.} 2004, \aj, 127, 3516

\bibitem[{{Hawley} {et~al.}(2002)}]{2002AJ....123.3409H}
{Hawley}, S.~L., {et~al.} 2002, \aj, 123, 3409

\bibitem[{{Joergens}(2008)}]{2008A&A...492..545J}
{Joergens}, V. 2008, \aap, 492, 545

\bibitem[{{Joergens} \& {M{\"u}ller}(2007)}]{2007ApJ...666L.113J}
{Joergens}, V., \& {M{\"u}ller}, A. 2007, \apjl, 666, L113

\bibitem[{{Kendall} {et~al.}(2004){Kendall}, {Delfosse}, {Mart{\'{\i}}n}, \&
  {Forveille}}]{2004A&A...416L..17K}
{Kendall}, T.~R., {Delfosse}, X., {Mart{\'{\i}}n}, E.~L., \& {Forveille}, T.
  2004, \aap, 416, L17

\bibitem[{{Kendall} {et~al.}(2007){Kendall}, {Jones}, {Pinfield}, {Pokorny},
  {Folkes}, {Weights}, {Jenkins}, \& {Mauron}}]{2007MNRAS.374..445K}
{Kendall}, T.~R., {Jones}, H.~R.~A., {Pinfield}, D.~J., {Pokorny}, R.~S.,
  {Folkes}, S., {Weights}, D., {Jenkins}, J.~S., \& {Mauron}, N. 2007, \mnras,
  374, 445

\bibitem[{{Kendall} {et~al.}(2003){Kendall}, {Mauron}, {Azzopardi}, \&
  {Gigoyan}}]{2003A&A...403..929K}
{Kendall}, T.~R., {Mauron}, N., {Azzopardi}, M., \& {Gigoyan}, K. 2003, \aap,
  403, 929

\bibitem[{{Kirkpatrick}(2005)}]{2005ARA&A..43..195K}
{Kirkpatrick}, J.~D. 2005, \araa, 43, 195

\bibitem[{{Kirkpatrick} {et~al.}(2000){Kirkpatrick}, {Reid}, {Liebert},
  {Gizis}, {Burgasser}, {Monet}, {Dahn}, {Nelson}, \&
  {Williams}}]{2000AJ....120..447K}
{Kirkpatrick}, J.~D., {Reid}, I.~N., {Liebert}, J., {Gizis}, J.~E.,
  {Burgasser}, A.~J., {Monet}, D.~G., {Dahn}, C.~C., {Nelson}, B., \&
  {Williams}, R.~J. 2000, \aj, 120, 447

\bibitem[{{Kirkpatrick} {et~al.}(1999)}]{1999ApJ...519..802K}
{Kirkpatrick}, J.~D., {et~al.} 1999, \apj, 519, 802

\bibitem[{{Kirkpatrick} {et~al.}(2008)}]{2008ApJ...689.1295K}
---. 2008, \apj, 689, 1295

\bibitem[{{Knapp} {et~al.}(2004)}]{2004AJ....127.3553K}
{Knapp}, G.~R., {et~al.} 2004, \aj, 127, 3553

\bibitem[{{Leggett} {et~al.}(2007){Leggett}, {Saumon}, {Marley}, {Geballe},
  {Golimowski}, {Stephens}, \& {Fan}}]{2007ApJ...655.1079L}
{Leggett}, S.~K., {Saumon}, D., {Marley}, M.~S., {Geballe}, T.~R.,
  {Golimowski}, D.~A., {Stephens}, D., \& {Fan}, X. 2007, \apj, 655, 1079

\bibitem[{{Leggett} {et~al.}(2000)}]{2000ApJ...536L..35L}
{Leggett}, S.~K., {et~al.} 2000, \apjl, 536, L35

\bibitem[{{Leggett} {et~al.}(2002)}]{2002ApJ...564..452L}
---. 2002, \apj, 564, 452

\bibitem[{{Liebert} \& {Burgasser}(2007)}]{2007ApJ...655..522L}
{Liebert}, J., \& {Burgasser}, A.~J. 2007, \apj, 655, 522

\bibitem[{{Liebert} \& {Gizis}(2006)}]{2006PASP..118..659L}
{Liebert}, J., \& {Gizis}, J.~E. 2006, \pasp, 118, 659

\bibitem[{{Liebert} {et~al.}(2003){Liebert}, {Kirkpatrick}, {Cruz}, {Reid},
  {Burgasser}, {Tinney}, \& {Gizis}}]{2003AJ....125..343L}
{Liebert}, J., {Kirkpatrick}, J.~D., {Cruz}, K.~L., {Reid}, I.~N., {Burgasser},
  A., {Tinney}, C.~G., \& {Gizis}, J.~E. 2003, \aj, 125, 343

\bibitem[{{Liu} {et~al.}(2008){Liu}, {Dupuy}, \&
  {Ireland}}]{2008ApJ...689..436L}
{Liu}, M.~C., {Dupuy}, T.~J., \& {Ireland}, M.~J. 2008, \apj, 689, 436

\bibitem[{{Liu} \& {Leggett}(2005)}]{2005ApJ...634..616L}
{Liu}, M.~C., \& {Leggett}, S.~K. 2005, \apj, 634, 616

\bibitem[{{Liu} {et~al.}(2006){Liu}, {Leggett}, {Golimowski}, {Chiu}, {Fan},
  {Geballe}, {Schneider}, \& {Brinkmann}}]{2006ApJ...647.1393L}
{Liu}, M.~C., {Leggett}, S.~K., {Golimowski}, D.~A., {Chiu}, K., {Fan}, X.,
  {Geballe}, T.~R., {Schneider}, D.~P., \& {Brinkmann}, J. 2006, \apj, 647,
  1393

\bibitem[{{Lodders}(2002)}]{2002ApJ...577..974L}
{Lodders}, K. 2002, \apj, 577, 974

\bibitem[{{Lodieu} {et~al.}(2007){Lodieu}, {Dobbie}, {Deacon}, {Hodgkin},
  {Hambly}, \& {Jameson}}]{2007MNRAS.380..712L}
{Lodieu}, N., {Dobbie}, P.~D., {Deacon}, N.~R., {Hodgkin}, S.~T., {Hambly},
  N.~C., \& {Jameson}, R.~F. 2007, \mnras, 380, 712

\bibitem[{{Looper} {et~al.}(2008{\natexlab{a}}){Looper}, {Gelino}, {Burgasser},
  \& {Kirkpatrick}}]{2008ApJ...685.1183L}
{Looper}, D.~L., {Gelino}, C.~R., {Burgasser}, A.~J., \& {Kirkpatrick}, J.~D.
  2008{\natexlab{a}}, \apj, 685, 1183

\bibitem[{{Looper} {et~al.}(2007){Looper}, {Kirkpatrick}, \&
  {Burgasser}}]{2007AJ....134.1162L}
{Looper}, D.~L., {Kirkpatrick}, J.~D., \& {Burgasser}, A.~J. 2007, \aj, 134,
  1162

\bibitem[{{Looper} {et~al.}(2008{\natexlab{b}})}]{2008ApJ...686..528L}
{Looper}, D.~L., {et~al.} 2008{\natexlab{b}}, \apj, 686, 528

\bibitem[{{Lucas} \& {Roche}(2000)}]{2000MNRAS.314..858L}
{Lucas}, P.~W., \& {Roche}, P.~F. 2000, \mnras, 314, 858

\bibitem[{{Luhman} {et~al.}(2007)}]{2007ApJ...654..570L}
{Luhman}, K.~L., {et~al.} 2007, \apj, 654, 570

\bibitem[{{Marley} {et~al.}(1996){Marley}, {Saumon}, {Guillot}, {Freedman},
  {Hubbard}, {Burrows}, \& {Lunine}}]{1996Sci...272.1919M}
{Marley}, M.~S., {Saumon}, D., {Guillot}, T., {Freedman}, R.~S., {Hubbard},
  W.~B., {Burrows}, A., \& {Lunine}, J.~I. 1996, Science, 272, 1919

\bibitem[{{Martin} {et~al.}(1999){Martin}, {Brandner}, \&
  {Basri}}]{1999Sci...283.1718M}
{Martin}, E.~L., {Brandner}, W., \& {Basri}, G. 1999, Science, 283, 1718

\bibitem[{{Mart{\'{\i}}n} {et~al.}(1999){Mart{\'{\i}}n}, {Delfosse}, {Basri},
  {Goldman}, {Forveille}, \& {Zapatero Osorio}}]{1999AJ....118.2466M}
{Mart{\'{\i}}n}, E.~L., {Delfosse}, X., {Basri}, G., {Goldman}, B.,
  {Forveille}, T., \& {Zapatero Osorio}, M.~R. 1999, \aj, 118, 2466

\bibitem[{{McCaughrean} {et~al.}(2004){McCaughrean}, {Close}, {Scholz},
  {Lenzen}, {Biller}, {Brandner}, {Hartung}, \& {Lodieu}}]{2004A&A...413.1029M}
{McCaughrean}, M.~J., {Close}, L.~M., {Scholz}, R.-D., {Lenzen}, R., {Biller},
  B., {Brandner}, W., {Hartung}, M., \& {Lodieu}, N. 2004, \aap, 413, 1029

\bibitem[{{McElwain} \& {Burgasser}(2006)}]{2006AJ....132.2074M}
{McElwain}, M.~W., \& {Burgasser}, A.~J. 2006, \aj, 132, 2074

\bibitem[{{McLean} {et~al.}(2003){McLean}, {McGovern}, {Burgasser},
  {Kirkpatrick}, {Prato}, \& {Kim}}]{2003ApJ...596..561M}
{McLean}, I.~S., {McGovern}, M.~R., {Burgasser}, A.~J., {Kirkpatrick}, J.~D.,
  {Prato}, L., \& {Kim}, S.~S. 2003, \apj, 596, 561

\bibitem[{{Metchev} \& {Hillenbrand}(2006)}]{2006ApJ...651.1166M}
{Metchev}, S.~A., \& {Hillenbrand}, L.~A. 2006, \apj, 651, 1166

\bibitem[{{Metchev} {et~al.}(2008){Metchev}, {Kirkpatrick}, {Berriman}, \&
  {Looper}}]{2008ApJ...676.1281M}
{Metchev}, S.~A., {Kirkpatrick}, J.~D., {Berriman}, G.~B., \& {Looper}, D.
  2008, \apj, 676, 1281

\bibitem[{{Mohanty} {et~al.}(2007){Mohanty}, {Jayawardhana}, {Hu{\'e}lamo}, \&
  {Mamajek}}]{2007ApJ...657.1064M}
{Mohanty}, S., {Jayawardhana}, R., {Hu{\'e}lamo}, N., \& {Mamajek}, E. 2007,
  \apj, 657, 1064

\bibitem[{{Muench} {et~al.}(2007){Muench}, {Lada}, {Luhman}, {Muzerolle}, \&
  {Young}}]{2007AJ....134..411M}
{Muench}, A.~A., {Lada}, C.~J., {Luhman}, K.~L., {Muzerolle}, J., \& {Young},
  E. 2007, \aj, 134, 411

\bibitem[{{Mugrauer} {et~al.}(2006){Mugrauer}, {Seifahrt}, {Neuh{\"a}user}, \&
  {Mazeh}}]{2006MNRAS.373L..31M}
{Mugrauer}, M., {Seifahrt}, A., {Neuh{\"a}user}, R., \& {Mazeh}, T. 2006,
  \mnras, 373, L31

\bibitem[{{Nakajima} {et~al.}(2004){Nakajima}, {Tsuji}, \&
  {Yanagisawa}}]{2004ApJ...607..499N}
{Nakajima}, T., {Tsuji}, T., \& {Yanagisawa}, K. 2004, \apj, 607, 499

\bibitem[{{Rayner} {et~al.}(2003){Rayner}, {Toomey}, {Onaka}, {Denault},
  {Stahlberger}, {Vacca}, {Cushing}, \& {Wang}}]{2003PASP..115..362R}
{Rayner}, J.~T., {Toomey}, D.~W., {Onaka}, P.~M., {Denault}, A.~J.,
  {Stahlberger}, W.~E., {Vacca}, W.~D., {Cushing}, M.~C., \& {Wang}, S. 2003,
  \pasp, 115, 362

\bibitem[{{Reid} {et~al.}(2001{\natexlab{a}}){Reid}, {Burgasser}, {Cruz},
  {Kirkpatrick}, \& {Gizis}}]{2001AJ....121.1710R}
{Reid}, I.~N., {Burgasser}, A.~J., {Cruz}, K.~L., {Kirkpatrick}, J.~D., \&
  {Gizis}, J.~E. 2001{\natexlab{a}}, \aj, 121, 1710

\bibitem[{{Reid} {et~al.}(2008){Reid}, {Cruz}, {Kirkpatrick}, {Allen},
  {Mungall}, {Liebert}, {Lowrance}, \& {Sweet}}]{2008AJ....136.1290R}
{Reid}, I.~N., {Cruz}, K.~L., {Kirkpatrick}, J.~D., {Allen}, P.~R., {Mungall},
  F., {Liebert}, J., {Lowrance}, P., \& {Sweet}, A. 2008, \aj, 136, 1290

\bibitem[{{Reid} {et~al.}(2001{\natexlab{b}}){Reid}, {Gizis}, {Kirkpatrick}, \&
  {Koerner}}]{2001AJ....121..489R}
{Reid}, I.~N., {Gizis}, J.~E., {Kirkpatrick}, J.~D., \& {Koerner}, D.~W.
  2001{\natexlab{b}}, \aj, 121, 489

\bibitem[{{Reid} {et~al.}(2000){Reid}, {Kirkpatrick}, {Gizis}, {Dahn}, {Monet},
  {Williams}, {Liebert}, \& {Burgasser}}]{2000AJ....119..369R}
{Reid}, I.~N., {Kirkpatrick}, J.~D., {Gizis}, J.~E., {Dahn}, C.~C., {Monet},
  D.~G., {Williams}, R.~J., {Liebert}, J., \& {Burgasser}, A.~J. 2000, \aj,
  119, 369

\bibitem[{{Reid} {et~al.}(2006){Reid}, {Lewitus}, {Burgasser}, \&
  {Cruz}}]{2006ApJ...639.1114R}
{Reid}, I.~N., {Lewitus}, E., {Burgasser}, A.~J., \& {Cruz}, K.~L. 2006, \apj,
  639, 1114

\bibitem[{{Saumon} \& {Marley}(2008)}]{2008ApJ...689.1327S}
{Saumon}, D., \& {Marley}, M.~S. 2008, \apj, 689, 1327

\bibitem[{{Schneider} {et~al.}(2002)}]{2002AJ....123..458S}
{Schneider}, D.~P., {et~al.} 2002, \aj, 123, 458

\bibitem[{{Sheppard} \& {Cushing}(2009)}]{2009AJ....137..304S}
{Sheppard}, S.~S., \& {Cushing}, M.~C. 2009, \aj, 137, 304

\bibitem[{{Siegler} {et~al.}(2007){Siegler}, {Close}, {Burgasser}, {Cruz},
  {Marois}, {Macintosh}, \& {Barman}}]{2007AJ....133.2320S}
{Siegler}, N., {Close}, L.~M., {Burgasser}, A.~J., {Cruz}, K.~L., {Marois}, C.,
  {Macintosh}, B., \& {Barman}, T. 2007, \aj, 133, 2320

\bibitem[{{Simons} \& {Tokunaga}(2002)}]{2002PASP..114..169S}
{Simons}, D.~A., \& {Tokunaga}, A. 2002, \pasp, 114, 169

\bibitem[{{Skrutskie} {et~al.}(2006)}]{2006AJ....131.1163S}
{Skrutskie}, M.~F., {et~al.} 2006, \aj, 131, 1163

\bibitem[{{Stassun} {et~al.}(2006){Stassun}, {Mathieu}, \&
  {Valenti}}]{2006Natur.440..311S}
{Stassun}, K.~G., {Mathieu}, R.~D., \& {Valenti}, J.~A. 2006, \nat, 440, 311

\bibitem[{{Stephens} {et~al.}(2009){Stephens}, {Leggett}, {Cushing}, {Marley},
  {Saumon}, {Geballe}, {Golimowski}, {Fan}, \& {Noll}}]{2009ApJ...702..154S}
{Stephens}, D.~C., {et~al.} 2009, \apj, 702, 154

\bibitem[{{Strauss} {et~al.}(1999)}]{1999ApJ...522L..61S}
{Strauss}, M.~A., {et~al.} 1999, \apjl, 522, L61

\bibitem[{{Stumpf} {et~al.}(2008){Stumpf}, {Brandner}, {Henning}, {Bouy},
  {Koehler}, {Hormuth}, {Joergens}, \& {Kasper}}]{2008arXiv0811.0556S}
{Stumpf}, M.~B., {Brandner}, W., {Henning}, T., {Bouy}, H., {Koehler}, R.,
  {Hormuth}, F., {Joergens}, V., \& {Kasper}, M. 2008, ArXiv e-prints

\bibitem[{{Testi} {et~al.}(2001){Testi}, {D'Antona}, {Ghinassi}, {Licandro},
  {Magazz{\`u}}, {Maiolino}, {Mannucci}, {Marconi}, {Nagar}, {Natta}, \&
  {Oliva}}]{2001ApJ...552L.147T}
{Testi}, L., {et~al.} 2001, \apjl, 552, L147

\bibitem[{{Tinney} {et~al.}(2003){Tinney}, {Burgasser}, \&
  {Kirkpatrick}}]{2003AJ....126..975T}
{Tinney}, C.~G., {Burgasser}, A.~J., \& {Kirkpatrick}, J.~D. 2003, \aj, 126,
  975

\bibitem[{{Tinney} {et~al.}(2005){Tinney}, {Burgasser}, {Kirkpatrick}, \&
  {McElwain}}]{2005AJ....130.2326T}
{Tinney}, C.~G., {Burgasser}, A.~J., {Kirkpatrick}, J.~D., \& {McElwain}, M.~W.
  2005, \aj, 130, 2326

\bibitem[{{Tokunaga} {et~al.}(2002){Tokunaga}, {Simons}, \&
  {Vacca}}]{2002PASP..114..180T}
{Tokunaga}, A.~T., {Simons}, D.~A., \& {Vacca}, W.~D. 2002, \pasp, 114, 180

\bibitem[{{Tsuji}(2002)}]{2002ApJ...575..264T}
{Tsuji}, T. 2002, \apj, 575, 264

\bibitem[{{Tsuji}(2005)}]{2005ApJ...621.1033T}
---. 2005, \apj, 621, 1033

\bibitem[{{Tsuji} \& {Nakajima}(2003)}]{2003ApJ...585L.151T}
{Tsuji}, T., \& {Nakajima}, T. 2003, \apjl, 585, L151

\bibitem[{{Tsuji} {et~al.}(2004){Tsuji}, {Nakajima}, \&
  {Yanagisawa}}]{2004ApJ...607..511T}
{Tsuji}, T., {Nakajima}, T., \& {Yanagisawa}, K. 2004, \apj, 607, 511

\bibitem[{{Tsuji} {et~al.}(1996){Tsuji}, {Ohnaka}, \&
  {Aoki}}]{1996A&A...305L...1T}
{Tsuji}, T., {Ohnaka}, K., \& {Aoki}, W. 1996, \aap, 305, L1+

\bibitem[{{Tsuji} {et~al.}(1999){Tsuji}, {Ohnaka}, \&
  {Aoki}}]{1999ApJ...520L.119T}
---. 1999, \apjl, 520, L119

\bibitem[{{Tsvetanov} {et~al.}(2000)}]{2000ApJ...531L..61T}
{Tsvetanov}, Z.~I., {et~al.} 2000, \apjl, 531, L61

\bibitem[{{Vacca} {et~al.}(2003){Vacca}, {Cushing}, \&
  {Rayner}}]{2003PASP..115..389V}
{Vacca}, W.~D., {Cushing}, M.~C., \& {Rayner}, J.~T. 2003, \pasp, 115, 389

\bibitem[{{Vrba} {et~al.}(2004)}]{2004AJ....127.2948V}
{Vrba}, F.~J., {et~al.} 2004, \aj, 127, 2948

\bibitem[{{Wilson} {et~al.}(2003){Wilson}, {Miller}, {Gizis}, {Skrutskie},
  {Houck}, {Kirkpatrick}, {Burgasser}, \& {Monet}}]{2003IAUS..211..197W}
{Wilson}, J.~C., {Miller}, N.~A., {Gizis}, J.~E., {Skrutskie}, M.~F., {Houck},
  J.~R., {Kirkpatrick}, J.~D., {Burgasser}, A.~J., \& {Monet}, D.~G. 2003, in
  IAU Symposium, Vol. 211, Brown Dwarfs, ed. E.~{Mart{\'{\i}}n}, 197--+

\bibitem[{{York} {et~al.}(2000)}]{2000AJ....120.1579Y}
{York}, D.~G., {et~al.} 2000, \aj, 120, 1579

\bibitem[{{Zapatero Osorio} {et~al.}(2004){Zapatero Osorio}, {Lane},
  {Pavlenko}, {Mart{\'{\i}}n}, {Britton}, \& {Kulkarni}}]{2004ApJ...615..958Z}
{Zapatero Osorio}, M.~R., {Lane}, B.~F., {Pavlenko}, Y., {Mart{\'{\i}}n},
  E.~L., {Britton}, M., \& {Kulkarni}, S.~R. 2004, \apj, 615, 958

\end{thebibliography}
\end{document}